\documentclass[10pt,floatfix,twocolumn,superscriptaddress,notitlepage,nofootinbib,nobibnotes,amssymb,amsmath,longbibliography]{revtex4-1}
\usepackage[table, dvipsnames]{xcolor}
\usepackage{slashed}
\usepackage{hyperref}
\usepackage{mathtools}
\usepackage{bm}
\usepackage{graphicx,graphics}
\graphicspath{ {./figures/} }
\usepackage[utf8]{inputenc}
\definecolor{lightblue}{RGB}{204,204,255}
\definecolor{verylightblue}{RGB}{235,235,255}
\definecolor{lightred}{RGB}{255,204,204}
\definecolor{verylightred}{RGB}{255,235,235}
\newcommand{\beqn}{\begin{eqnarray}}
\newcommand{\eeqn}{\end{eqnarray}}
\newcommand{\eq}[1]{(\ref{#1})}

\newcommand{\fl}{{\mathrm{fl}}}
\newcommand{\ch}{{\mathrm{ch}}}
\newcommand{\conf}{{\mathrm{conf}}}

\newcommand{\RE}{{\mathrm{Re}\,}}

\newcommand{\cZ}{{\cal Z}}

\newcommand{\Tr}{{\mathrm{Tr}\,}}

\newcommand{\Z}{{\mathbb Z}}
\newcommand{\R}{{\mathbb R}}

\newcommand{\bs}{\boldsymbol}

\newcommand{\avr}[1]{{\left\langle #1 \right\rangle}}
\newcommand{\bsi}{\bar{\psi}}

\newcommand{\bum}{$\blacktriangleright$}

\hypersetup{
    unicode=false,          
    pdftoolbar=true,        
    pdfmenubar=true,        
    pdffitwindow=false,     
    pdfstartview={FitH},    
    pdftitle={My title},    
    pdfauthor={Author},     
    pdfsubject={Subject},   
    pdfcreator={Creator},   
    pdfproducer={Producer}, 
    pdfkeywords={keyword1, key2, key3}, 
    pdfnewwindow=true,      
    colorlinks=true,       
    linkcolor=black,          
    citecolor=black,        
    filecolor=magenta,      
    urlcolor=blue           
}

\usepackage{color}
\definecolor{purple}{rgb}{0.8,0,0.6}

\begin{document}

\title{Finite-density QCD transition in magnetic field background}

\author{V. V.~Braguta}
\email[]{braguta@itep.ru}
\affiliation{Moscow Institute of Physics and Technology, Dolgoprudny, 141700 Russia}
\affiliation{Institute for Theoretical and Experimental Physics NRC ``Kurchatov Institute'', Moscow, 117218 Russia} 
\affiliation{Laboratory of Physics of Living Matter, Far Eastern Federal University, Sukhanova 8, Vladivostok, 690950, Russia}
\affiliation{Bogoliubov Laboratory of Theoretical Physics, Joint Institute for Nuclear Research, Dubna, 141980 Russia}

\author{M. N. Chernodub}
\email[]{maxim.chernodub@idpoisson.fr}
\affiliation{Laboratory of Physics of Living Matter, Far Eastern Federal University, Sukhanova 8, Vladivostok, 690950, Russia}
\affiliation{Institut Denis Poisson, CNRS UMR 7013, Universit\'e de Tours -- Universit\'e d'Orl\'eans, Tours 37200 France}

\author{A. Yu.~Kotov}
\email[]{andrey.kotov@phystech.edu}
\affiliation{Moscow Institute of Physics and Technology, Dolgoprudny, 141700 Russia}
\affiliation{Institute for Theoretical and Experimental Physics NRC ``Kurchatov Institute'', Moscow, 117218 Russia} 
\affiliation{Bogoliubov Laboratory of Theoretical Physics, Joint Institute for Nuclear Research, Dubna, 141980 Russia}  

\author{A. V. Molochkov}
\email[]{amurg@mail.ru}
\affiliation{Laboratory of Physics of Living Matter, Far Eastern Federal University, Sukhanova 8, Vladivostok, 690950, Russia}

\author{A. A.~Nikolaev}
\email[]{aleksandr.nikolaev@swansea.ac.uk}
\affiliation{Department of Physics, College of Science, Swansea University, Swansea SA2 8PP, United Kingdom}

\date{September 20, 2019}

\begin{abstract}
Using numerical simulations of lattice QCD with physical quark masses, we reveal the influence of magnetic-field background on chiral and deconfinement crossovers in finite-temperature QCD at low baryonic density. In the absence of thermodynamic singularity, we identify these transitions with inflection points of the approximate order parameters: normalized light-quark condensate and renormalized Polyakov loop, respectively. We show that the quadratic curvature of the chiral transition temperature in the ``temperature--chemical potential'' plane depends rather weakly on the strength of the background magnetic field. At weak magnetic fields, the thermal width of the chiral crossover gets narrower as the density of the baryon matter increases, possibly indicating a proximity to a real thermodynamic phase transition. Remarkably, the curvature of the chiral thermal width flips its sign at $eB_{\fl} \simeq 0.6\,\mathrm{GeV}^2$, so that above the flipping point $B > B_{\fl}$, the chiral width gets wider as the baryon density increases. Approximately at the same strength of magnetic field, the chiral and deconfining crossovers merge together at $T \approx 140\,\mathrm{MeV}$. The phase diagram in the parameter space ``temperature-chemical potential-magnetic field'' is outlined, and single-quark entropy and single-quark magnetization are explored. The curvature of the chiral thermal width allows us to estimate an approximate position of the chiral critical endpoint at zero magnetic field: $(T_c^{\text{CEP}}, \mu_B^{\text{CEP}})= (100(25)\, \text{MeV},\ 800(140)\,\text{MeV})$.
\end{abstract}

\maketitle

\section{Introduction}

Strongly interacting fundamental particles, quarks and gluons, form a plasma state at sufficiently high temperature. The quark-gluon plasma (QGP), which existed at certain stage of the evolution of the early Universe, may also be created in relativistic heavy-ion collisions. The QGP has been studied at Relativistic Heavy Ion Collider (RHIC) at Brookhaven National Laboratory, at the Large Hadron Collider (LHC) at CERN, and will also be subjected to further investigation at Nuclotron Ion Collider fAcility (NICA) at JINR in Dubna, and the Facility for Antiproton and Ion Research (FAIR) in Darmstadt ~\cite{Busza:2018rrf}.

These experiments offer a unique tool to investigate the QCD phase diagram in a range of increasing baryon densities. A collision of heavy ions creates a QGP fireball which expands, locally thermalizes, cools down, passes through the confining/chiral QCD transition and then (re)hadronizes into final-state colorless states, hadrons. Noncentral collisions also generate a very strong magnetic field which may affect, at least at the early stages, the evolution of the QGP fireball ~\cite{Heinz:2004qz}. Despite that the whole process evolves in an out-of-equilibrium regime, certain features of the expanding QGP at zero or sufficiently low baryon density can be determined by its properties in the thermodynamic equilibrium, which are accessible in numerical lattice simulations of QCD. 

At vanishing magnetic field and zero baryon density, the equilibrium QCD experiences a broad crossover transition~\cite{Aoki:2006we} which incorporates the transitions associated with the restoration of the chiral symmetry and the loss of the color confinement in the high-temperature regime. 

The crossover has a noncritical character, with both phases being analytically connected. The chiral and deconfining transitions need not to happen precisely at the same point. Moreover, due to the non-singular nature of the crossover, the concrete value of the ``pseudo-critical'' temperature depends on the operator which is used to define it. The most recent studies indicate that the chiral crossover transition, determined via the inflection point of the light-quark chiral condensate, takes place at $T^{\mathrm{ch}}_c = 156.5(1.5) \,\mathrm{MeV}$~\cite{Bazavov:2018mes}. The deconfinement transition, identified as the inflection point of the Polyakov loop, appears at substantially higher temperature value, $T^\conf_c = 171(3) \,\mathrm{MeV}$~\cite{Aoki:2006we}. Alternatively, one may also use the susceptibilities of these order parameters which would give slightly different crossover transitions even in the thermodynamic limit. 

Among many possible options, we define the pseudo-critical temperatures of the chiral and deconfining transitions via the inflection points of the light-quark chiral condensate and the Polyakov loop, respectively. These quantities are the order parameters of QCD with quarks of zero masses (the chiral limit of QCD) and with quarks of infinite masses (the pure Yang-Mills theory), where the associated symmetries are not broken explicitly. 

Due to the analyticity of the transition, the continuity arguments suggest that the pseudo-critical nature of the transition persists in a low-density region at small values of the baryon chemical potential~$\mu_B$. Thus, at sufficiently low baryon density, the transition temperature may be expanded over even powers of $\mu_B$:
\beqn
T_c(\mu_B) = T_c(0) - A_2 \mu_B^2 + A_4 \mu_B^4 + O(\mu_B^6),
\label{eq:Tc}
\eeqn
where $A_2$ and $A_4$ are the first two curvature coefficients of the pseudo-critical transition line. The general form of the polynomial~\eq{eq:Tc} is supported by the analyticity arguments at $\mu_B = 0$ along with the invariance of thermodynamic properties of the system under the charge reflection, $\mu_B \to - \mu_B$: due to charge conjugation symmetry, the transition temperature of an equilibrium QGP is an even function of the baryon chemical potential~$\mu_B$. 

Lattice simulations of the $T$--$\mu_B$ phase diagram give the first-principles determination of the transition line~\eq{eq:Tc}, which may be confronted with the results of the heavy-ion experiments on the chemical freeze--out line. The freeze-out line corresponds to another curve in the $T$--$\mu_B$ plane at which the hadron abundances, that encode the chemical composition of the expanding plasma, get stabilized and thus leave an imprint in the experimentally measured hadronic spectra. It is expected that the chemical freeze--out of the expanding quark-gluon plasma takes place right after the completion of the (re)hadronization process, so that the chemical freeze-out temperatures of a generic QGP fireball lies below the pseudo-critical temperature curve~\eq{eq:Tc}. The observed momenta of hadrons provide more details on the thermal freeze-out stage that happen at later stages after the chemical freeze-out~\cite{Bzdak:2019pkr}. The chemical freeze-out temperature may well be described by a polynomial fit similar to the crossover temperature~\eq{eq:Tc} \cite{Cleymans:2005xv}.

We study hot strongly interacting matter at low baryo\-nic density subjected to a classical strong magnetic field background. These environmental parameters match the quark-gluon plasma created in the noncentral collisions at the LHC. Due to computational constraints, we do not consider inhomogeneous effects of the high vorticity which is an inevitable feature of plasma created in noncentral collisions with large initial angular momentum~\cite{STAR:2017ckg}.

In the first--principles lattice simulations, the effects of the strong magnetic field (${\bs B} \neq 0$), low baryonic densities ($\mu \neq 0$) and high temperatures ($T \sim T_c$) were studied, so far, in different combinations. At zero magnetic field, the presence of the baryonic matter lowers the pseudo-critical temperature of the QCD crossover transition in the region of low baryon densities. This property is rigidly established in numerical simulations of lattice QCD with imaginary baryonic chemical potential $\mu_I \equiv i \mu_B$~\cite{Nagata:2011yf,Bonati:2014rfa,Bazavov:2018mes} and is also well understood in effective modes of nonperturbative QCD~\cite{Klevansky:1992qe, Andreichikov:2017ncy, Abramchuk:2019lso}. A review of recent lattice results may be found in Ref.~\cite{DElia:2018fjp}.

At zero baryonic density, the strengthening of the magnetic-field background leads to a smooth decrease of the QCD transition temperature~\cite{Bali:2011qj}. This phenomenon, known as the inverse magnetic catalysis, is not well understood.\footnote{A difficulty of the theoretical description of the inverse magnetic catalysis, observed at low quark masses, exhibits itself in the very fact that a set of standard effective models predict exactly the opposite phenomenon~\cite{Mizher:2010zb,Shovkovy:2012zn,Andersen:2014xxa}, the usual magnetic catalysis, provided the parameters of the mo\-dels are not fine-tuned to specific functions of the magnetic field.} The strength of the thermodynamic crossover transition was found to increase with the magnetic-field background, possibly indicating the existence of a magnetic-field induced phase transition endpoint at zero baryon density~\cite{Endrodi:2015oba}. Notice that the effect of the background magnetic field on transition temperature depends on the masses of the dynamical quarks: at relatively large (unphysical) quark masses, the magnetic catalysis phenomenon is observed: the transition temperature slightly raises with the strength of the magnetic field~\cite{DElia:2010abb,Ilgenfritz:2012fw}\footnote{We would like to mention that QCD properties in external magnetic field 
were studied in different effective models (see, for instance, works \cite{Galilo:2011nh, Orlovsky:2013aya, Andreichikov:2017dap}).}. In this article, we consider QCD with physical quark masses. 

Thus, both baryonic density ($\mu_B \neq 0$) and the magnetic field background (${\bs B} \neq 0$), considered separately, force the temperature of the crossover transition $T_c$ to drop. Hence, it would be natural to expect that the combined effect of both these factors, $\mu_B$ and ${\bs B}$, should enhance each other and lead to a much stronger decrease of the crossover temperature.

One of the results of our article is that we confirm the mentioned qualitative expectations. We will also show that the magnetic field affects the magnitude of the leading curvature $A_2 = A_2(B)$ of the transition temperature $T_c = T_c(\mu_B,B)$, Eq.~\eq{eq:Tc}. However, we will see that the combined effect of the magnetic field and the baryon density leads to unexpected effects such as strengthening (weakening) of the finite-temperature chiral crossover transition at low (high) magnetic field, with the change of the regime at the magnetic strength of the order of the (vacuum) mass rho-meson squared. An interplay of the wide deconfinement crossover and the narrow chiral crossover is discussed in details. The single-quark magnetization is studied for the first time.

The structure of the paper is as follows. In Section~\ref{sec:simulation} we discuss particularities of the lattice model and describe technical details of our numerical simulations, which were performed on $N_t = 6,8$ lattices generated with a Symanzik improved gluons and stout-improved 2+1 flavor staggered fermions at imaginary baryonic chemical potential with subsequent analytical continuation. The properties of the chiral and deconfinement crossovers, uncovered via the chiral condensate of light quarks and the Polyakov loop, are presented, respectively, in Sect.~\ref{sec:chiral:crossover} and Sect.~\ref{sec:deconfinement}. We discuss the pseudo-critical temperatures and the thermal widths of both transitions, as well as the effects of the magnetic field and the imaginary chemical potential on these quantities. In Sect.~\ref{sec:entropy} we use the renormalized Polyakov loop to calculate the single-quark entropy and the single-quark magnetization. The differences between the properties of the magnetization of the bulk quarks and the single-quark magnetization are outlined. The last section is devoted to the discussion of the overall picture of the crossover transitions and to conclusions.

\section{Details of numerical simulations}
\label{sec:simulation}

\subsection{Quark densities and chemical potentials}

We consider the lattice QCD with three, $N_f=2+1$, quark flavors: two light, up ($u$) and down ($d$), quarks and one heavier, strange ($s$), quark. The total number of quarks ${\mathcal N}_f = \int d^4 x {\bar \psi}_f \gamma^0 \psi_f$ of the definite flavor $f=u, d, s$ is controlled by the set of the chemical potentials $\mu_f$, via the direct coupling in the density part of the action, $\sum_f \mu_f {\mathcal N}_f$. The conserved quantities -- the baryon number $B$, the electric charge $Q$ and the strangeness $S$ -- are determined by the corresponding chemical potentials $\mu_q$ with $q=B,Q,S$, and are related to the quark numbers as follows:
\beqn
B &=& ({\mathcal N}_u + {\mathcal N}_d + {\mathcal N}_s)/3, \nonumber \\
Q &=& (2\, {\mathcal N}_u - {\mathcal N}_d - {\mathcal N}_s)/3, \\
S &=& - {\mathcal N}_s. \nonumber
\label{eq:BQS}
\eeqn
Each quark, irrespective of its flavor, carries one-third baryonic charge, and their electric charges are $q_u =2/3e$ and $q_d = q_s = - 1/3e$, where $e = |e|$ is the elementary charge.  The strangeness of the $s$ quark is $S = - 1$, while $u$ and $d$ quarks carry zero strangeness.

Comparing the density part of the action in the basis of the quark numbers and in the basis of the conserved charges, $\sum_q \mu_q q = \sum_f \mu_f {\mathcal N}_f$, we find the following relations between all six chemical potentials:
\beqn
\mu_u & = & \mu_B/3 + 2 \mu_Q/3,  \qquad\quad\, \ \ \mu_B = \mu_u + 2 \mu_d, \nonumber \\
\mu_d & = & \mu_B/3 - \mu_Q/3,  \qquad\qquad\ \mu_Q = \mu_u - \mu_d, \\
\mu_s & = & \mu_B/3 - \mu_Q/3 -\mu_S,   \qquad \mu_S = \mu_d - \mu_s .\nonumber
\label{eq:BQS:uds}
\eeqn

In our simulations we take equal potentials for the light quarks and zero chemical potential for the strange quark:
\beqn
\mu_s =0, \qquad\ \mu_u =\mu_d = \mu \equiv \frac{\mu_B}{3}.
\label{eq:mus}
\eeqn
With this setup, the chemical potentials for the light quarks, $\mu$, and for the baryon charge, $\mu_B$, are related via Eqs.~\eq{eq:BQS:uds}, $\mu_B = 3 \mu$. The chemical potential for the electric charge is zero, $\mu_Q = 0$. Controversially, the strange chemical potential takes its value from the one of the light quarks, $\mu_S = \mu$. However, in the absence of a strong electromagnetic background, which would otherwise distinguish between the up and down (strange) quarks due to the difference in their electric charges $q_f$, this choice of the chemical potentials corresponds to near-equal densities of the light-quarks and vanishing strange quark density in the quark-gluon plasma phase. Such quark-gluon plasma should necessarily possess a nonzero (positive) electric charge, but so do the colliding ions. Therefore the choice of the quark content~\eq{eq:mus} is considered to be a natural one, and it is used in many numerical simulations of the quark-gluon plasma~\cite{Kaczmarek:2011zz,Endrodi:2011gv,Bonati:2014rfa}. In any case, the dependence of the curvature of the phase transition on the chemical potential of the relatively heavy $s$ quark is negligible~\cite{Bonati:2015bha,Cea:2014xva}, so that we may safely set the chemical potential of the strange quark $\mu_s$ to zero.

\subsection{Lattice partition function}

In our numerical simulations, we partially follow the numerical setup of Ref.~\cite{Bonati:2014rfa}. We perform lattice simulations of QCD with $N_f=2+1$ flavors in the presence of purely imaginary quark chemical potentials, $\mu_f=i\mu_{f,I},\ \mu_{f,I}\in \R$, with $f=u, d, s$, subjected to a strong magnetic field background. We work with the following Euclidean partition function of the discretized theory:
\beqn
\cZ =\int DU \,e^{- S_{\mathrm{YM}}[U]} \prod_{f=u, d, s} \det{\left({M^{f}_{\mathrm{st}}[u,{U},\mu_{f,I}]}\right)^{\frac{1}{4}}},
\label{eq:Z}
\eeqn
where the functional integration is performed over the SU(3) gauge link fields $U_{x\mu}$ with the tree-level-improved Symanzik action for the gluon fields~\cite{Weisz:1982zw,Curci:1983an}
\beqn
S_{\mathrm{YM}} [U] = - \frac{\beta}{3}\sum_{x, \mu \neq \nu} \left( \frac{5}{6} W^{1 \times 1}_{x; \mu\nu} - \frac{1}{12} W^{1 \times 2}_{x; \mu\nu} \right).
\label{eq:S:YM}
\eeqn
The lattice coupling $\beta$ is related to the continuum gauge coupling $g$ in the standard way, $\beta  = 6/g^2$. The action~\eq{eq:S:YM} is given by the sum over the traces of the flat $n\times m$-sized Wilson lines $W^{n \times m}_{x;\mu\nu} \equiv W^{n \times m}_{x;\mu\nu}[U]$ labelled by the plane vectors $\mu$ and $\nu$, and by the starting point $x$.

The quark degrees of freedom enter the partition function~\eq{eq:Z} via the product of the determinants of the staggered Dirac operators:
\beqn
& & (M^f_{\mathrm{st}}[u,{U},\mu_{f,I}])_{x, y} = a m_f \delta_{x, y} \nonumber \\
& & \hskip 20mm + \sum_{\nu=1}^{4}\frac{\eta_{x; \nu}}{2}
\left[e^{i a \mu_{f,I}\delta_{\nu,4}} u^f_{x,\nu}{ U}^{(2)}_{x; \nu}\delta_{x, y - \hat{\nu}} \right.  \quad\
\label{eq:M:fermion}
\\
& & \hskip 20mm - \left. e^{-i a \mu_{f,I}\delta_{\nu,4}} u^{f,*}_{x - \hat\nu; \nu}{U}^{(2)\dagger}_{x - \hat\nu; \nu}\delta_{x, y + \hat\nu}  \right],
\nonumber
\eeqn
constructed from the two-times stout-smeared links ${U}^{(2)}_{x; \nu} \equiv U^{(2)}_{x; \nu}[{U}]$ following the method of Ref.~\cite{Morningstar:2003gk} with the isotropic smearing parameters $\rho_{\mu\nu}=0.15$ for $\mu\neq \nu$. Here $a = a(\beta)$ is the lattice spacing. The stout smearing improvement is a standard technique used to ameliorate the systematics related to the effects of finite lattice spacing and reduce taste symmetry violations~\cite{Bazavov:2010pg}. Following similar approaches~\cite{Bonati:2014rfa,Bonati:2015bha,Bazavov:2009bb,Kaczmarek:2011zz,Endrodi:2011gv,Borsanyi:2012cr,Cea:2014xva}, we use the rooting procedure in the partition function~\eq{eq:Z} in order to remove a residual, fourth degeneracy of the lattice Dirac operator~\eq{eq:M:fermion}.

The Dirac operator~\eq{eq:M:fermion} corresponds to the quarks with the imaginary chemical potential $\mu_{f,I}$ subjected to the magnetic field background $B$. The chemical potential enters the Dirac operator~\eq{eq:M:fermion} via the additional phases $e^{+ i a \mu_{f,I}}$ and $e^{- i a \mu_{f,I}}$ associated with the temporal links in, respectively, forward and backward directions. The magnetic field appears in the quark operator~\eq{eq:M:fermion} of the $f$-th flavor via the composite link field, ${\widetilde U}^f_{x,\mu} = u^f_{x,\mu} \cdot U^{(2)}_{x,\mu}$, where $U^{(2)}_{x,\mu}$ is the usual (stout-smeared) SU(3) gauge field while the $u^f_{x,\mu}$ prefactor represents the classical $U(1)$ gauge field corresponding to the uniform magnetic-field background. We consider the classical magnetic background so that the kinetic term of the Abelian field $u^f_{x,\mu}$ is absent.

In a finite volume with periodic boundary conditions, the total magnetic flux through any lattice plane must be an integer number in units of the elementary magnetic flux~\cite{tHooft:1979rtg,AlHashimi:2008hr}. For our lattice geometry $N_s^3 \times N_t$, this property leads to quantization of the strength of the uniform magnetic field $B$, acting on the quarks of the $f$-th flavor:
\beqn 
B = \frac{1}{q_f} \frac{2\pi n}{N_s^2 a^2}.
\label{eq:B:quant:pre} 
\eeqn 
Here the integer quantity $n \in \Z$ counts the number of total magnetic fluxes. Given the fact that the quark electric charges are not the same, one takes the minimal charge, $q_f \equiv |q_d| = e/3$, so that the quantization~\eq{eq:B:quant}  gives a consistent field for all three quarks:
\beqn 
e B = \frac{6\pi n}{N_s^2 a^2}, \qquad n \in \Z, \qquad   0 \leqslant n \leqslant N_s^2.
\label{eq:B:quant} 
\eeqn

For the uniform magnetic field $B_i = \delta_{i3} B$ directed along the third axis, the Abelian link field $u^f_{x,\mu} \equiv u^f_\mu(x)$, acting on the quark of the flavor $f$, may be chosen in the following explicit form~\cite{Ilgenfritz:2012fw}:
\begin{equation}
\begin{split}
& u^f_1(x_1, x_2, x_3, x_4) = e^{-i a^2 q_f B x_2 / 2}, \quad\ x_1\ne N_s-1, \\
& u^f_1(N_s-1, x_2, x_3, x_4) = e^{-i a^2 q_f B (N_s+1) x_2/2}, \\
& u^f_2(x_1, x_2, x_3, x_4) = e^{i a^2 q_f B x_1 /2}, \quad\ x_2\ne N_s-1, \\
& u^f_2(x_1, N_s-1, x_3, x_4) = e^{i a^2 q_f B (N_s+1) x_1/2} \\
& u^f_3(x) = u^f_4(x) = 1, \\
\end{split}
\label{eq:u:links}
\end{equation}
where $x\equiv(x_1,x_2,x_3,x_4)$ is the four-coordinate with the elements running through $x_\nu=0\ldots N_s-1$. The magnetic field $B$ is given by Eq.~\eq{eq:B:quant}.

Due to the periodic structure of the Abelian field ~\eq{eq:u:links}, the magnetic field cannot be larger then maximal value, determined by the flux number $n_{\mathrm{max}} = \lfloor N_s^2/2 \rfloor$, where $\lfloor x \rfloor$ gives the greatest integer less than or equal to~$x$. Thus, the nonzero lattice magnetic field $B$ may only be imposed in the range $ 6 \pi/(e N_s^2 a^2) \leqslant B \lesssim 3 \pi/(e a^2)$, where the strongest value of the field may lead to strong ultraviolet artifacts. To avoid these discretization artifacts, we take $n \ll N_s^2/2$ in our numerical simulations.

\subsection{Observables}

\subsubsection{Chiral sector}

The chiral condensate $\avr{\bsi\psi}$ is the most straightforward characteristic of the dynamical chiral symmetry breaking in the system of fermions $\psi$. The condensate vanishes in the phase with unbroken chiral symmetry, $\psi \to e^{i \gamma_5 \omega} \psi$ and $\bsi \to \bsi e^{i \gamma_5 \omega}$,  while its deviation from zero signals the violation of the chiral symmetry.  The chiral condensate corresponds to an order parameter of the spontaneous chiral symmetry breaking of massless fermions, for which the group of chiral transformations is an exact symmetry group of the classical Lagrangian.

In QCD, the nonzero masses of quarks, $m_f \neq 0$, break the chiral symmetry explicitly, in all phases. Therefore, the chiral condensate, in a strict mathematical sense, is not an order parameter. However, the condensate of light up and down quarks, with masses well below the characteristic QCD energy scale $m_u \sim m_d \ll \Lambda_{\mathrm{QCD}}$ may still serve as an approximate order parameter and thus effectively probe the chiral dynamics. 

The chiral condensate of the quark flavor $f$ is given by the partial derivative of the partition function~\eq{eq:Z} with respect to the quark's mass:
\beqn
\avr{\bsi\psi}_f = \frac{T}{V}\frac{\partial \log \cZ}{\partial m_f},
\eeqn
where $V$ is the spatial volume of the system. 

In our $N_f=2+1$ simulations the masses of the light $u$ and $d$ quarks are degenerate, $m_l\equiv m_u=m_d$. Therefore, it is convenient to introduce the common light quark condensate given by the sum:
\beqn
\avr{\bsi\psi}_l=\frac{T}{V}\frac{\partial \log \cZ}{\partial m_l} = \langle\bar{u}u\rangle+\langle\bar{d}d\rangle.
\label{eq:condensate:light}
\eeqn

The chiral condensate of $f$-th flavor,
\beqn
\avr{\bsi_f\psi_f} = \frac{T}{4V} \avr{\Tr M_f^{-1}},
\eeqn
is evaluated as the trace over the negative power of the Dirac operator~\eq{eq:M:fermion}. Numerically, this calculation is performed with the help of the noisy estimators which comprise $O(10)$ random vectors for each fixed flavor.

The finite-temperature renormalization of the light-quark condensate~\eq{eq:condensate:light} in the presence of the condensate $\avr{\bar{s}s}$ of the third, heavier quark $s$, is implemented following the prescription of Ref.~\cite{Cheng:2007jq}:
\beqn
\avr{\bsi\psi}^r_l (B,T,\mu_I) \equiv \frac{\left[
\avr{\bsi\psi}_l - 2 \frac{m_{l}}{m_s} \avr{\bar{s}s}\right](B,T,\mu_I)}{
\left[\avr{\bsi\psi}_l - 2 \frac{m_{l}}{ m_s}\avr{\bar{s}s}\right](0,0,0)}, \qquad
\label{eq:condensate:renorm}
\eeqn
where $m_s$ is the bare mass of the strange quark $s$. 

The condensate entering the denominator of the renormalized condensate~\eq{eq:condensate:renorm} is computed in the vacuum state, i.e. at zero magnetic field $B=0$, zero temperature $T=0$, and zero (imaginary) chemical potential $\mu_I=0$. We took the data for this quantity from (interpolated, when needed) results of Ref.~\cite{Bonati:2014rfa}. Other possible renormalization prescriptions may be found in Refs.~\cite{Kaczmarek:2011zz,Endrodi:2011gv}.

\subsubsection{Gluon sector}

The nonperturbative dynamics of the gluon sector gives rise to the confinement of color: the formation of the colorless hadronic states, mesons and baryons, in the low-temperature QCD. At high temperatures, these states melt, and the system enters the quark-gluon plasma phase with unconfined quarks and gluons. The order parameter of the quark confinement is the Polyakov loop, which may suitably be formulated in the Euclidean QCD as follows:
\beqn
P = \frac{1}{V} \sum_{{\bs x}} \frac{1}{3} \Tr \left(  \prod_{x_4 = 0}^{N_t-1} U_{{\bs x}, x_4; 4} \right).
\label{eq:Polyakov:loop}
\eeqn
The Polyakov loop operator is averaged over the spatial volume $V = N_s^3$ with the spatial coordinate ${\bs x}$.

In a purely gluonic Yang-Mills theory, the vacuum expectation value of the Polyakov loop~\eq{eq:Polyakov:loop} vanishes in the confining, low-temperature phase, and differs from zero in the high-temperature phase that corresponds to the quark-gluon plasma regime. In a purely gluonic theory, the Polyakov loop~\eq{eq:Polyakov:loop} is an exact order parameter associated with the spontaneous breaking of the global $\Z_3$ center symmetry, $P \to Z P$, where $Z = e^{2 \pi n i/3}$, $n=0,1,2$ are the elements of the center subgroup $\Z_3$ of the $SU(3)$ group. In the presence of light dynamical quarks, the Polyakov loop represents an approximate order parameter of the quark confinement.  

For practical reasons of studies of the deconfinement phenomenon, it is convenient to consider the real part of the Polyakov loop:
\beqn
L = \RE P.
\label{eq:Polyakov:loop:Re}
\eeqn

\subsection{Parameters}

We perform numerical simulations at finite temperature around the phase transition using mainly $N_s^3 \times N_t = 24^3 \times 6$ lattice. In order to estimate the magnitude of the lattice artifacts related to the ultraviolet cutoff effects, we also repeated certain runs on another, $32^3 \times 8$ lattice with the same ratio $N_t/N_s = 1/4$. The comparison of the selected set of results with the ones obtained on the third lattice geometry, $32^3 \times 6$, gives us an opportunity to estimate the robustness of our data with respect to the finite-volume effects.

The zero-temperature data, used in the renormalization of the condensate~\eq{eq:condensate:renorm}, were taken from simulations on a $32^4$ lattice of Ref.~\cite{Bonati:2014rfa}. The physical temperature $T=1/(a(\beta)N_t)$ is controlled by the lattice coupling constant $\beta$. The lattice spacing varied from $a = 0.113\,\mathrm{fm}$ at our largest coupling $\beta = 3.7927$ till $a = 0.253\,\mathrm{fm}$ at the lowest coupling $\beta = 3.4949$.

The bare (lattice) masses of the quarks, $m_l$ and $m_s$, are fine-tuned at each value of the lattice coupling $\beta$ in order to keep the pion mass at its physical value, $m_\pi \simeq 135$\,MeV, and maintain, at the same time, the physical ratio of the quark masses, $m_s/m_{l}=28.15$. This line of constant physics is well-known phenomenologically from the numerical simulations of Refs.~\cite{Aoki:2009sc,Borsanyi:2010cj,Borsanyi:2013bia}. 

We simulated the lattice QCD at the physical point at seven values of the background magnetic field in the interval $eB= (0.1- 1.5)\,\mathrm{GeV}^2$. We took eight points of the imaginary chemical potential of the light quarks, $\mu_I \equiv \mu_{l,I}$, in the range from a zero value up to $\mu_I/ (\pi T) = 0.275$. 

\section{Chiral crossover}
\label{sec:chiral:crossover}

\subsection{Chiral condensate and discretization errors}

We have performed the numerical calculations of the chiral condensate at the wide range of the external magnetic fields, $eB/{\mathrm{GeV}}^2 = 0.1, 0.5, 0.6, 0.8, 1.0, 1.5$, and at a dense set of the imaginary chemical potentials $\mu_I/(\pi T) = 0, 0.1, 0.14, 0.17, 0.2, 0.22, 0.24, 0.275$. 

The data for the renormalized chiral condensate may be excellently described by the following function:
\beqn
\avr{\bsi \psi}^r_l (T) = C_0 + C_1 \arctan \frac{T - T_c^\ch}{\delta T_c^\ch},
\label{eq:condensate:fit}
\eeqn
which has also been used to study the condensate at zero magnetic field in Ref.~\cite{Bonati:2014rfa}. The fitting function~\eq{eq:condensate:fit} contains four free parameters: two amplitudes $C_0$ and $C_1$, which describe the scale of the condensate and the degree of its variation over in the crossover region, as well as the pseudo-critical transition temperature $T_c^\ch$ and the width of the crossover $\delta T_c^\ch$. All four fitting parameters in Eq.~\eq{eq:condensate:fit} are the functions of the magnetic field $B$ and the imaginary chemical potential $\mu_I$.

\begin{figure}[!htb]
\begin{center}
\hskip -4mm \includegraphics[scale=0.77,clip=true]{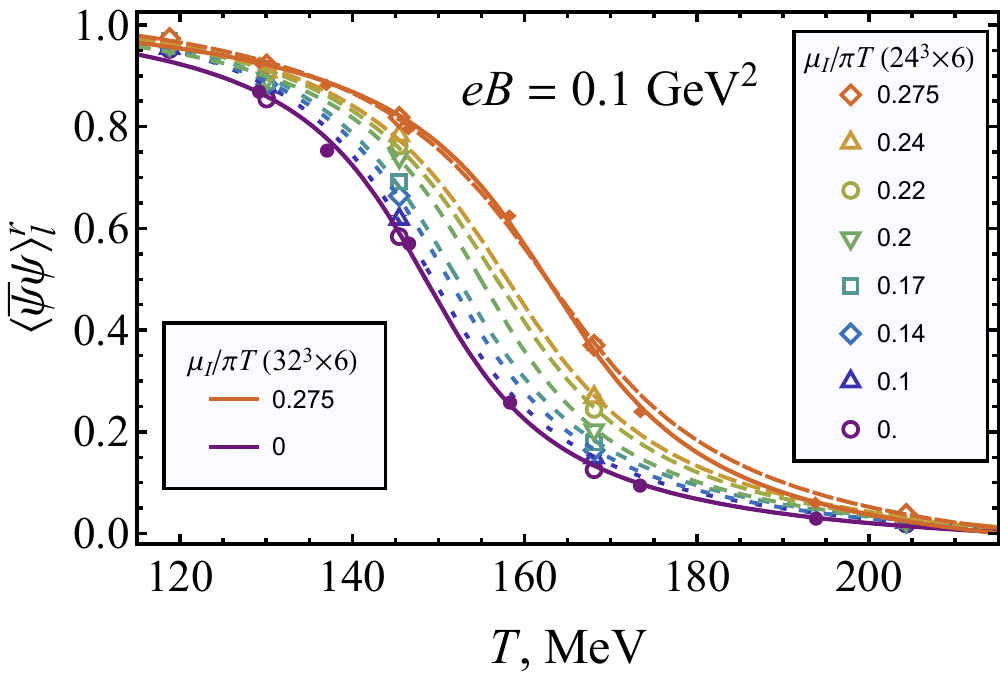} \\[3mm]
\includegraphics[scale=0.50,clip=true]{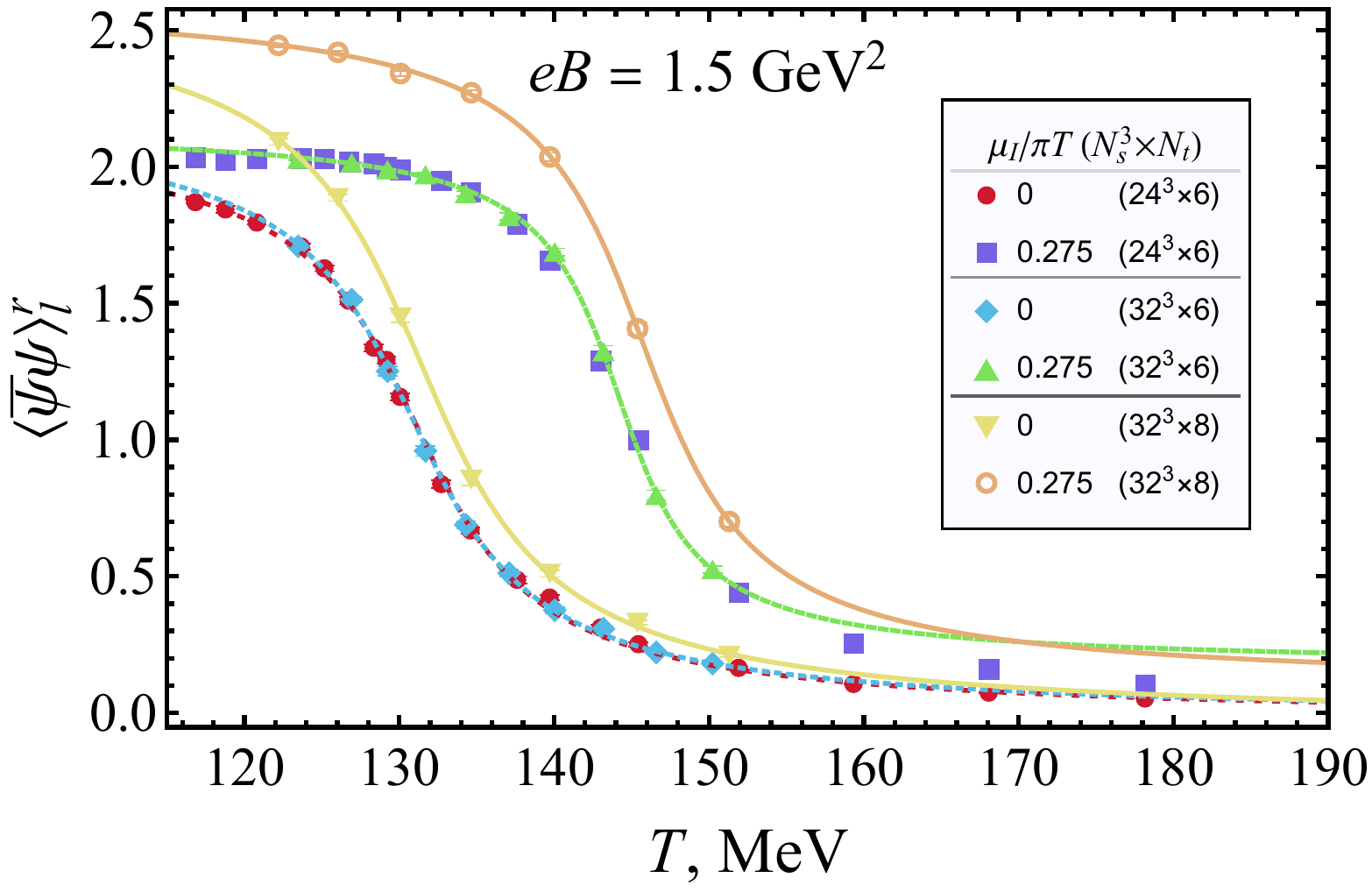} 
\end{center}
\vskip 0mm 
\caption{The light quark condensate as the function of temperature at fixed imaginary chemical potentials $\mu_I$ in the background of the weakest, $eB = 0.1\,\mathrm{GeV}^2$ (the upper plot) and the strongest, $eB = 1.5\,\mathrm{GeV}^2$ (the lower plot) magnetic fields. The lines are the best fits by the function~\eq{eq:condensate:fit}. The condensate for the weakest field is shown for all available values of the imaginary chemical potential $\mu_I$ at $24^3 \times 6$ lattice. The strongest field is represented by the lowest and largest imaginary chemical potentials, $\mu_I/(\pi T) = 0, 0.275$ for $24^3 \times 6$, $32^3 \times 6$ and $32^3 \times 8$ lattices.}
\label{fig:condensate:fits}
\end{figure}

The numerical data for the condensates and their fits are shown in Fig.~\ref{fig:condensate:fits} for a set of imaginary chemical potentials at smallest nonzero and largest values of the magnetic field. On a qualitative level, the data clearly demonstrate the well-known effect of the inverse magnetic catalysis: the stronger the magnetic field $B$ the smaller the chiral crossover temperature $T_c^\ch$. They also show that at fixed magnetic field, the increase of the {\it{imaginary}} chemical potential $\mu_I$ leads, as expected, to increase of the critical crossover temperature. 

Before going to the quantitative description of the main results, we estimate the influence of effects of ultraviolet and infrared artifacts of the lattice discretization. At zero magnetic field, these effects were investigated in Refs.~\cite{Bonati:2014rfa,Bonati:2015bha}, and we extend the study to the case of the strongest magnetic field, $eB=1.5\,\mathrm{GeV}^2$, shown in the bottom plot of Fig.~\ref{fig:condensate:fits} for the lowest and largest available imaginary chemical potentials, $\mu_I/(\pi T) = 0$ and $0.275$. The analysis of lattices with different spatial volumes, $N_s = 24, 32$ at fixed temporal extension $N_t=6$ ensures us that the volume-dependent infrared effects are almost negligible. The inspection of the lattices with $N_s = 24, 32$ and fixed ratio $N_t/N_s = 1/4$ demonstrates that while the ultraviolet discretization effects on the condensate are noticeable, the effect of varying lattice spacing on the transition temperature is rather small.

In order to quantify these assertions, we show in Table~\ref{ref:artifacts} the critical temperature $T_c$ at both vanishing and largest studied chemical potentials $\mu_I$ at lattices of all mentioned geometries. The critical temperature, obtained with the fits~\eq{eq:condensate:fit} shown in the bottom plot of Fig.~\ref{fig:condensate:fits}, indicate that the variations of the chiral crossover temperature are of the order of $1\,\mbox{MeV}$, i.e. less than one percent.

\begin{table}[htb]
\begin{tabular}{|c|c|c|c|c|}
\hline
\multicolumn{5}{|c|}{$eB = 1.5$\,GeV${}^2$} \\
\hline
\ Lattice \   &    $\ \mu_I/\pi T\ $ &           $T^\ch_c$, MeV  & $\delta T^\ch_c$, MeV & $\chi^2/$d.o.f\\
\hline
$24^3 \times 6$ &         0         &   \         130.5(2) & 5.2(4) & 1.2  \\
$32^3 \times 6$ &         0         &             130.8(1) & 5.2(3) & 0.8 \\
$32^3 \times 8$ &         0         &             131.3(1) & 5.7(2)  & 0.3 \\
\hline
$24^3 \times 6$ &        0.275   &              144.5(2) & 4.7(2) & 1.0 \\
$32^3 \times 6$ &        0.275   &              144.7(6) & 4.1(3) & 1.2 \\
$32^3 \times 8$ &        0.275    &             145.6(5) & 4.8(7) & 1.6 \\
\hline
\end{tabular}
\caption{Illustration of finite-size and finite-volume effects on the chiral crossover temperature at strongest studied magnetic field $eB = 1.5\,\mathrm{GeV}^2$, at both vanishing and largest available values of the chemical potential~$\mu_I$.}
\label{ref:artifacts}
\end{table}

We would like to notice that at low (but nonzero) values of the background magnetic fields, there is a particular property of the lattice system which leads to large systematic errors of certain computed quantities. Due to lattice discretization effects, the quantization of the magnetic field~\eq{eq:B:quant} limits the number of temperature points at fixed value of magnetic-field strength. Narrowing the study to the crossover region imposes further restrictions thus reducing the quality of the data. We will see below that the data at low magnetic fields has a tendency to possess larger errors at compared to the data at stronger fields.

\subsection{Chiral crossover temperature and its width}

\subsubsection{General picture}

The quantitative analysis of the fits of the chiral condensate gives us the important information how the chiral crossover temperature evolves with increase of the imaginary chemical potential in the magnetic-field background. While the behaviour of the critical temperature is known both at zero chemical potential $\mu_I = 0$, Ref.~\cite{Bali:2011qj} and at zero magnetic field $B = 0$, Ref.~\cite{Bonati:2014rfa}, the studies in the full $(B, \mu_I)$ plane are performed here for the first time. In addition, we would like to clarify the influence of magnetic field on the thermal crossover width $\delta T^\ch$ in the finite-density QCD. This question is important in view of the fact that the role of the magnetic field on the strength of the QCD (phase) transition even at zero chemical potential, $\mu=0$, has historically been evolving via a set of controversies~\cite{Mizher:2010zb,DElia:2010abb,Bali:2011qj}.

\begin{figure}[!thb]
\begin{center}
\includegraphics[scale=0.45,clip=true]{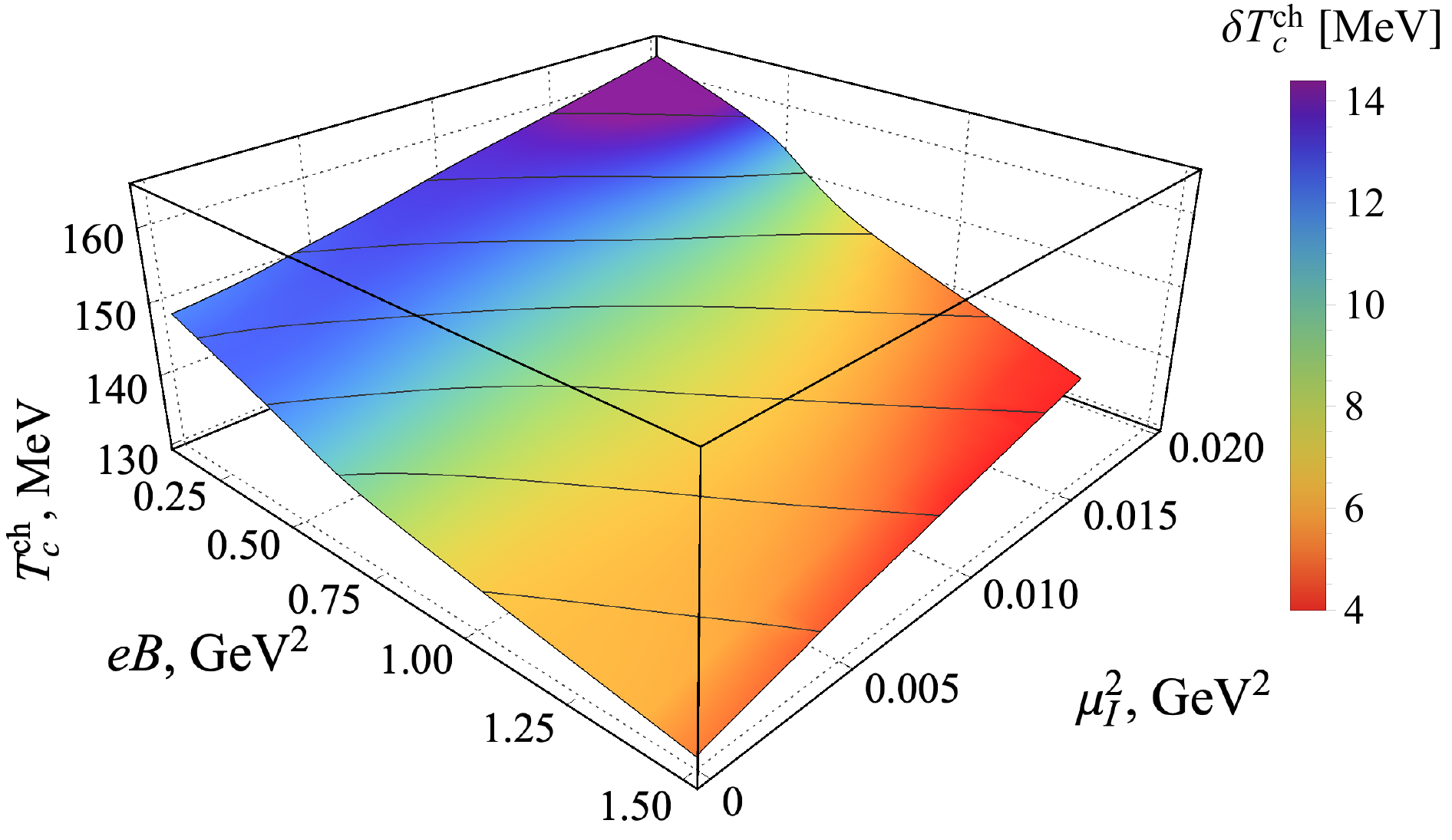}
\end{center}
\vskip 0mm 
\caption{The critical temperature $T^{\ch}_c$ of the chiral crossover as the function of the magnetic-field strength $B$ and the imaginary chemical potential squared $\mu_I^2$. The color encodes the width $\delta T^\ch_c$ of the chiral crossover transition.}
\label{fig:Tc:3d}
\end{figure}

In Fig.~\ref{fig:Tc:3d} we show the spline-interpolated data for the critical temperature of the chiral crossover in the plane of magnetic field $B$ and the squared imaginary chemical potential $\mu_I^2$. One may clearly see that the increase of the imaginary chemical potential, at fixed magnetic field $B$, leads to the enhancement of the critical temperature for all studied values of $B$. On the other hand, the strengthening of the magnetic field at fixed imaginary chemical potential $\mu_I$ gives rise to the decrease of the critical temperature. 

The equitemperature curves in Fig.~\ref{fig:Tc:3d} are close to the straight, almost-parallel lines. These properties indicate, respectively, that at small baryon densities (i) the critical temperature of the chiral crossover $T^\ch_c$ at fixed magnetic field $B$ is a quadratic function of the chemical potential~$\mu_I^2$; (ii) the strength of the quadratic dependence does not depend significantly on the strength of the magnetic field. In terms of the baryonic potential, $\mu_B^2 = - (3\mu_I)^2$, we conclude that the slope $A_2$ of the chiral crossover temperature~\eq{eq:Tc} is a positive nonvanishing quantity which moderately depends on the value of magnetic field.

\begin{figure}[!thb]
\begin{center}
\includegraphics[scale=0.45,clip=true]{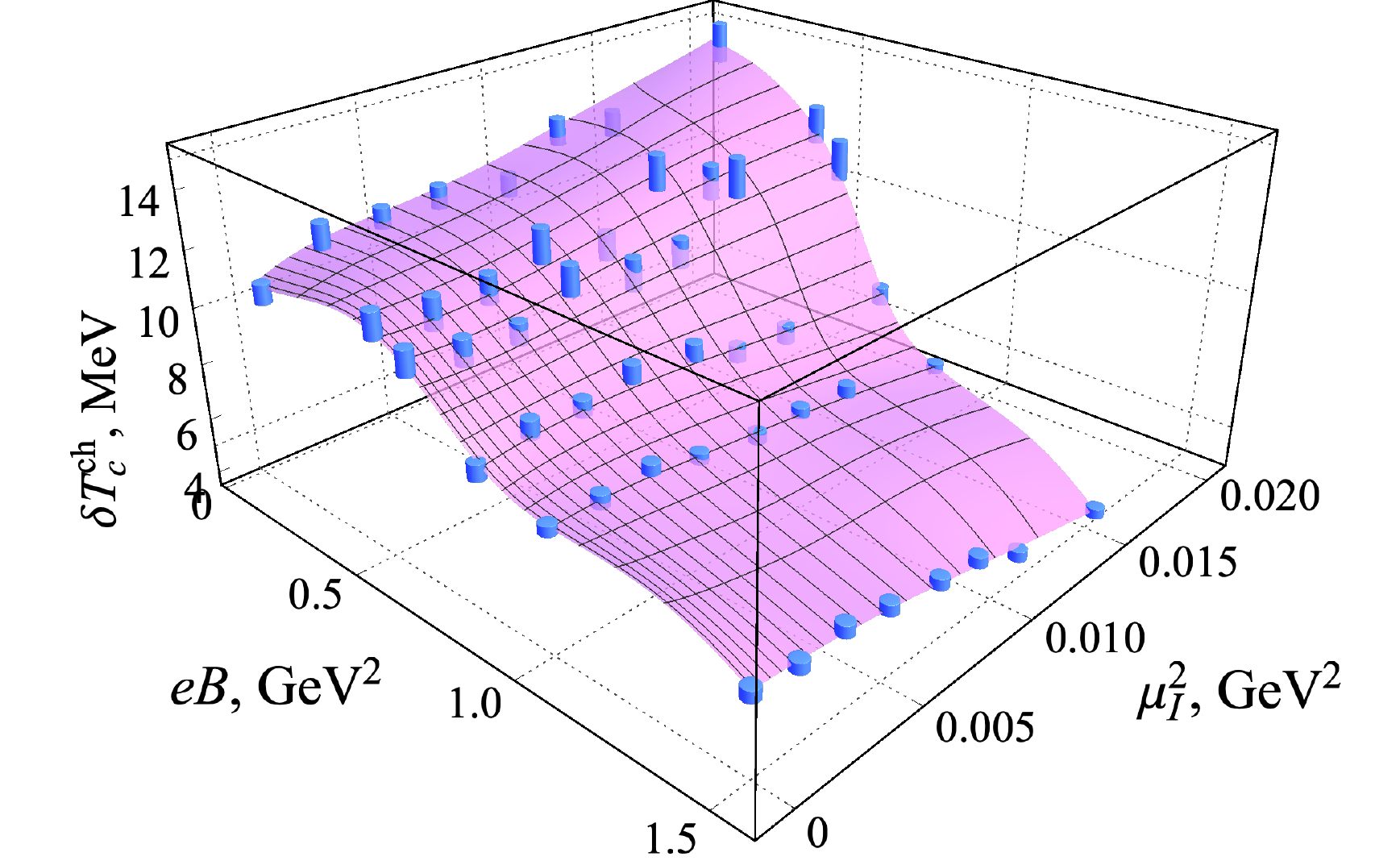}
\end{center}
\vskip 0mm 
\caption{The width of the chiral crossover $\delta T_c^\ch$ as the function of the magnetic field~$B$ and the imaginary chemical potential~$\mu_I$ squared. The height of the cylinders represents the error bars of the data, and the smooth surface corresponds to a spline interpolation.}
\label{fig:Tc:delta}
\end{figure}

The thermal width of the chiral crossover transition (the ``chiral thermal width'') is encoded in the color of the surface in the same Fig.~\ref{fig:Tc:3d}. The chiral width exhibits a weak, but still noticeable, dependence on the imaginary chemical potential. However, the influence of the magnetic field on the chiral thermal width $\delta T^\ch$ is much more pronounced: the stronger magnetic field $B$ the narrower transition. This behaviour is well seen in the spline representation of the thermal width  in Fig.~\ref{fig:Tc:delta}. Interestingly, the magnetic field has a qualitative effect on the behaviour of the chiral width: at weak (strong) magnetic field, the chiral thermal width is an increasing (decreasing) function of the {\it{imaginary}} chemical potential~$\mu_I$.

\subsubsection{Chiral transition temperature and its curvature}

At small values of the imaginary chemical potential $\mu_I$, the behavior of thermodynamic quantities is necessarily analytic in $\mu_I$ due to the absence of a thermodynamic singularity in the vicinity the $\mu_I = 0$ point. The Taylor series of the observable (real-valued) quantities must therefore run over the even powers of the chemical potential, which makes it possible to use the trivial relation between the imaginary and real baryonic chemical potentials, $\mu_I^2 \equiv - (\mu_B/3)^2$. Therefore, the behavior $T_c = T_c(\mu_B,B)$ of the critical crossover temperature~\eq{eq:Tc} of the finite-density QCD may be restored from the series of $T_c(\mu_I,\mu_B)$ at small imaginary chemical potential $\mu_I$:
\beqn
\frac{T_c^\ch(\mu_I,B)}{T_c^\ch(B)} & = & 1 + \kappa_2^\ch(B) \left( \frac{3 \mu_I}{T^\ch_c(B)} \right)^2 \nonumber \\
& & + \kappa_4^\ch(B) \left( \frac{3 \mu_I}{T^\ch_c(B)} \right)^4 + O\left(\frac{\mu_I^6}{T_c^6}\right)\!. \qquad
\label{eq:Tc:I}
\eeqn
where we used the notation $T_c(B) \equiv T_c(\mu_B = 0,B)$.

In analogy with the lattice studies with a vanishing magnetic field, we deduce that at $B > 0$ the curvature $A_2$ of the critical transition~\eq{eq:Tc} at nonzero baryon density $\mu_B$ is related to the dimensionless curvature coefficient~$\kappa_2$ at the imaginary chemical potential~$\mu_I$ in Eq.~\eq{eq:Tc:I} as:
\beqn
A_2^\ch(B) = \frac{\kappa_2^\ch(B)}{T_c^\ch(\mu_B=0,B)}.
\label{eq:A2}
\eeqn
Equations~\eq{eq:Tc}, \eq{eq:Tc:I} and \eq{eq:A2} have rather universal character and can be equally applied to both chiral and deconfining transitions.

\begin{figure}[!thb]
\begin{center}
\includegraphics[scale=0.45,clip=true]{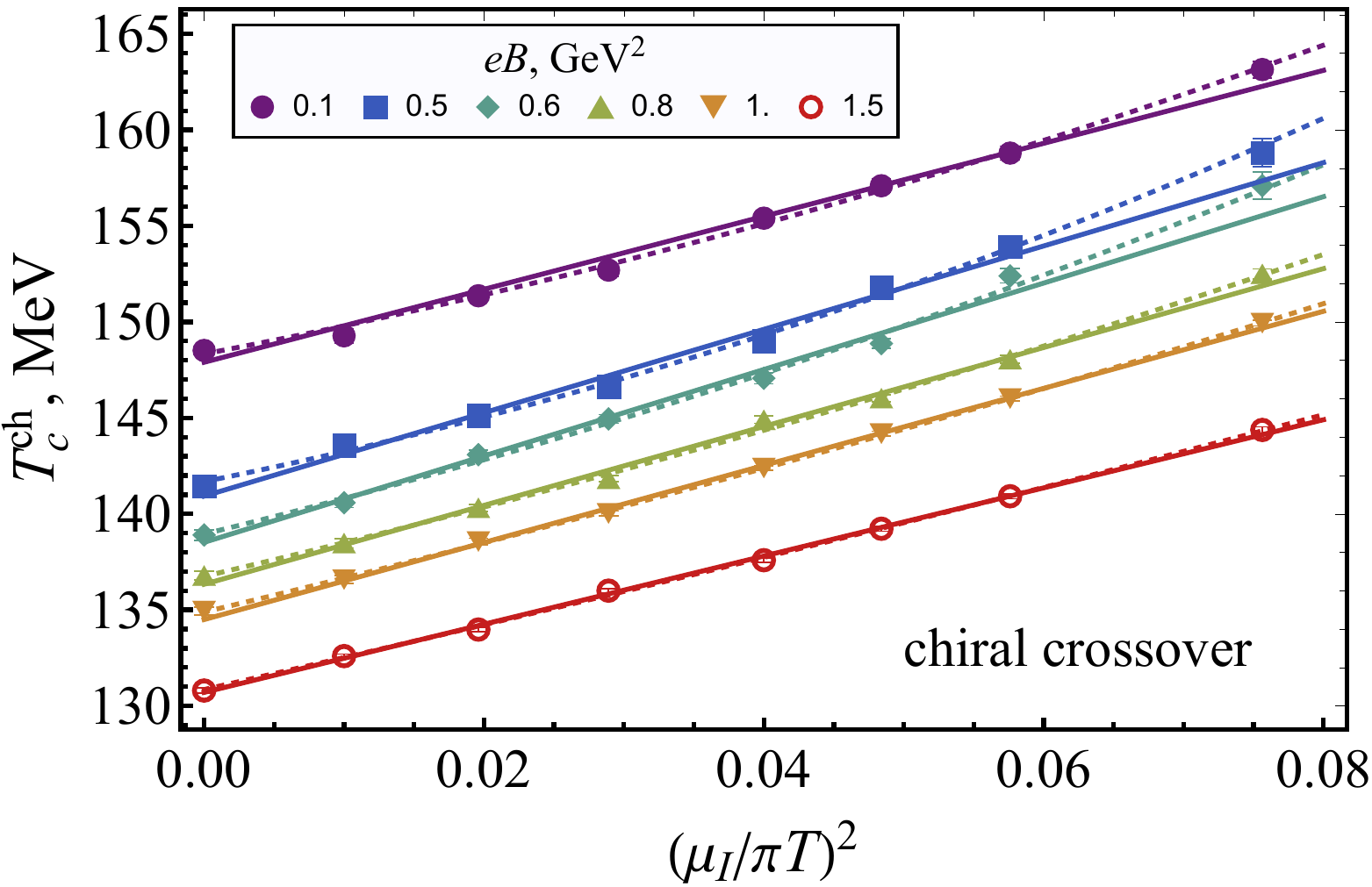}
\end{center}
\vskip 0mm 
\caption{The critical temperature $T_c^\ch$ of the chiral crossover transition as the function of the imaginary chemical potential squared at a set of values of the magnetic field $B$. The translucent (opaque) lines correspond to quadratic (quartic) truncations by the fitting function~\eq{eq:Tc:I}. }
\label{fig:Tc:crossover:fit}
\end{figure}

In Fig.~\ref{fig:Tc:crossover:fit} we show the fits of the critical temperature $T_c = T_c(\mu_B,\mu_I)$ by the polynomial~\eq{eq:Tc:I}. We fix the magnetic field $B$ and consider the critical temperature as a function of the dimensionless ratio $\mu_I/T$. All three fitting parameters $T_c(B)$, $\kappa_{2}(B)$ and $\kappa_{4}(B)$ are treated as functions of the magnetic field $B$. We use both quadratic (with $\kappa_4 \equiv 0$) and quartic (with $\kappa_4$ being a fit parameter) fits. 

\begin{figure*}[!thb]
\begin{center}
\begin{tabular}{ccc}
\includegraphics[scale=0.4,clip=true]{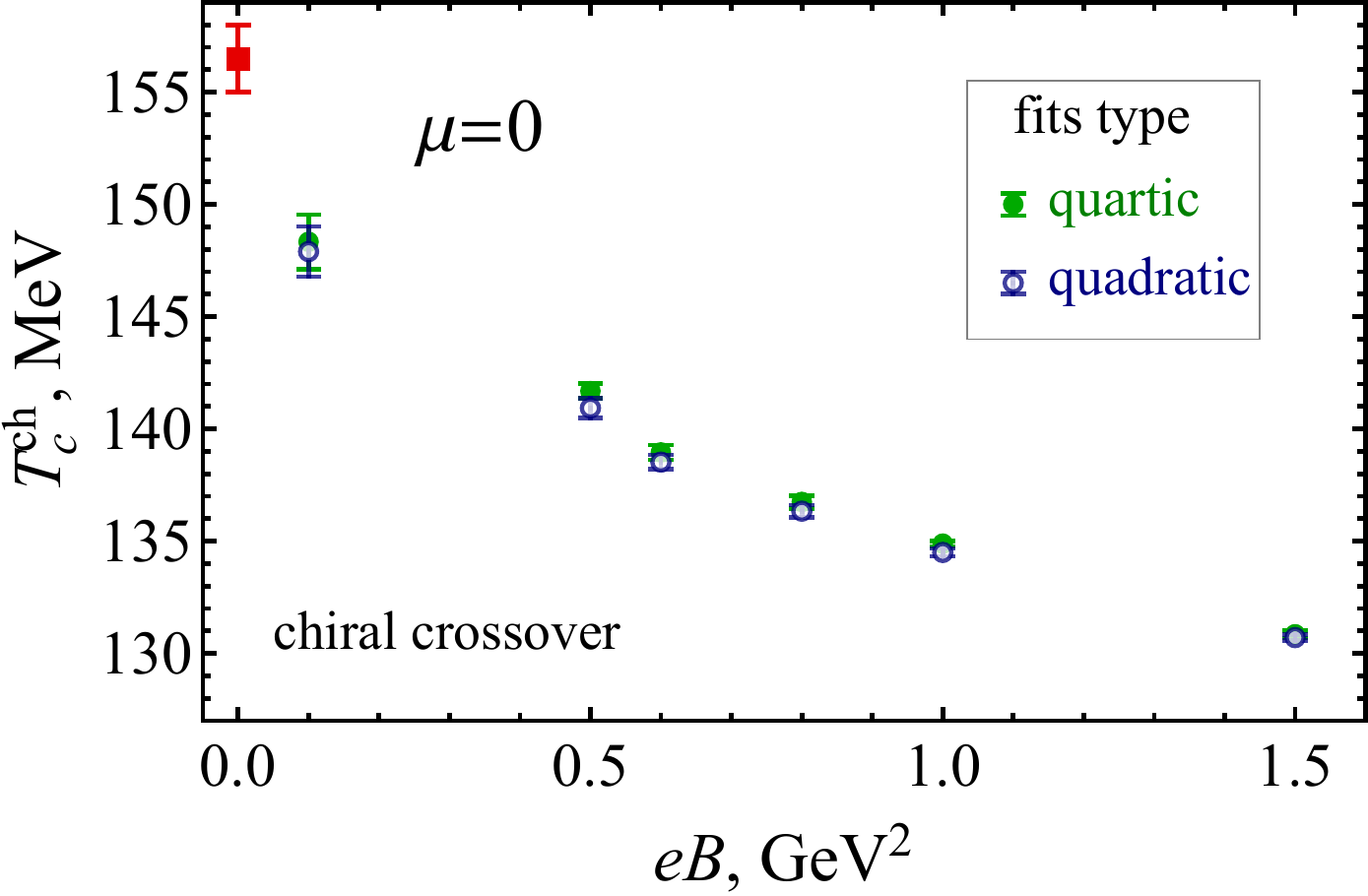} &
\hskip 2mm \includegraphics[scale=0.415,clip=true]{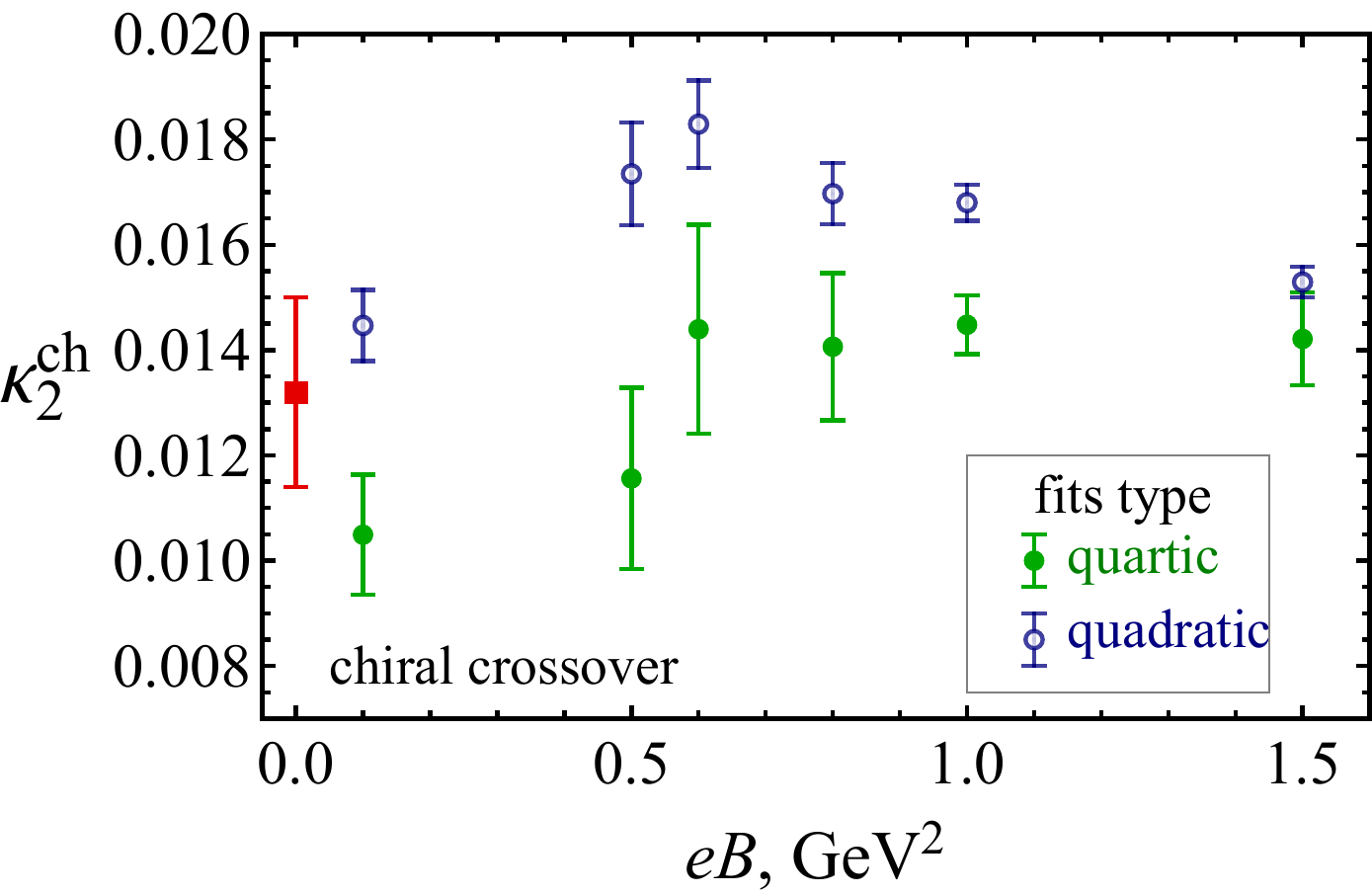} & 
\hskip 2mm \includegraphics[scale=0.4,clip=true]{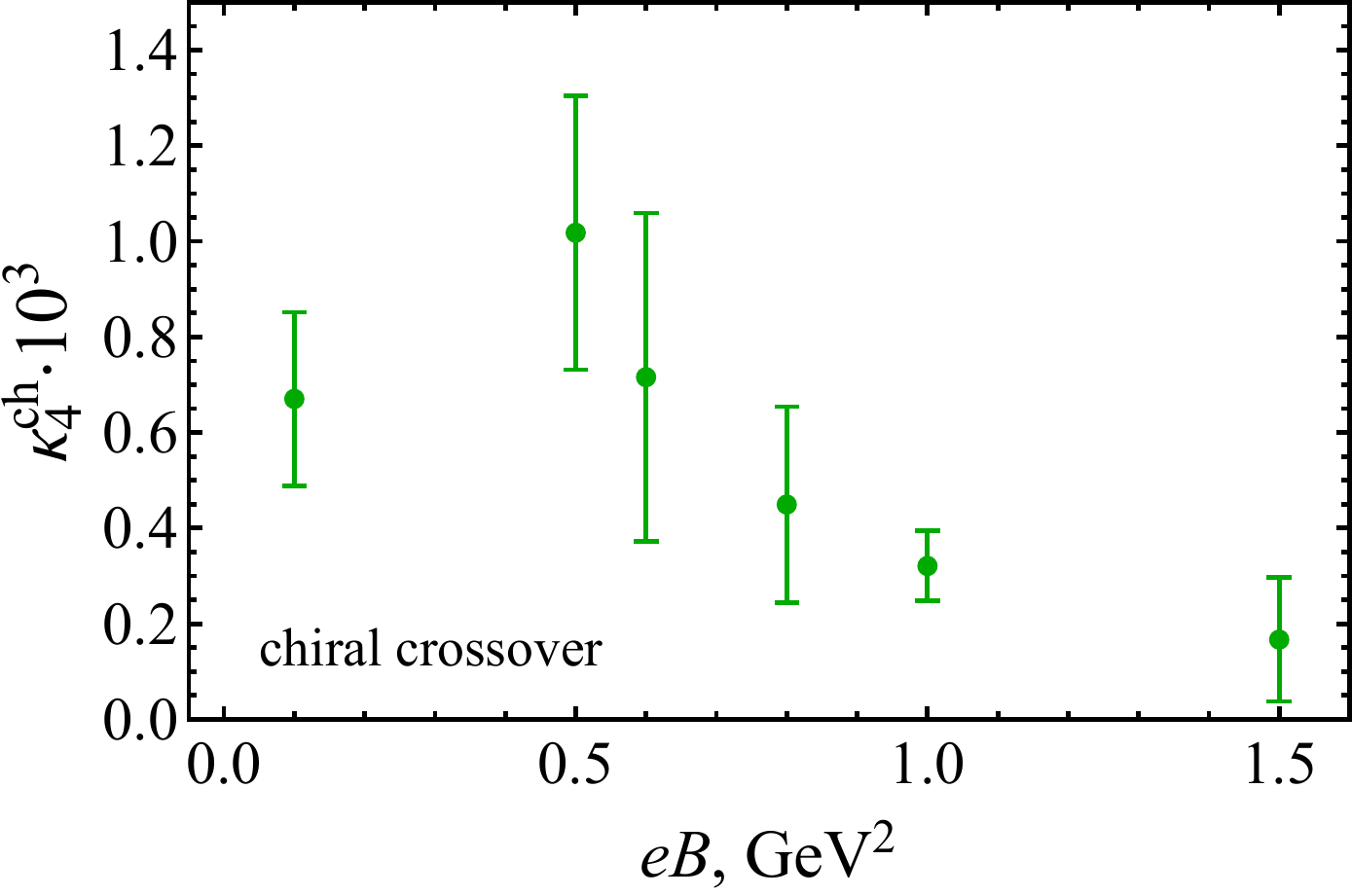}\\
(a) & (b) & (c)
\end{tabular}
\end{center}
\vskip 0mm 
\caption{(a) The critical temperature $T_c^\ch$ of the chiral crossover at $\mu_B=0$, as well as the curvatures (b) $\kappa_{2}$ and (c) $\kappa_4$ as functions of the magnetic-field strength~$B$. The first two quantities are obtained with the help of both quadratic and quartic versions of the fitting function~\eq{eq:Tc:I}, with fits shown in Fig.~\ref{fig:Tc:crossover:fit}. The red points in plots (a) and (b) correspond to the known results at $B=0$. They are taken from Refs.~\cite{Bazavov:2018mes} and \cite{Bonati:2014rfa}, respectively.}
\label{fig:Tc:fits:chiral}
\end{figure*}

The fitting results for the chiral crossover are presented in Fig.~\ref{fig:Tc:fits:chiral}. We conclude that
\begin{itemize}

\item The fits allow us to estimate the critical temperature $T^\ch_c(B)$ at zero baryon chemical potential, $\mu_B = 0$, subjected to a strong magnetic-field background, Fig.~\ref{fig:Tc:fits:chiral}(a). The critical temperature decreases with the magnetic field, in an agreement with the inverse magnetic catalysis~\cite{Bali:2011qj}. In the zero-field limit, our data converge well to the known result $T^{\mathrm{ch}}_c = 156.5(1.5) \,\mathrm{MeV}$ of Ref.~\cite{Bazavov:2018mes}, shown by the red square in Fig.~\ref{fig:Tc:fits:chiral}(a).

\item Both for quadratic and quartic fits~\eq{eq:Tc:I}, the quadratic curvature coefficient $\kappa_2 = \kappa_2(B)$ is largely insensitive to the strength of the magnetic field, Fig.~\ref{fig:Tc:fits:chiral}(b). These fits give qualitatively consistent results, all of which are in agreement with the $B=0$ result $\kappa_2 = 0.0132(18)$ obtained in Ref.~\cite{Bonati:2014rfa}, and shown by the red square in Fig.~\ref{fig:Tc:fits:chiral}(b).

\item According to Fig.~\ref{fig:Tc:fits:chiral}(b), the quartic curvature coefficient $\kappa_4 = \kappa_4(B)$ raises with increase of the magnetic field until it reaches the peak around $eB_{\fl} \simeq (0.5 -0.6)\,\mathrm{GeV}^2$.  Eq.~\eq{eq:eBc}. At higher magnetic fields, the quartic coefficient $\kappa_4$ decreases, and almost vanishes around $eB \simeq 1.5\,\mathrm{GeV}^2$. These conclusions have a preliminary character as our numerical results for $\kappa_4$ possess rather large statistical errors. Below, we will exclude this coefficient from our analysis, and concentrate on the quadratic truncation of the curvature polynomial~\eq{eq:Tc:I}.

\end{itemize}

\begin{figure}[!thb]
\begin{center}
\includegraphics[scale=0.45,clip=true]{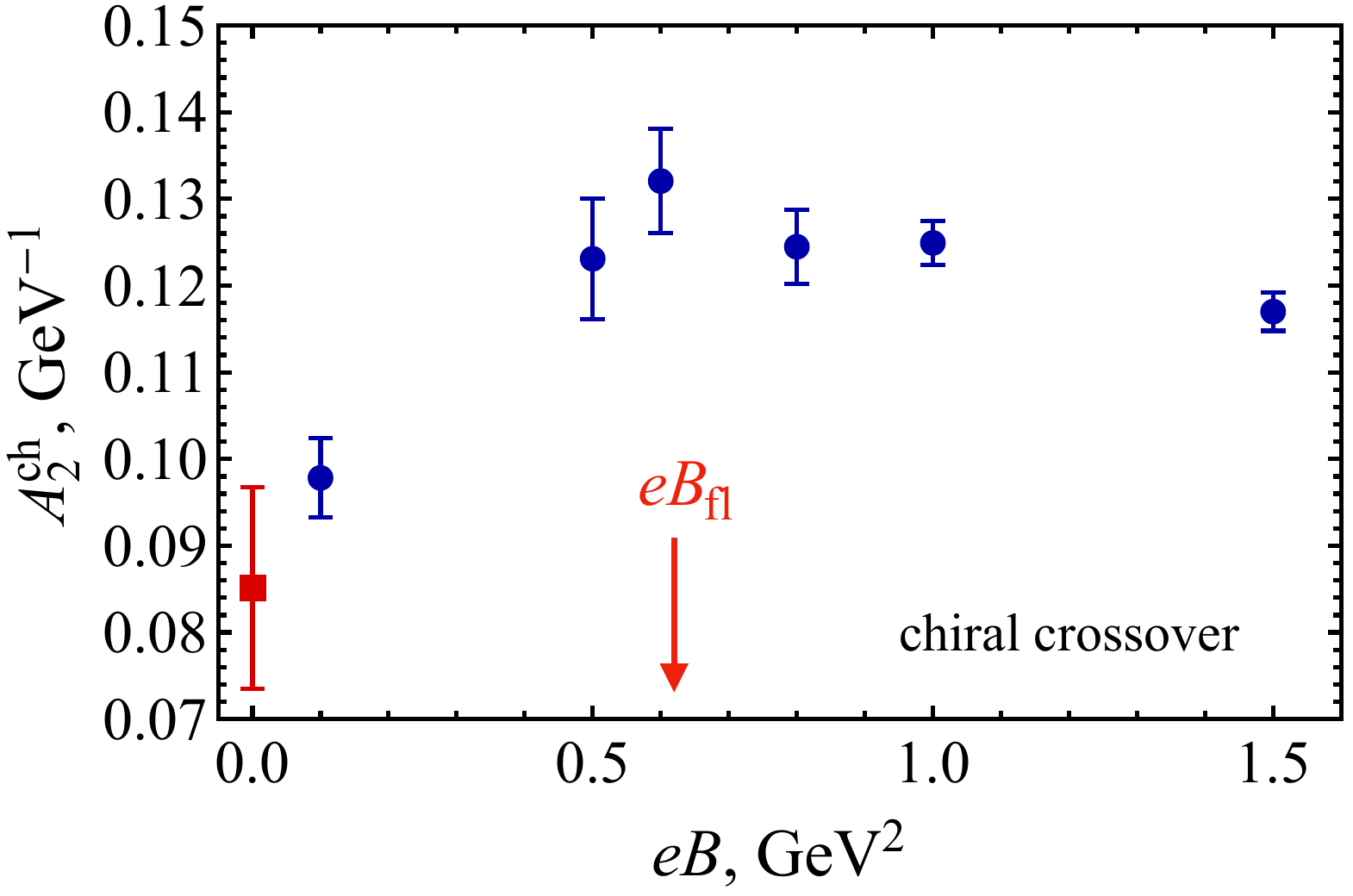}
\end{center}
\vskip 0mm 
\caption{The quadratic curvature $A_2^\ch$ of the chiral crossover temperature~\eq{eq:Tc} at nonzero magnetic field $B$, calculated via Eq.~\eq{eq:A2}. The red data point is obtained with the help of the $B=0$ data of Refs.~\cite{Bazavov:2018mes,Bonati:2014rfa}. The arrow marks the magnetic flipping point for the width of the chiral crossover~\eq{eq:eBc}.}
\label{fig:A2}
\end{figure}

The physical curvature $A_2^\ch$ of the chiral crossover temperature~\eq{eq:Tc} for the real-valued chemical potential $\mu_B$ can be obtained with the help of the analytical continuation~\eq{eq:A2}. The curvature, shown in Fig.~\ref{fig:A2}, seems to exhibit a wide maximum at the magnetic-field strength $eB \sim 0.6\,\mathrm{GeV}^2$. Unfortunately, the substantial statistical errors of our data do not allow us to determine the presence (and, the position) of this maximum with sufficient certainty. However we will see below that this particular value of the magnetic field marks another interesting effect in the low-density QCD.

To summarize, we observed the effect of the inverse magnetic catalysis both at zero and finite densities. The increasing magnetic field affects the curvature~$A_2^\ch$ of the chiral crossover transition, making it larger compared to the zero-field value, Fig.~\ref{fig:A2}. We found the presence of the baryonic matter enhances the effect of the inverse magnetic catalysis in a sense that the combined effect of both these factors, $\mu_B$ and ${\bs B}$, leads to a stronger decrease of the crossover temperature.

\subsubsection{Chiral thermal width and its curvature}

As we have already seen, the thermal width of the chiral crossover transition $\delta T_c^\ch = \delta T^\ch_c(\mu_I,B)$ has a set of interesting features in the parameter space of the magnetic field $B$ and the imaginary chemical potential $\mu_I$, as illustrated in Fig.~\ref{fig:Tc:delta}. What do these properties mean for the crossover transition in the dense QCD with a real-valued baryonic chemical potential~$\mu_B$? In order to answer this question we notice that the thermal chiral width $\delta T_c^\ch$ -- which is, essentially, a difference in temperatures corresponding to opposite sides of the crossover -- may be analytically continued to the real chemical potentials, similarly to the critical temperature $T_c^\ch$ itself. To this end, we define the quadratic curvature $\delta \kappa_2^\ch$ of the chiral thermal width $\delta T_c^\ch$ as follows:
\beqn
\frac{\delta T_c^\ch(\mu_I,B)}{\delta T^\ch_c(B)} = 1 + \delta \kappa^\ch_2 \left( \frac{3 \mu_I}{T^\ch_c(B)} \right)^2 \hskip -2mm + O\left(\biggl(\frac{\mu_I}{T_c^\ch}\biggr)^4 \right)\!, \qquad
\label{eq:delta:Tc:I}
\eeqn
where $T^\ch_c(B) \equiv T^\ch_c(\mu_I = 0, B)$.

Similarity to Eqs.~\eq{eq:Tc}, \eq{eq:Tc:I} and \eq{eq:A2}, the thermal width may be analytically continued to the real-valued baryonic potential as follows:
\beqn
\delta T_c^\ch(\mu_I,B) = \delta T_c^\ch(0,B) - \delta A^\ch_2(B) \mu_B^2 + O(\mu_B^4), \quad
\label{eq:delta:Tc:phys}
\eeqn
where
\beqn
\delta A_2^\ch(B) = \frac{\delta T^\ch_c(\mu_B=0, B) \delta \kappa^\ch_2(B)}{\left(T^\ch_c(\mu_B=0,B)\right)^2},
\label{eq:delta:A2}
\eeqn
is the curvature of the thermal width in the ``temperature-baryon chemical potential'' plane.

\begin{figure}[!thb]
\begin{center}
\includegraphics[scale=0.475,clip=true]{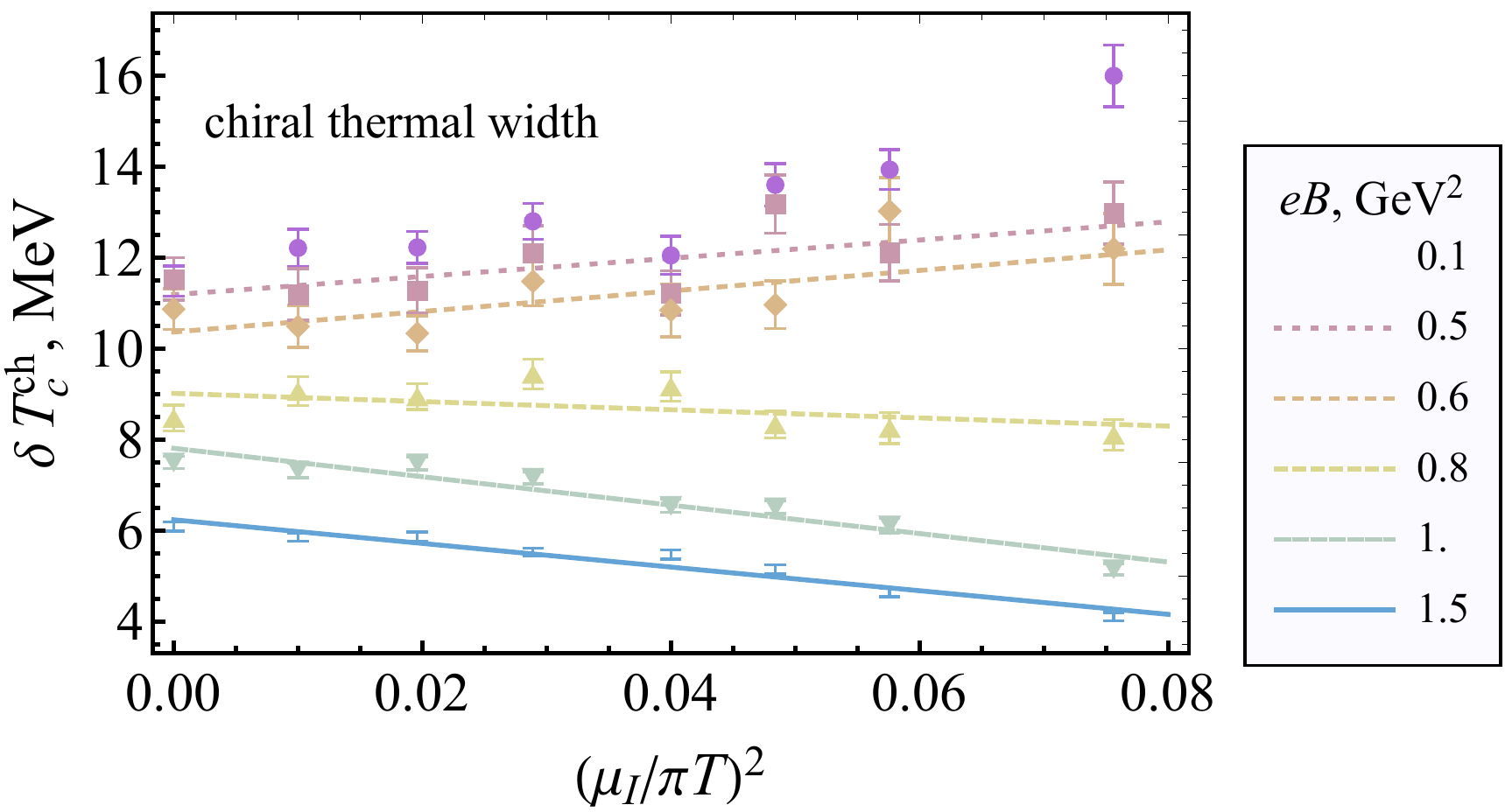}
\end{center}
\vskip 0mm 
\caption{The chiral thermal width $\delta T_c^\ch$ as the function of the imaginary chemical potential $\mu_I$ squared. The lines represent the quadratic fits~\eq{eq:delta:Tc:I}.}
\label{fig:delta:T:chiral}
\end{figure}

\begin{figure*}[!thb]
\begin{center}
\begin{tabular}{ccc}
\includegraphics[scale=0.38,clip=true]{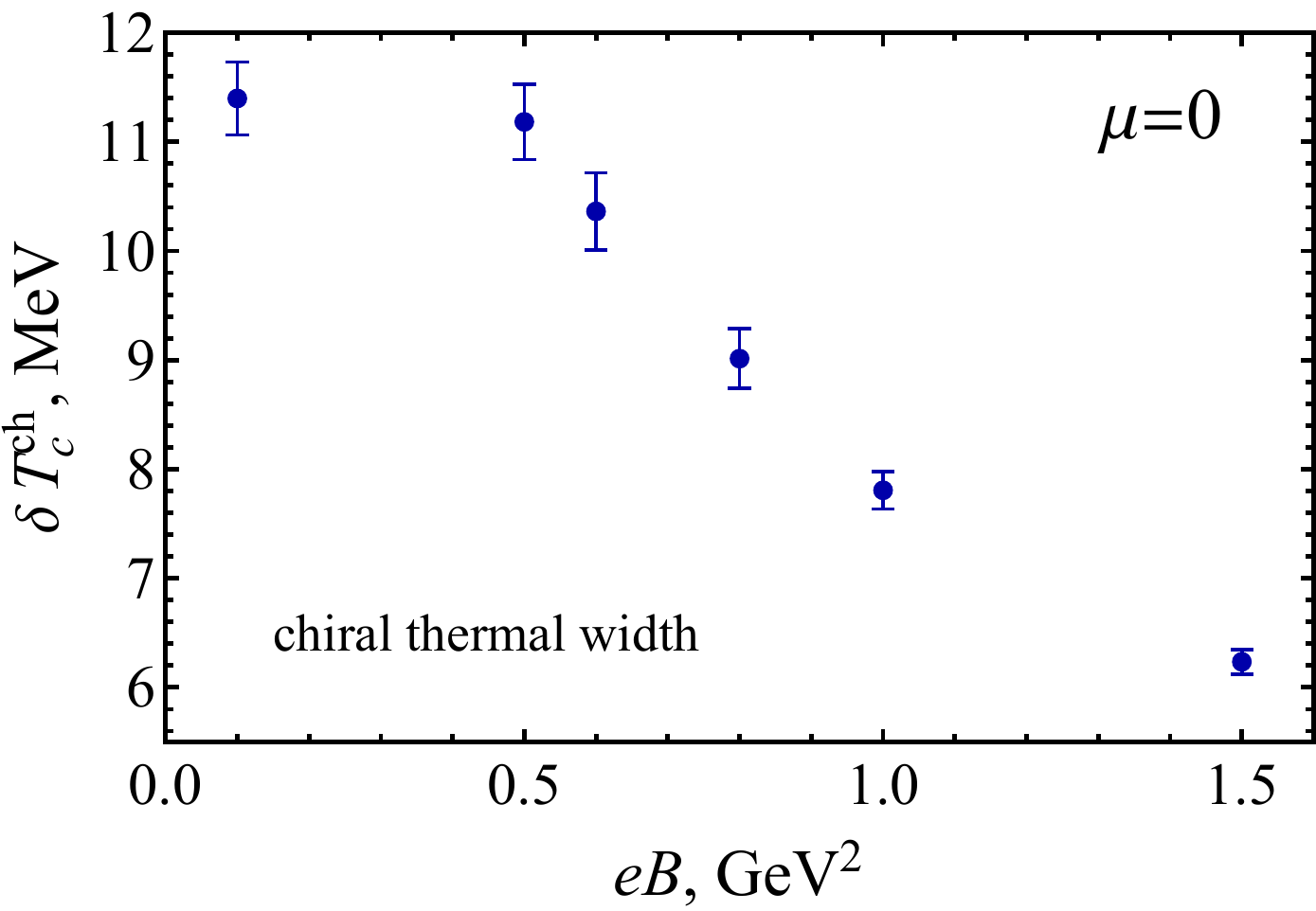} &
\hskip 2mm \includegraphics[scale=0.38,clip=true]{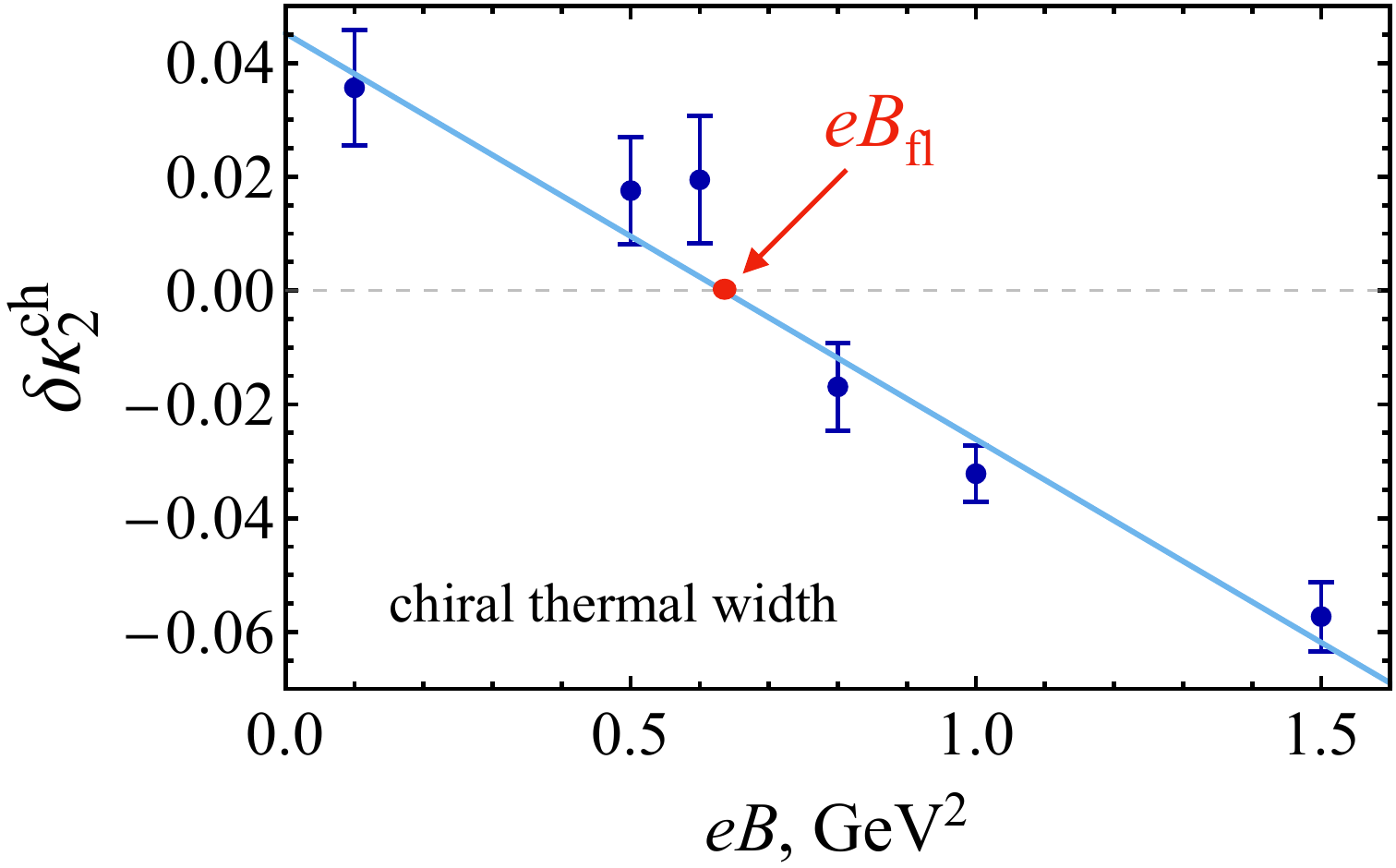} & 
\hskip 2mm \includegraphics[scale=0.37,clip=true]{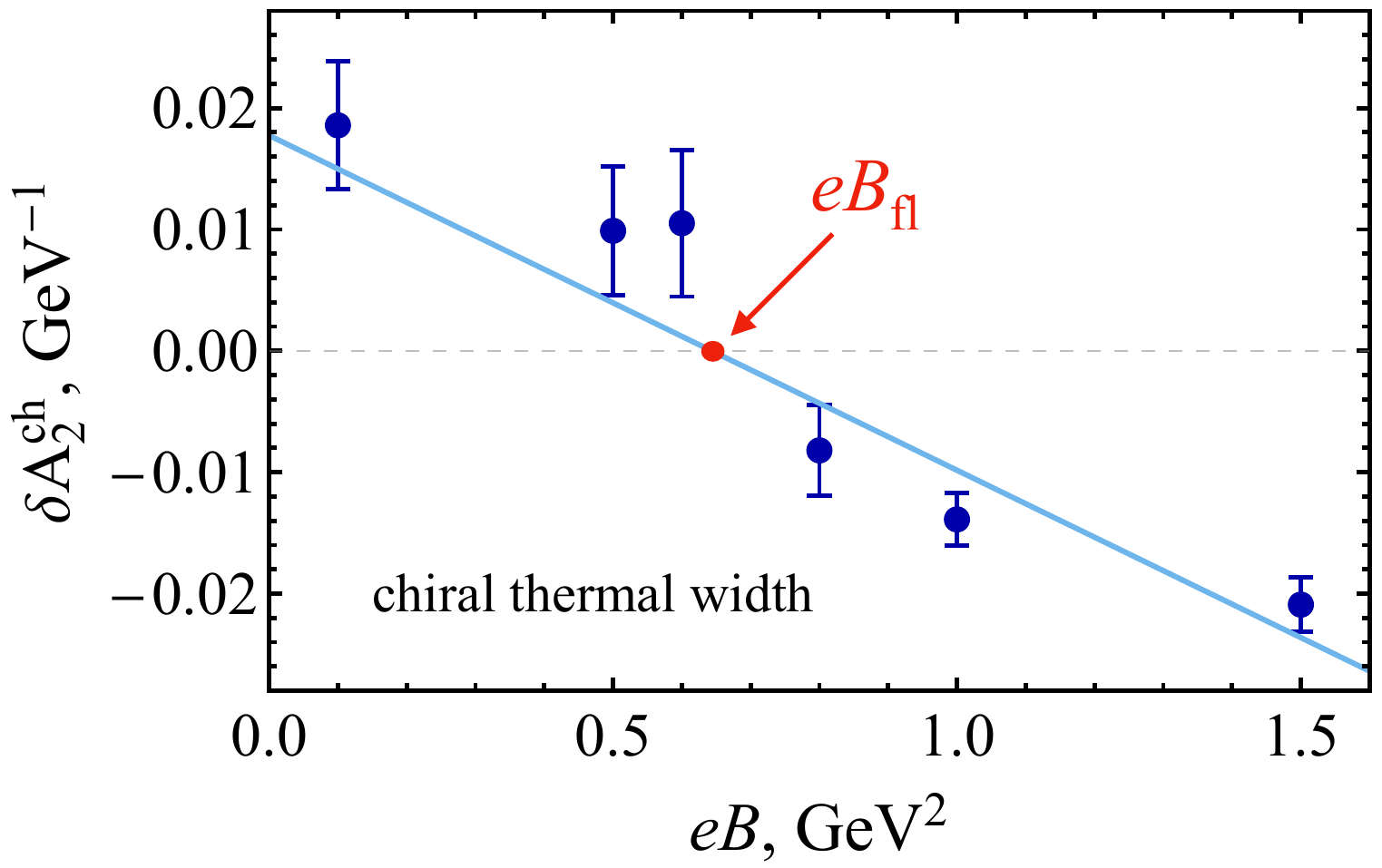}\\
(a) & (b) & (c)
\end{tabular}
\end{center}
\vskip 0mm 
\caption{(a) The chiral thermal width $\delta T_c^\ch$, the corresponding quadratic curvature $\delta \kappa_2^\ch$ in dimensionless (b) and physical (c) units. In plot (c), the solid line shows the best fit of the data by the linear function~\eq{eq:A2:loopar}, and the red arrow marks the critical value of magnetic field~\eq{eq:eBc} where the curvature of the chiral thermal width vanishes.}
\label{fig:delta:Tc:A2}
\end{figure*}

In Fig.~\ref{fig:delta:T:chiral} we demonstrate that the numerical data for the chiral thermal width can be well described by the quadratic function~\eq{eq:delta:Tc:I}. 
The fits give us the chiral thermal width at zero chemical potential, $\mu_B = 0$,  shown in Fig.~\ref{fig:delta:Tc:A2}(a). The plot suggests that the chiral thermal width is insensitive to the magnetic field until the field reaches the value $eB_{\fl} \sim 0.5\mathrm{GeV}^2$, and then the width start to decrease slowly. 

The effect of the magnetic field background on the curvature of the chiral thermal width is shown in dimensionless, $\delta \kappa_2^\ch$, and physical $\delta A_2^\ch$ units in Figs.~\ref{fig:delta:Tc:A2}(b) and (c), respectively. It turns out that the curvature $\delta A_2^\ch$ of the chiral thermal width $\delta T_c^\ch$ may be well approximated by the linear function of the background magnetic field~$B$ for both quantities:
\begin{subequations}
\beqn
\delta \kappa_2^\ch(B) = \delta \kappa^{(0)}_2 + \delta \kappa^{(1)}_2 eB, \\
\delta A_2^\ch(B) = \delta A^{(0)}_2 + \delta A^{(1)}_2 eB.
\eeqn
\label{eq:A2:loopar}
\end{subequations}
The best linear fits are shown in Fig.~\ref{fig:delta:Tc:A2}(c) by the solid lines. The corresponding best fit parameters are the thermal width at a vanishing magnetic field, 
$\delta \kappa^{(0)}_2 \equiv \delta \kappa_2(0) = 0.045(9)$ and 
$\delta A^{(0)}_2 \equiv \delta A_2(0) = 0.018(6)\,{\mathrm{GeV}}^{-1}$, and the linear slopes:
$\delta \kappa^{(1)}_2 = - 0.071(9)$ and $\delta A^{(1)}_2 = - 0.028(5)\,{\mathrm{GeV}}^{-3}$,
respectively.

From Figs.~\ref{fig:delta:Tc:A2}(b) and ~\ref{fig:delta:Tc:A2}(c) we readily notice the interesting feature of the chiral width: at certain strength of the magnetic field,
\beqn
eB_{\fl} = 0.63(6)\, {\mathrm{GeV}}^{-2},
\label{eq:eBc}
\eeqn
the curvature of the chiral thermal width flips the sign from positive to negative values. We call the value~\eq{eq:eBc} as ``the magnetic flipping point''.

Qualitatively, one can understand the effect of the sign flip of the curvature $\delta A_2^\ch$ as follows. The magnetic flipping point~\eq{eq:eBc} separates two regimes: at weaker magnetic fields, $B < B_{\fl}$, the quadratic curvature is positive, $\delta A_2 (B < B_{\fl}) > 0$, and the thermal width of the crossover temperature gets narrower~\eq{eq:delta:Tc:phys} with the rise of the baryon chemical potential. At stronger magnetic fields, $B > B_{\fl}$, the thermal width become wider, $\delta A_2 (B < B_{\fl}) < 0$ as the density of the baryonic medium increases.

Numerically, the strength of the magnetic field at the flipping point~\eq{eq:eBc} coincides with the (vacuum) mass of the $\rho$ meson squared, $eB_{\fl} \simeq m_\rho^2 \simeq 0.601\,\mathrm{GeV}^2$. At this value of the flipping magnetic field the $\rho$ mesons were proposed to form a superconducting condensate at low enough temperature~\cite{Chernodub:2010qx,Chernodub:2011mc}. While this statement is subjected to critical debates~\cite{Hidaka:2012mz,Andreichikov:2013zba,Li:2013aa}, we notice that thermal effects contribute to the $\rho$ meson mass and are likely to destroy the $\rho$-meson condensate should it be formed at low temperature. 

Nevertheless, the closeness of the magnetic flipping point~\eq{eq:eBc} to the mass of the mass scale of the $\rho$ meson suggests that the latter may play a particular role. One could suggest that the mechanism behind the appearance of the magnetic flipping point may be related to the vector meson dominance model~\cite{Sakurai:1960ju}. This model proposes that the electromagnetic field interacts with the quark matter via the creation of the quark--anti-quark pairs with the quantum numbers of photons. The lightest such pairs correspond to the neutral rho mesons.  

Numerical lattice calculations and effective analytical models suggest that the mass of the neutral meson slowly raises with the strengthening of the magnetic field~\cite{Hidaka:2012mz,Andreichikov:2013zba,Andreichikov:2016ayj,Bali:2017ian}. Moreover,  at the crossover temperature, thermal fluctuations slightly increase the mass of the $\rho$ meson as well~\cite{Gale:1990pn}. Quantitatively, we expect that the combined temperature and magnetic-field effects at the crossover shift the mass by about 20\% from its vacuum value so that the magnetic flipping point~\eq{eq:eBc} is approximately given by the scale of the rho-meson mass.

In order to shed more light on the sign flip of the chiral thermal width, Fig.~\ref{fig:delta:Tc:A2}(c), in the next section we study the confining properties of dense QCD in the magnetic field background.
The vector dominance hypothesis suggests that the photons interact with the hadronic medium predominantly via the neutral $\rho$ mesons, which are colorless states that do not couple directly to gluons. As we will see below, the sign-flip phenomenon does not exist in the gluonic sector.

We close this section by noticing that our findings on the chiral crossover at zero baryonic density agree well with already known properties of the system. As the strength of the magnetic field increases, the chiral crossover temperature becomes lower, Fig.~\ref{fig:Tc:fits:chiral}(a) while the transition itself becomes stronger, Fig.~\ref{fig:delta:T:chiral}(a), in agreement with  Refs.~\cite{Bali:2011qj} and \cite{Endrodi:2015oba}, respectively.

In addition, our data on the chiral thermal width of the crossover raise an interesting possibility that the parameter plane of the imaginary chemical potential and temperature may contain a thermodynamic phase transition in the limit of large baryonic density at low magnetic field, $B < B_{\fl}$. At stronger magnetic field, $B > B_{\fl}$, the increase of the baryon density leads to the softening of the phase transition. 

The chiral pseudo-critical temperature and the thermal width of the chiral crossover, as well the their curvatures in the $(\mu,T)$ plane are summarized in Table~\ref{ref:parameters:transitions}.

\subsubsection{Shrinking chiral width and critical chiral endpoint}

We would like to finish this section by the following curious observation. As we mentioned above, the width of the chiral transition shrinks in the presence of the baryonic density. Although our numerical simulations are done in the region of relatively low baryon density, we notice that the observation of the shrinking chiral width is consistent with the expectation that at a higher baryon density the crossover turns into a critical endpoint (CEP) of the second order which, at even higher densities, becomes a transition line of the first order.

We may estimate the position of the endpoint as a value of the baryonic chemical potential $\mu_B = \mu_B^{\mathrm{CEP}}$ at which the width of the phase transition $\delta T_c^{ch}(\mu_B,0)$ vanishes. To this end, we use Eq.~\eq{eq:delta:Tc:phys} along with the the results for the chiral thermal width $\delta T_c^{ch}(0,0)$ and its curvature $\delta A_2(0)$ to get for the baryonic density at the CEP: 
\begin{equation}
    \mu_B^{\text{CEP}}=\sqrt{\frac{\delta T_c(0,0)}{\delta A_2(0)}}=  800(140) \text{MeV}. 
\label{eq:cep}
\end{equation}
where we neglected the corrections of the order $O(\mu_B^4)$ and higher.

The result~\eq{eq:cep} is obtained at zero magnetic field. Notice that with the strengthening of magnetic field $B$, the curvature of the chiral width $\delta A_2(B)$ quickly diminishes towards zero, Fig.~\ref{fig:delta:Tc:A2}(c), while the width $\delta T_c^{ch}(\mu_B=0,B)$ at zero density $\mu_B=0$ drops down less dramatically, Fig.~\ref{fig:delta:Tc:A2}(a). Therefore, we may expect that $\mu_B^{\text{CEP}}(B)$ is an increasing function of the magnetic field $B$. However, at the flipping point~\eq{eq:eBc}, our estimation of the CEP~\eq{eq:cep} formally gives infinite value of the CEP baryonic chemical potential $\mu_B^{\text{CEP}}$, which shows the limitation of our approach and importance of the higher-order terms, $O(\mu_B^4)$, which were neglected in our simple analysis based on quadratic curvature width~\eq{eq:delta:Tc:phys}.

The temperature of the critical endpoint may be obtained with the help of Eq.~(\ref{eq:Tc}) which takes into account the curvature of the chiral crossover temperature:  $T^{\text{CEP}} \equiv T_c(\mu_B^{\mathrm{CEP}})=100(25)$ MeV. Together with the result~\eq{eq:cep}, this result gives us a very naive estimation of the position of the critical end point in the $T-\mu$ plane of the phase diagram:
\begin{equation}
    (T_c^{\text{CEP}}, \mu_B^{\text{CEP}})= \bigl(100(25)\, \text{MeV},\ 800(140)\,\text{MeV} \bigr). 
\label{eq:cep:T:mu}
\end{equation}
Curiously, these numbers come quite close to the recent estimation  of the location of the critical end point $(T^{\text{CEP}},\mu_B^{\text{CEP}})=(107,635)$ MeV obtained by means of functional renormalization group \cite{Fu:2019hdw}, as well as to other estimations (we refer a reader to Ref.~\cite{Fu:2019hdw} for a detailed review).

\section{Deconfining crossover}
\label{sec:deconfinement}

\subsection{Renormalized Polyakov loop}

The deconfinement (phase) transition is associated with the dynamics of gluons. The corresponding order parameter, in purely gluonic Yang-Mills theory, is the Polyakov loop~\eq{eq:Polyakov:loop}. For the sake of convenience, we study the real part~\eq{eq:Polyakov:loop:Re} of the Polyakov loop, which is renormalized with the help of the gradient-flow approach following Ref.~\cite{Petreczky:2015yta}. The details of the renormalization, and the scheme dependence of the gradient-flow procedure are discussed in the Appendix~\ref{app:L_renorm}.

\begin{figure}[!thb]
\begin{center}
\includegraphics[scale=0.55,clip=true]{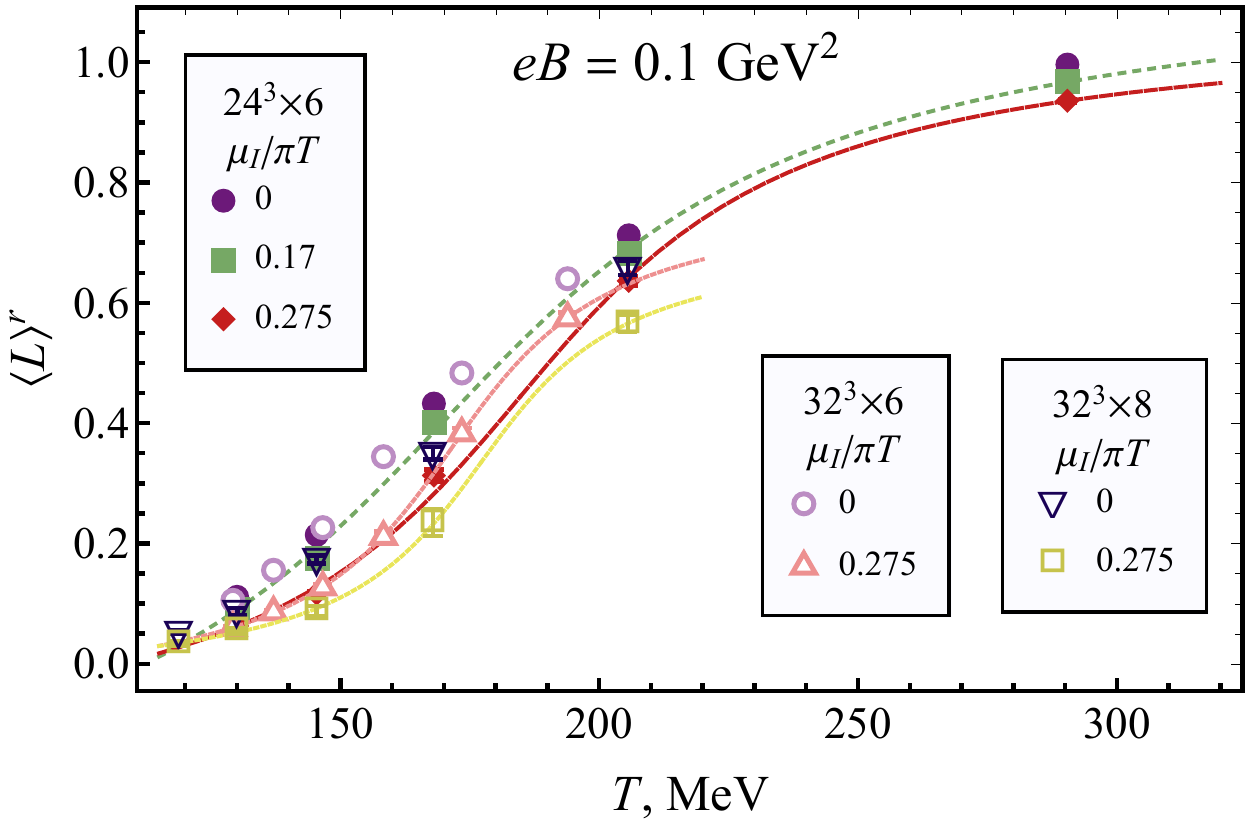}\\[3mm]
\includegraphics[scale=0.55,clip=true]{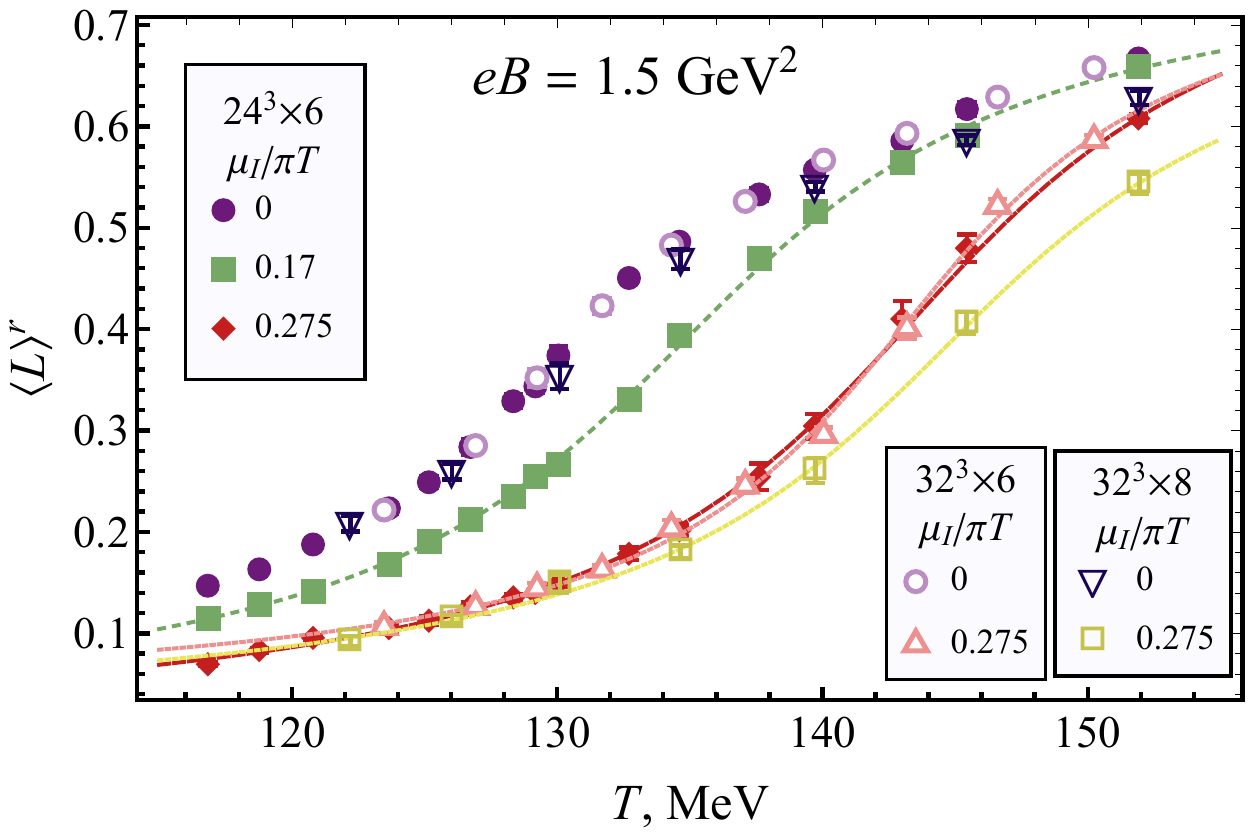}
\end{center}
\vskip 0mm 
\caption{The renormalized Polyakov loop~\eq{eq:Polyakov:loop} as the function of temperature at various fixed values of $\mu_I/(\pi T)$ in the background of the weakest ($eB = 0.1\,\mathrm{GeV}^2$) and the strongest ($eB = 1.5\,\mathrm{GeV}^2$) magnetic fields on the lattice $24^3 \times 6$. For comparison, we also show the renormalized Polyakov loop on the lattices $32^3 \times 6$ and $32^3 \times 8$, for the lowest and largest imaginary chemical potentials, $\mu_I/(\pi T) = 0,\,0.275$. The lines are the best fits by the function~\eq{eq:L:fit}.}
\label{fig:L:vs:T}
\end{figure}

In the vicinity of the crossover, the renormalized Polyakov loop may be well described by the same functional behaviour as the chiral condensate~\eq{eq:condensate:fit},
\beqn
\avr{L}^r (T) = C_2 + C_3 \arctan \frac{T - T_c^\conf}{\delta T_c^\conf},
\label{eq:L:fit}
\eeqn
where the fitting parameters $C_{2}$ and $C_3$ determine the value of the Polyakov loop at both sides of the crossover region, and $T_c^\conf$ is the pseudo-critical temperature of the deconfinement transition with the deconfining thermal width $\delta T_c^\conf$. 
Some selected fits at lowest ($eB = 0.1 \, \mathrm{GeV}^2$) and highest ($eB = 1.5 \, \mathrm{GeV}^2$) values of magnetic field are shown in Fig.~\ref{fig:L:vs:T} for zero, moderate and highest imaginary chemical potentials, $\mu_I = (0, 0.17, 0.275) \pi T$, respectively.

Similarly to the case of chiral condensate of light quarks, we check the robustness of our results with respect to the volume variations, Fig.~\ref{fig:L:vs:T}. In addition of the main results obtained on a $24^3 \times 6$ lattice, we also show the plots of the renormalized Polyakov loop calculated at a $32^3 \times 6$ lattice of a higher volume. The visual comparison of the results indicates that the increasing imaginary chemical potential leads to stronger volume dependence at low magnetic field, while at the low density and/or in the strong magnetic field, the sensitivity of the renormalized Polyakov loop to the infrared effects is almost unnoticeable. 

The reason for the emergence of these volume effects has a simple systematic origin which is not directly related to the dynamical volume effects. Due to the quantization of magnetic field~\eq{eq:B:quant}, the number of numerical points, available for the fit~\eq{eq:L:fit}, is very much limited for weak magnetic fields as compared to stronger magnetic fields. Instead, the magnetic field is varied by the discrete flux variable $n=1,2,\dots$, which needs to be counter-weighted by the variation of the lattice spacing $a = a(\beta)$ in Eq.~\eq{eq:B:quant}. The variation of the latter, in turn, affects the temperature $T = 1/(a N_s)$, which quickly goes out of the interesting temperature interval of the deconfining crossover. Therefore we are faced with an artificial limitation of the number of points that could be used in the fit~\eq{eq:L:fit}, thus bringing a large systematic error to our results. Due to these reasons, we do not discuss below the lattices other than $24^3\times 6$ (noticing, at the same time, that the formal low-field $B \to 0$ limit agrees with the known $B=0$ results). At larger magnetic fields, the flux variable $n$ may run over larger sets of points and this problem does not exist.

\subsection{Deconfining temperature and its thermal width}

\begin{figure}[!thb]
\begin{center}
\includegraphics[scale=0.6,clip=true]{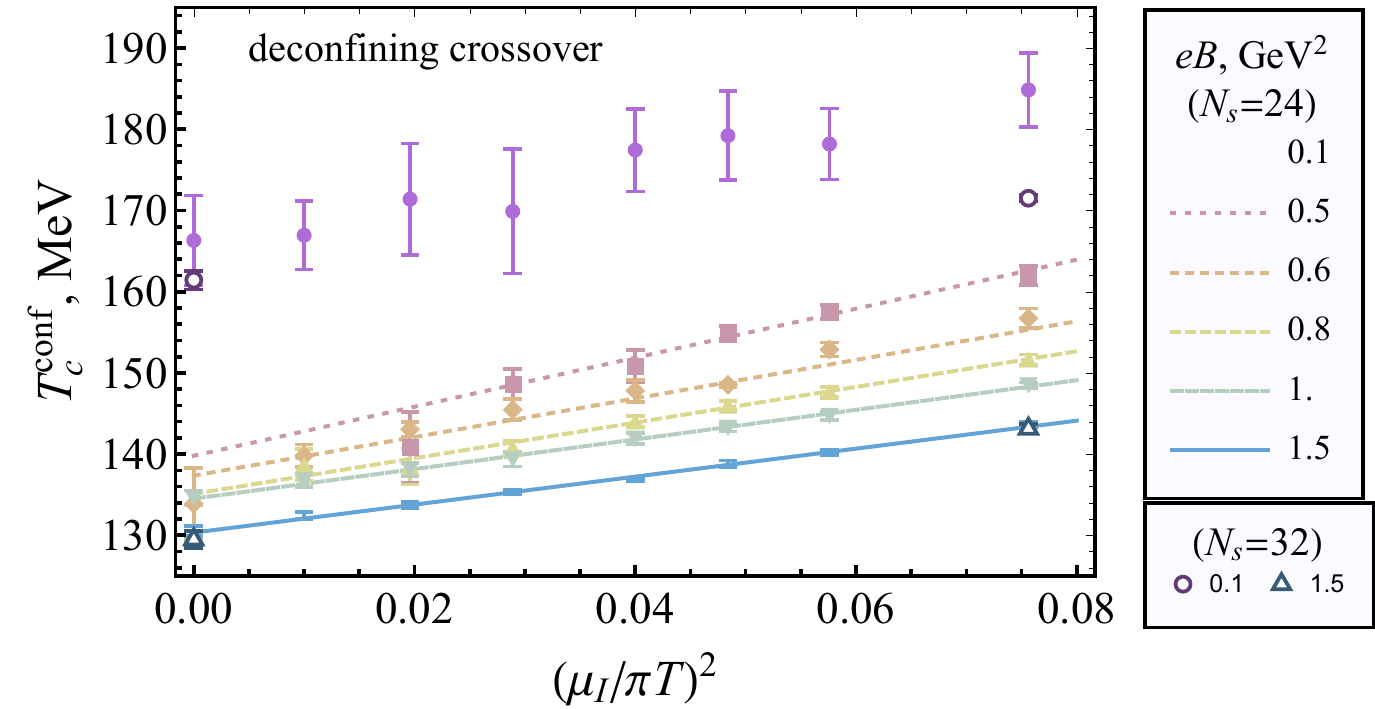}
\end{center}
\vskip 0mm 
\caption{The deconfinement crossover temperature $T_c^\conf$ determined via the fits~\eq{eq:L:fit} of the renormalized Polyakov loop. The lines represent the best fits by the quadratically truncated Eq.~\eq{eq:Tc:I}.}
\label{fig:Tc:deconf}
\end{figure}

In Fig.~\ref{fig:Tc:deconf} we show the pseudo-critical temperature of the deconfining crossover as the function of the imaginary chemical potential for the whole set of the available magnetic fields. It turns out that the dependence of the deconfining crossover temperature on imaginary chemical potential can well be fitted by the (quadratically truncated) Taylor series~\eq{eq:Tc:I} almost at all values of the magnetic field. Notice that the large error bars at the lowest magnetic strength as well as the difference of the results on two lattice sizes $N_s=24$ and $N_s=32$ have the systematic origin mentioned above.

\begin{figure*}[!thb]
\begin{center}
\begin{tabular}{ccc}
\includegraphics[scale=0.4,clip=true]{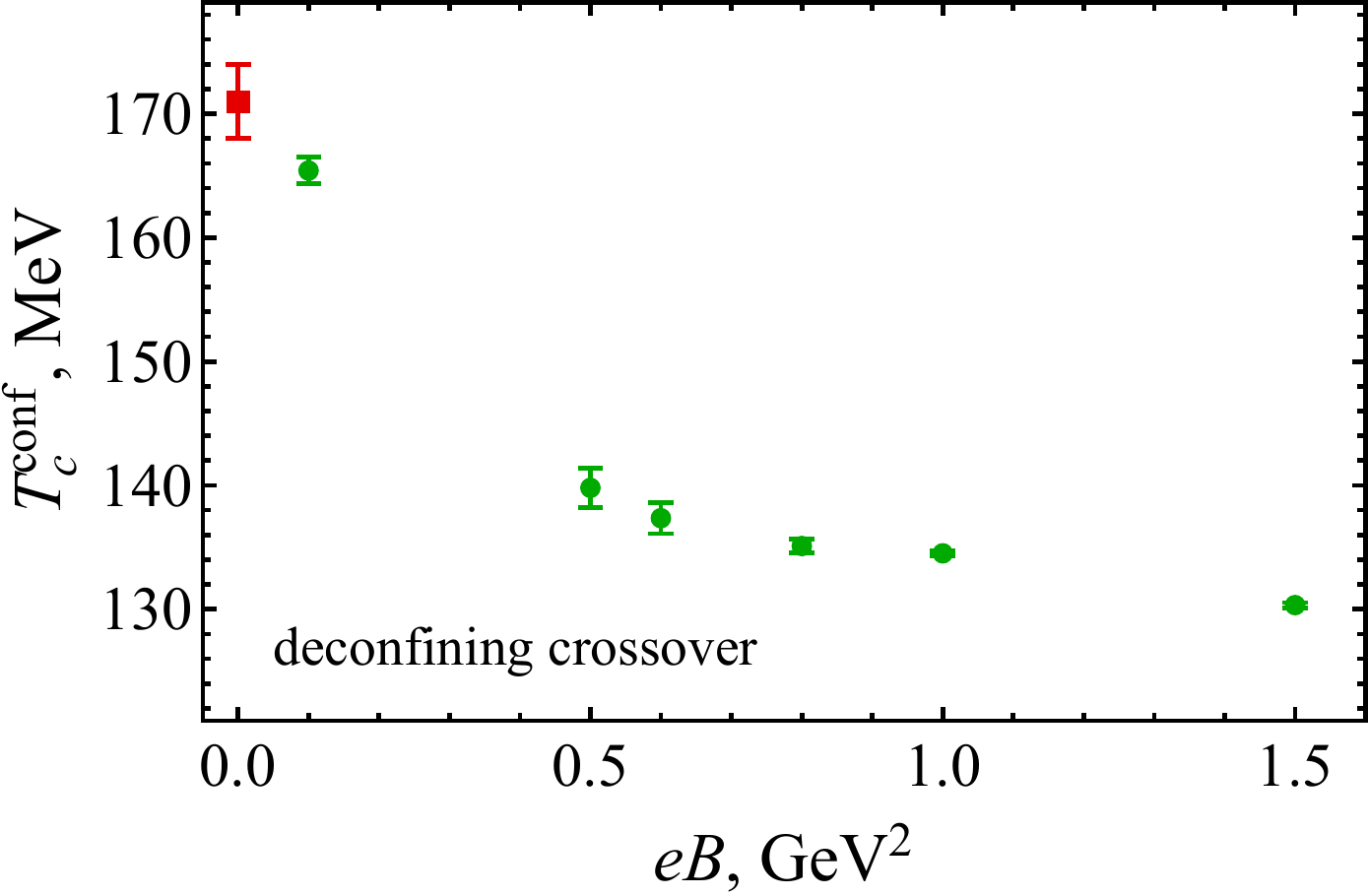} & 
\hskip 2mm \includegraphics[scale=0.42,clip=true]{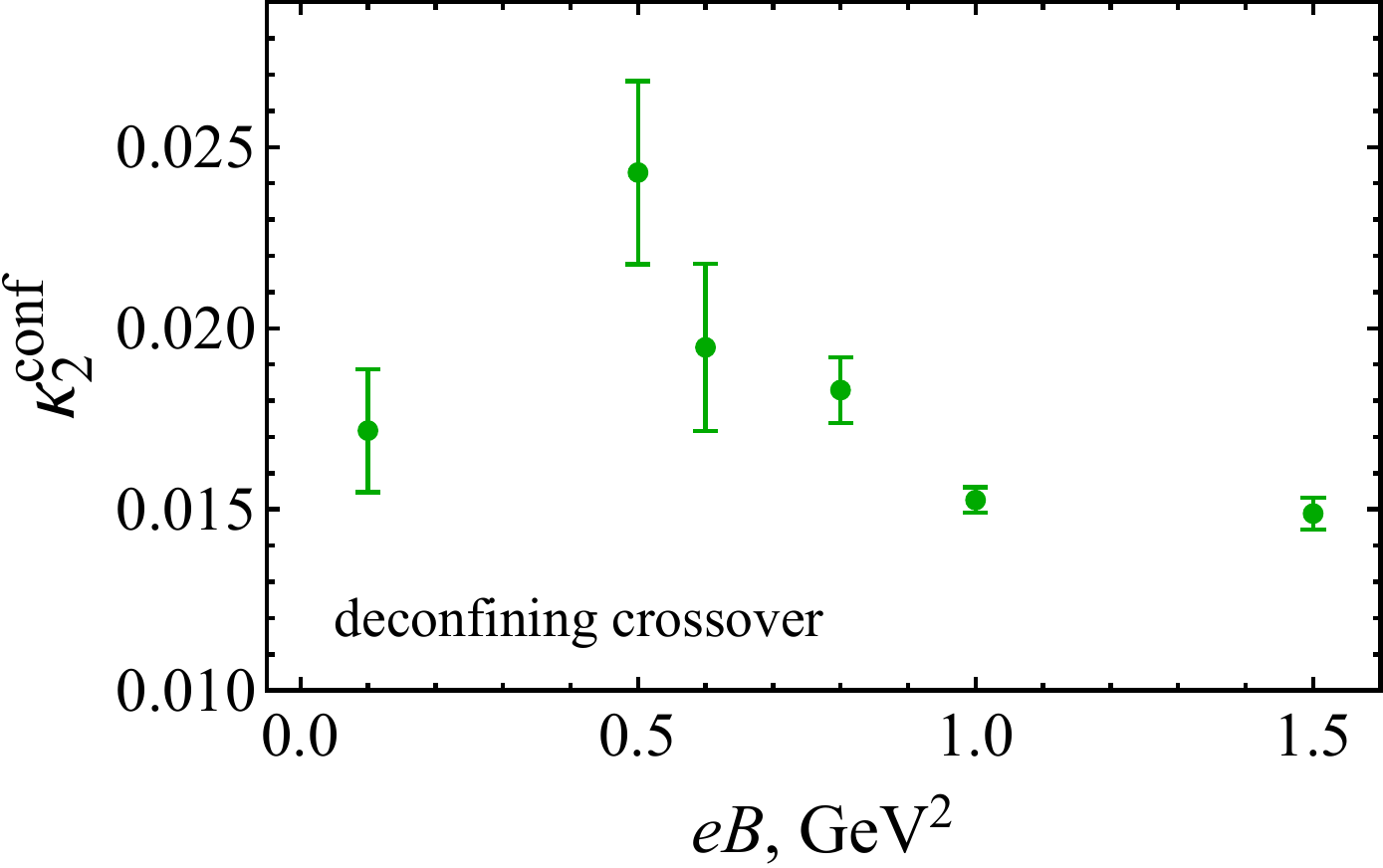} & 
\hskip 2mm \includegraphics[scale=0.4,clip=true]{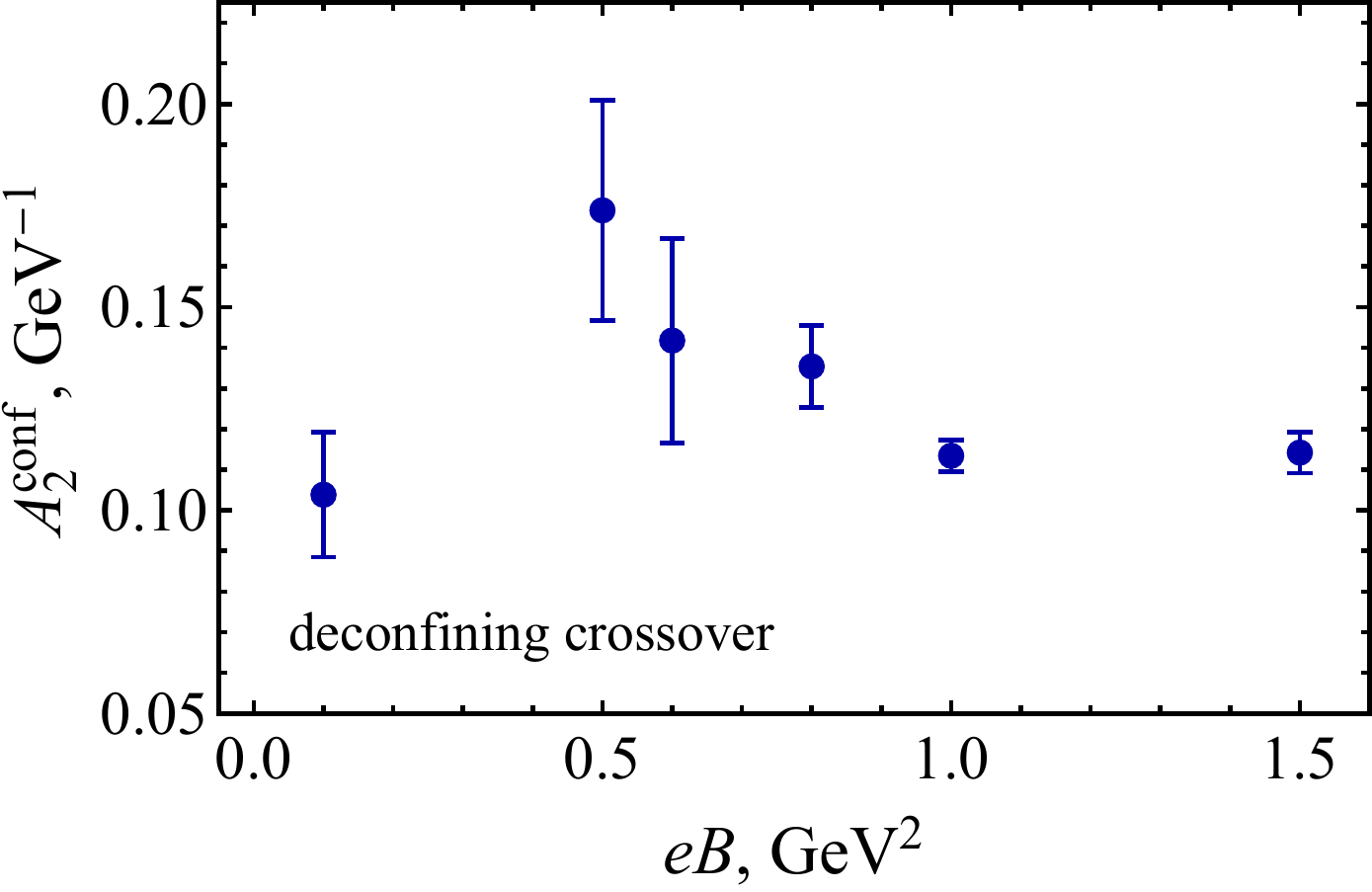} \\
(a) & 
(b) & 
(c)
\end{tabular}
\end{center}
\vskip 0mm 
\caption{(a) The pseudo-critical temperature $T^\conf_c$ of the deconfining crossover at zero chemical potential $\mu_B=0$, as well as the curvature in (b) dimensionless units  $\kappa^\conf_{2}$, and (c) physical units $A^\conf_2$ vs. the magnetic-field strength~$B$. The red data point in plot (a) corresponds to the zero field $B=0$ obtained in Ref.~\cite{Aoki:2006we}.}
\label{fig:Tc:fits:results}
\end{figure*}

The temperature $T_c$ of the deconfining crossover at zero chemical potential, obtained with the help of the quadratic fits, is shown in Fig.~\ref{fig:Tc:fits:results}(a).
Similarly to the chiral crossover temperature, the deconfining crossover temperature is a diminishing function of the magnetic field. This property agrees well with the earlier observation that the gluonic degrees of freedom, as probed by the gluon action, experience the inverse magnetic catalysis similarly to the light quark condensates~\cite{Bali:2013esa}.

According to Fig.~\ref{fig:Tc:fits:results}(a), in the limit of weak magnetic fields, the pseudo-critical temperature of the deconfining transition agrees well with the known $B=0$ result, $T_c = 171(3)\,\mathrm{MeV}$, obtained in Ref.~\cite{Aoki:2006we}.\footnote{For consistency reasons, we do not include the systematic error from Ref.~\cite{Aoki:2006we} for the deconfining pseudo-critical temperature.} At strong magnetic fields the pseudo-critical line of the deconfining transition, shown in Fig.~\ref{fig:Tc:fits:results}(a), overlaps with the line of the chiral crossover, Fig.~\ref{fig:Tc:fits:chiral}(a). This fact will be clearer in the last Section, where we discuss the overall phase diagram.

The quadratic curvature of the deconfining transition is shown in Fig.~\ref{fig:Tc:fits:results}(b) as the dimensionless quantity $\kappa^\conf_2$ and in Fig.~\ref{fig:Tc:fits:results}(c) as the physical curvature $A^\conf_2$, calculated via Eq.~\eq{eq:A2}. It seems to have a peak around the magnetic flipping field~\eq{eq:eBc} which is, however, determined with a substantial uncertainty due to large statistical errors. Still, the curvature $A^\conf_2$ is a positive quantity so that the pseudo-critical temperature of the deconfining crossover diminishes in the dense QCD matter. This fact means that the presence of the baryon density enhances the ``inverse magnetic catalysis'' effect for the deconfining phase transition: the presence of matter makes the deconfinement crossover transition happening at lower temperatures.

\begin{figure}[!thb]
\begin{center}
\includegraphics[scale=0.475,clip=true]{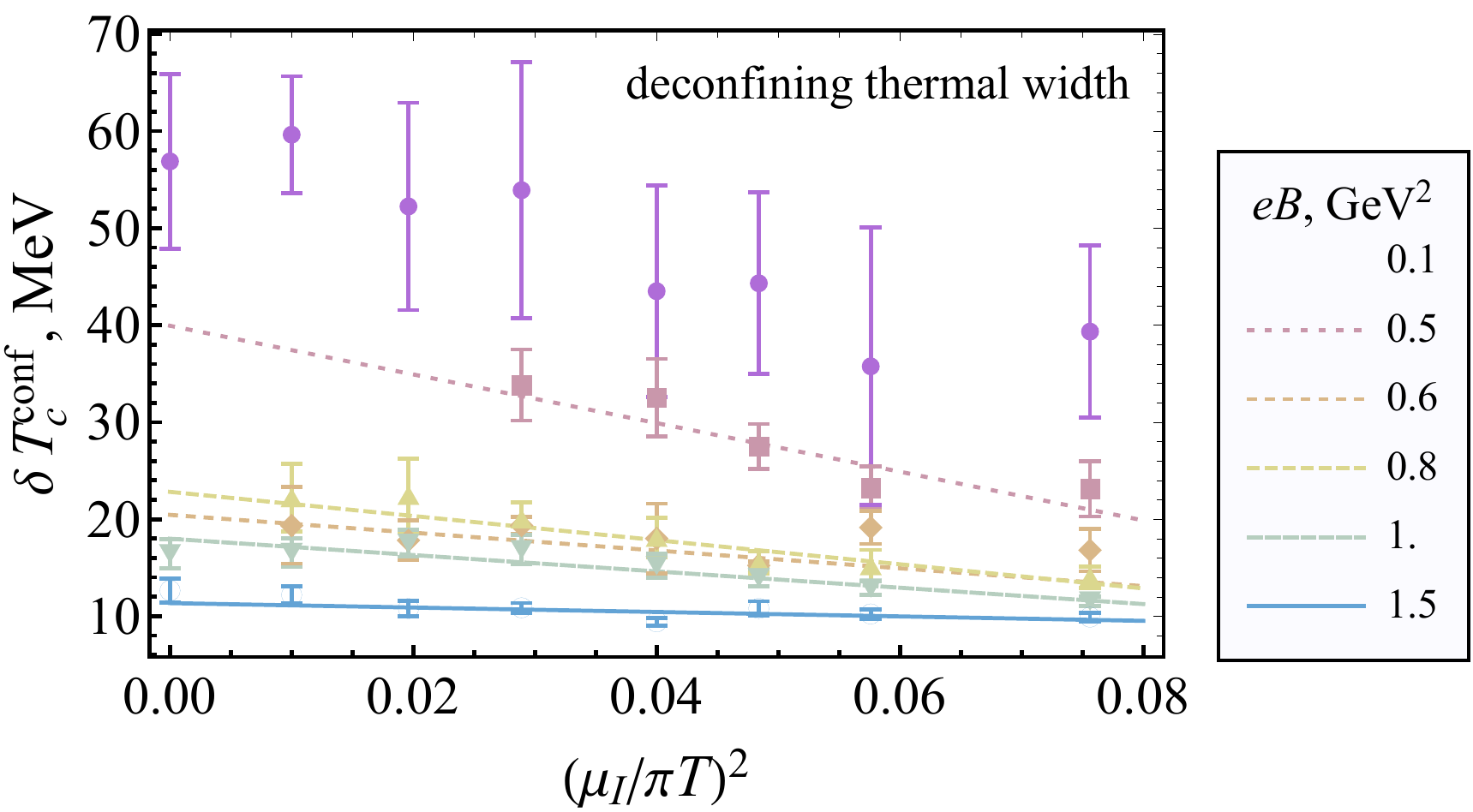}
\end{center}
\vskip 0mm 
\caption{The same as in Fig.~\ref{fig:Tc:deconf} but for the deconfining thermal width.}
\label{fig:delta:Tc:deconf}
\end{figure}

The thermal width of the deconfining transition may be analytically continued to the real-valued baryonic potential similarly to its chiral counterpart~\eq{eq:delta:Tc:phys}. In Fig.~\ref{fig:delta:Tc:deconf} we demonstrate that the numerical data for the deconfining thermal width can be well described by the quadratic function~\eq{eq:delta:Tc:I}. In this figure, we dropped a few points with very large error bars which practically do not contribute to the fits while making the figure less readable.

\begin{figure*}[!bht]
\begin{center}
\begin{tabular}{ccc}
\includegraphics[scale=0.39,clip=true]{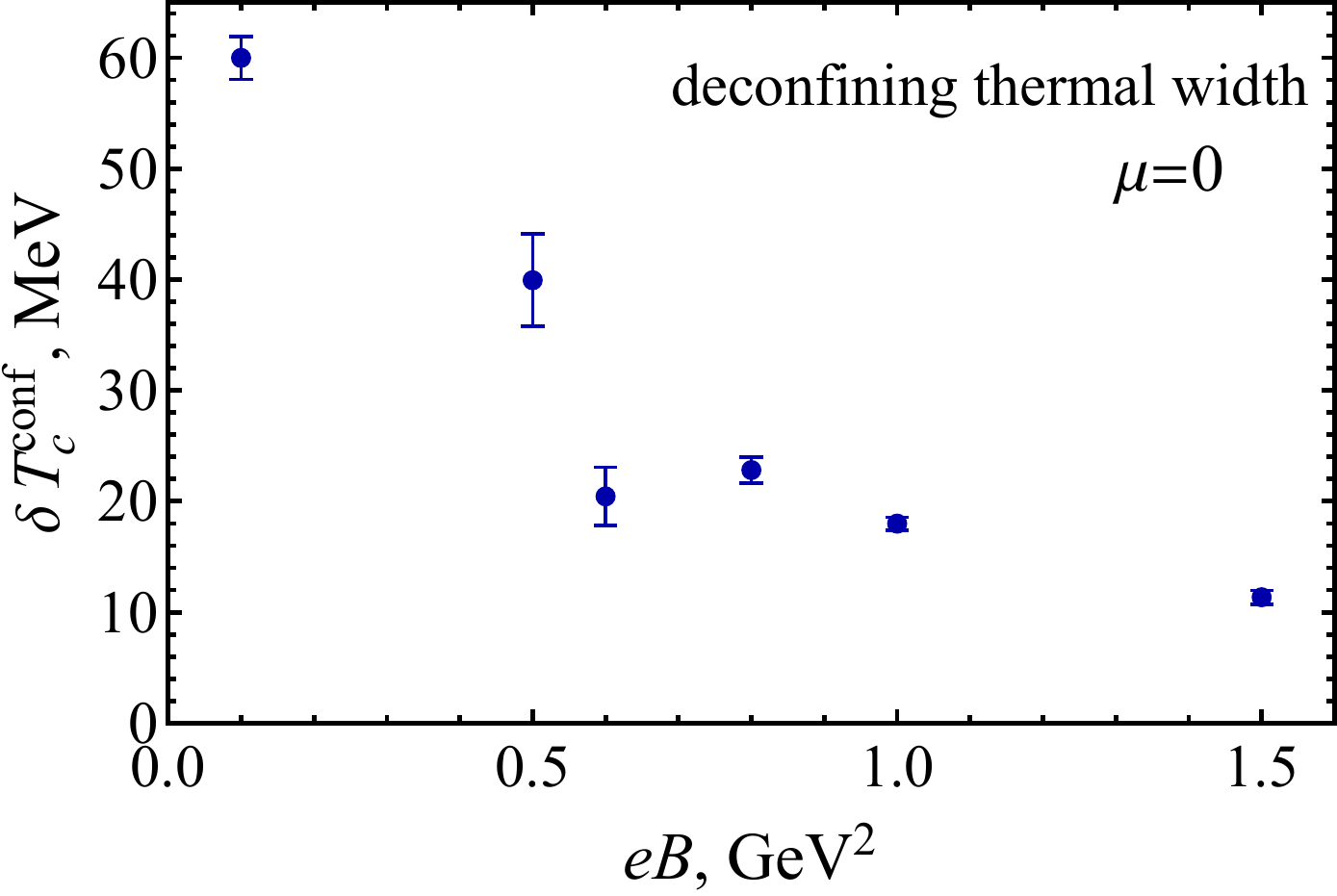} & 
\hskip 2mm \includegraphics[scale=0.415,clip=true]{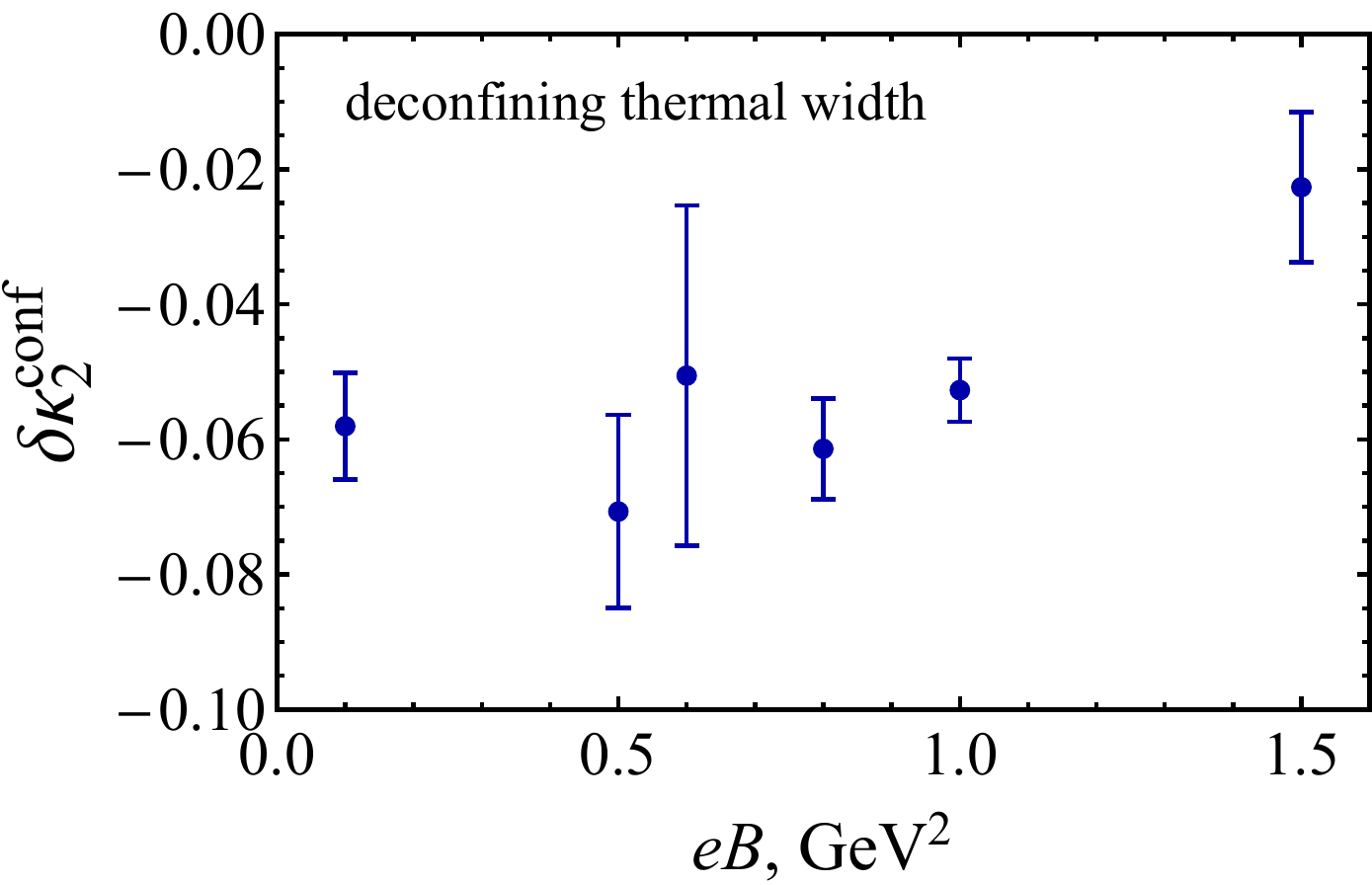} & 
\hskip 2mm \includegraphics[scale=0.41,clip=true]{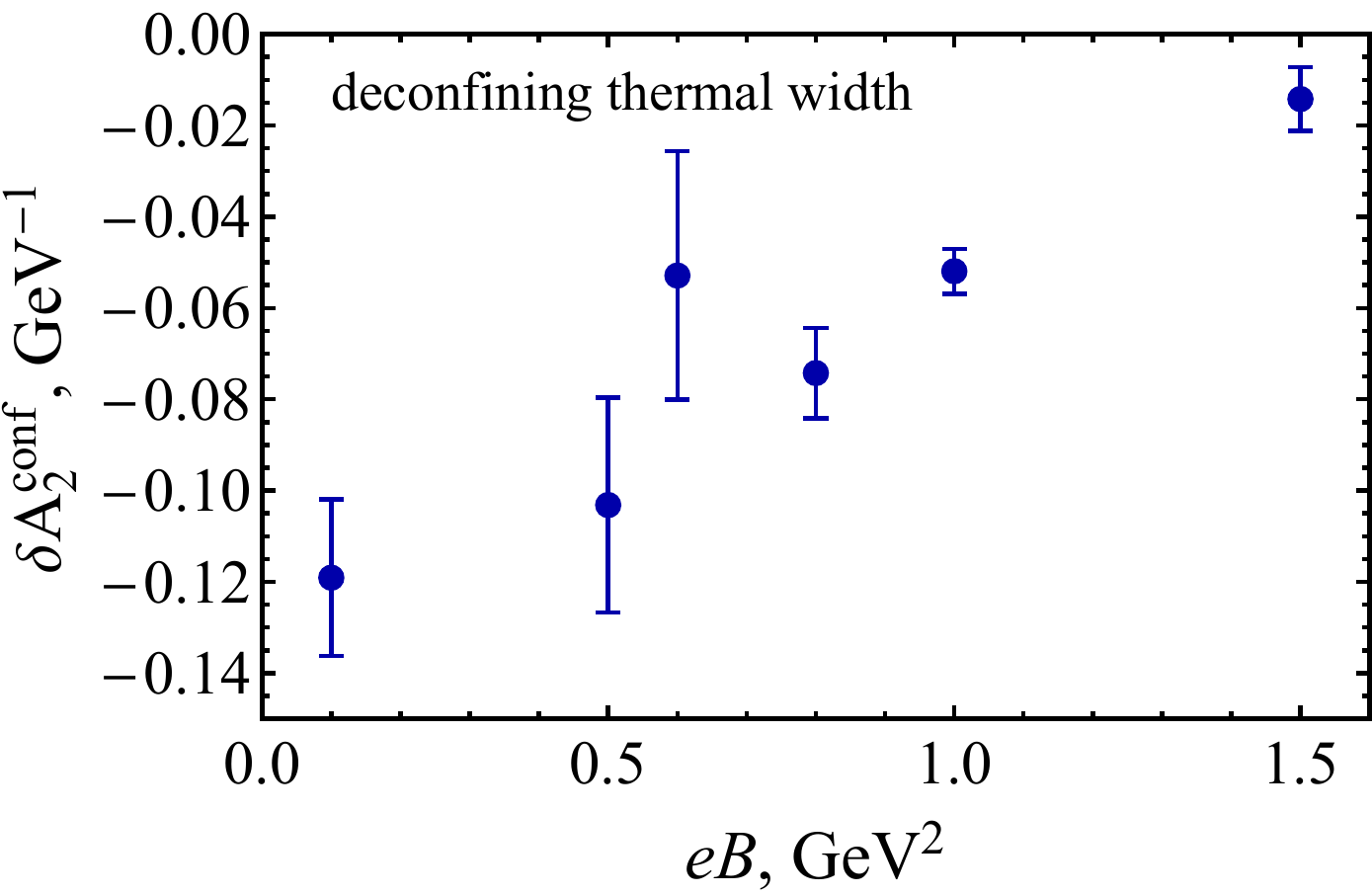} \\
(a) & 
(b) & 
(c)
\end{tabular}
\end{center}
\vskip 0mm 
\caption{(a) The thermal width $\delta T_c$ of the deconfining crossover at zero chemical potential $\mu_B=0$, as well as the curvature of the deconfining thermal width in (b) the dimensionless units  $\delta \kappa_{2}$, and (c) physical units $\delta A_2$ vs. the magnetic-field strength~$B$.}
\label{fig:delta:Tc:fits:results}
\end{figure*}

The width of the deconfining crossover is shown in Fig.~\ref{fig:delta:Tc:fits:results}(a). At low magnetic field, the deconfining crossover is very wide, $\delta T^\conf \simeq 60\,\mathrm{MeV}$ as compared with the $\delta T^\ch \simeq 11\,\mathrm{MeV}$ of the chiral crossover shown in Fig.~\ref{fig:delta:Tc:A2}(a). At the magnetic flipping point~\eq{eq:eBc} the width suddenly drops down. At largest studied magnetic field $eB = 1.5 \, \mathrm{GeV}^2$ the deconfining thermal width $\delta T^\conf \simeq 11\,\mathrm{MeV}$ becomes comparable with the chiral thermal width $\delta T^\ch \simeq 5\,\mathrm{MeV}$. 

The curvature of the deconfining thermal width is a {\it negatively}-valued quantity, as it is shown in dimensionless, Fig.~\ref{fig:delta:Tc:fits:results}(b), and physical, Fig.~\ref{fig:delta:Tc:fits:results}(c), units. The latter has been obtained with the help of Eq.~\eq{eq:delta:A2} but for the deconfining crossover.

We would like to stress that the presence of the baryonic matter makes the deconfining thermal width wider, thus softening the deconfining transition in the whole studied range of magnetic field.

The deconfining temperature and its thermal width, as well the their curvatures in the $(\mu,T)$ plane are summarized in Table~\ref{ref:parameters:transitions} below. We will discuss the general picture of the chiral and deconfining crossover transitions in the last section.

\section{Thermodynamics properties of heavy quarks}
\label{sec:entropy}

\subsection{Polyakov loop and thermodynamic potential}
The expectation value of the Polyakov loop~\eq{eq:Polyakov:loop} determines grand-canonical thermodynamic potential of the static quark $\Omega_Q$:
\beqn
|\avr{P}| = e^{-\Omega_Q/T}.
\label{eq:P}
\eeqn
Free quarks do not exist in the confining phase of QCD. Consequently, at low temperatures, the energy of an individual quark is large and the Polyakov loop is a small quantity. Notice that in the pure Yang-Mills theory the Polyakov loop is an exact order parameter of the quark confinement (the Polyakov loop vanishes in the confinement phase, $\avr{P} = 0$) while in QCD the expectation value of the Polyakov loop does not vanish exactly due to the presence of dynamical quarks. 

In the deconfining phase the free energy of a single quark is a finite quantity. However, the free energy suffers from ultraviolet divergences due to large perturbative contributions. In order to give a physical meaning to the free energy, it needs to be renormalized. In our paper, we use the gradient flow method to renormalize of the Polyakov loop~\cite{Petreczky:2015yta}. 

In a general thermodynamic system of a volume $V$, the grand-canonical thermodynamic potential $\Omega$ is related to the pressure $P$ as follow, $\Omega = - P V$. The differential of the potential is defined as follows
\beqn
d \Omega = - S d T - N d \mu - M d B\,,
\label{eq:d:Omega}
\eeqn
where entropy $S$, particle number $N$ and magnetization $M$, determine the response of the grand potential to the variations in temperature $T$, chemical potential $\mu$ and magnetic field $B$, respectively. These quantities may also be defined for a single static quark introduced into the system by the Polyakov loop operator~\eq{eq:Polyakov:loop}. They have a sense of variation in, respectively, the entropy, the (light) quark number and the magnetization of the overall system in a response of adding one infinitely-heavy, static quark. We use the subscript ``Q'' for these thermodynamic quantities in order to highlight their single-quark meaning, 

There is an important feature of our numerical approach: we perform the simulations at a fixed ratio of the imaginary chemical potential $\mu_I$ to temperature $T$ instead of fixing these quantities separately. Thus for convenience we define the ratios
\beqn
f = \frac{\mu}{T} = \frac{i \mu_I}{T}, 
\qquad
f_I = \frac{\mu_I}{T} \equiv - i f, 
\label{eq:f:fI}
\eeqn
and rewrite the differential of the free energy~\eq{eq:d:Omega} as
\beqn
d \Omega_Q = - \left( S_Q + f N_Q \right)d T - T N_Q d f - M_Q d B\,.
\label{eq:d:Omega:f}
\eeqn
Then it is easy to obtain the following relations in terms of $(f,\,T,\,B)$ variables:
\beqn
S_Q & = & - {\left( \frac{\partial \Omega_Q}{\partial T} \right)}_{f,B} + \frac{f}{T} {\left( \frac{\partial \Omega_Q}{\partial f} \right)}_{T,B}, 
\label{eq:S:f}\\ 
M_Q & = & - {\left( \frac{\partial \Omega_Q}{\partial B} \right)}_{T,f}.
\label{eq:M:f}
\eeqn
It is worth noticing that the baryon number, determined by a differentiation of the free energy with respect to the baryon chemical potential, is formally an imaginary quantity in our case. However, its physical meaning remains the same, and the baryon number may, in principle, be analytically continued to the domain of the real chemical potential. We do not analyze this quantity in the paper because the accuracy of our numerical data does not allow us to extract the baryon number density unambiguously.

\subsection{Single-quark entropy}
\label{subsec:S_Q}

Despite the imaginary nature of the chemical potential $\mu_I$, the entropy~\eq{eq:S:f} may be determined reliably in the region where the thermodynamic potential is an analytic function of the chemical potential $\mu$. Indeed, Eq.~\eq{eq:f:fI} implies the relation $f \partial_f \equiv f_I \partial_{f_I}$ which may be used to compute the last term of the entropy of the single quark~\eq{eq:S:f}. In this case the entropy of the single quark may be directly expressed via the renormalized Polyakov loop $\avr{P}^r$:
\beqn
S_Q = \ln |\avr{P}^r| + \frac{\partial \ln |\avr{P}^r|}{\partial \ln T}  - \frac{\partial \ln |\avr{P}^r|}{\partial \ln f_I},
\label{eq:S:Q}
\eeqn
where the dimensionless ratio $f_I$ is equal to $\mu_I / T$. 

Although the quark entropy~\eq{eq:S:Q} is not sensitive to the $\Z_3$ center symmetry, it is expected to pinpoint a (phase) transition between the low-temperature and high-temperature regions. The entropy of the single heavy quark has a peak -- a local maximum as a function of temperature at other parameters fixed -- which is close to the pseudo-critical temperature of the chiral crossover~\cite{Weber:2016fgn}.

\begin{figure*}[!thb]
\begin{center}
\begin{tabular}{ccc}
\includegraphics[scale=0.37,clip=true]{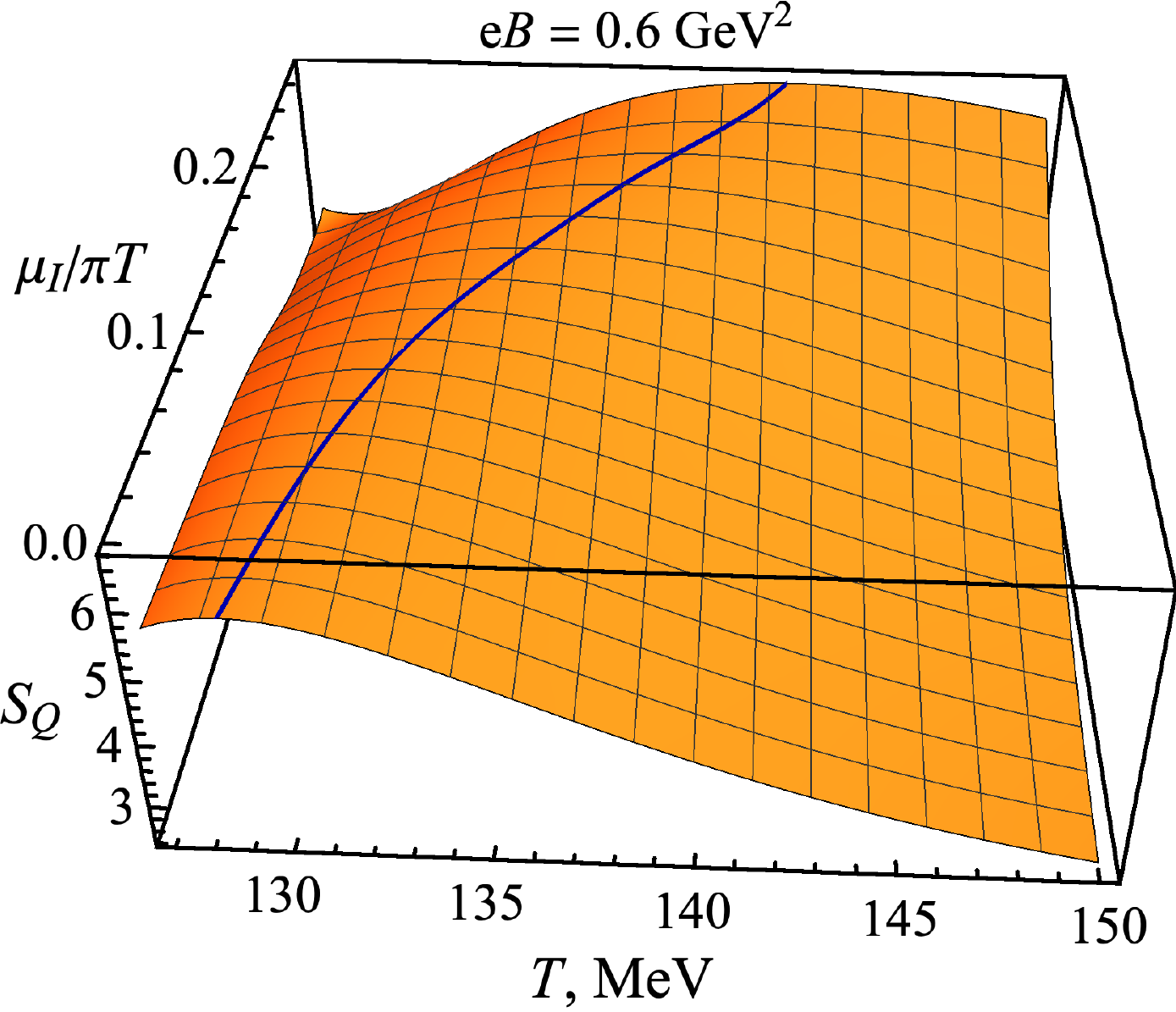} &
\hskip 5mm \includegraphics[scale=0.37,clip=true]{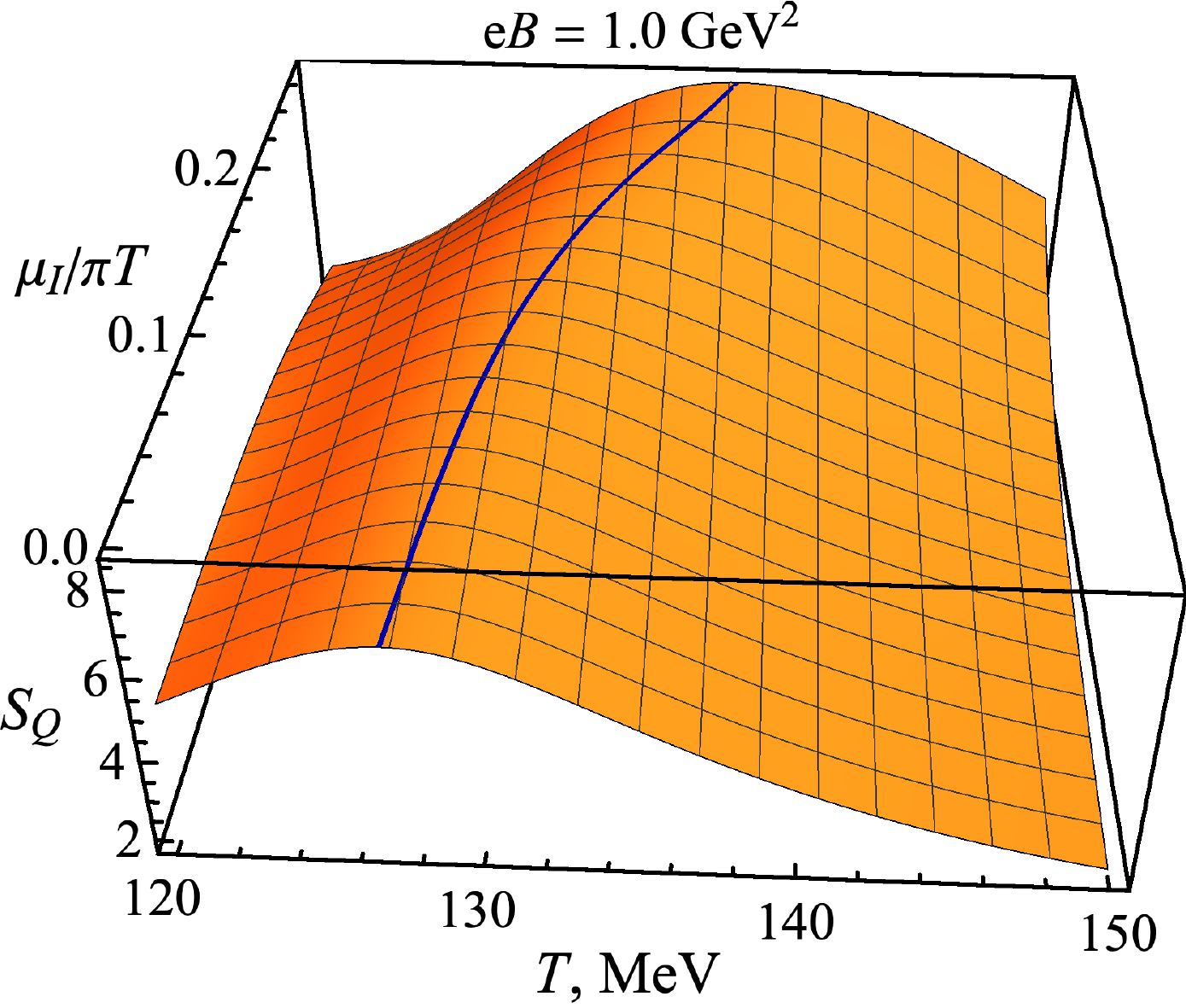} & 
\hskip 5mm \includegraphics[scale=0.37,clip=true]{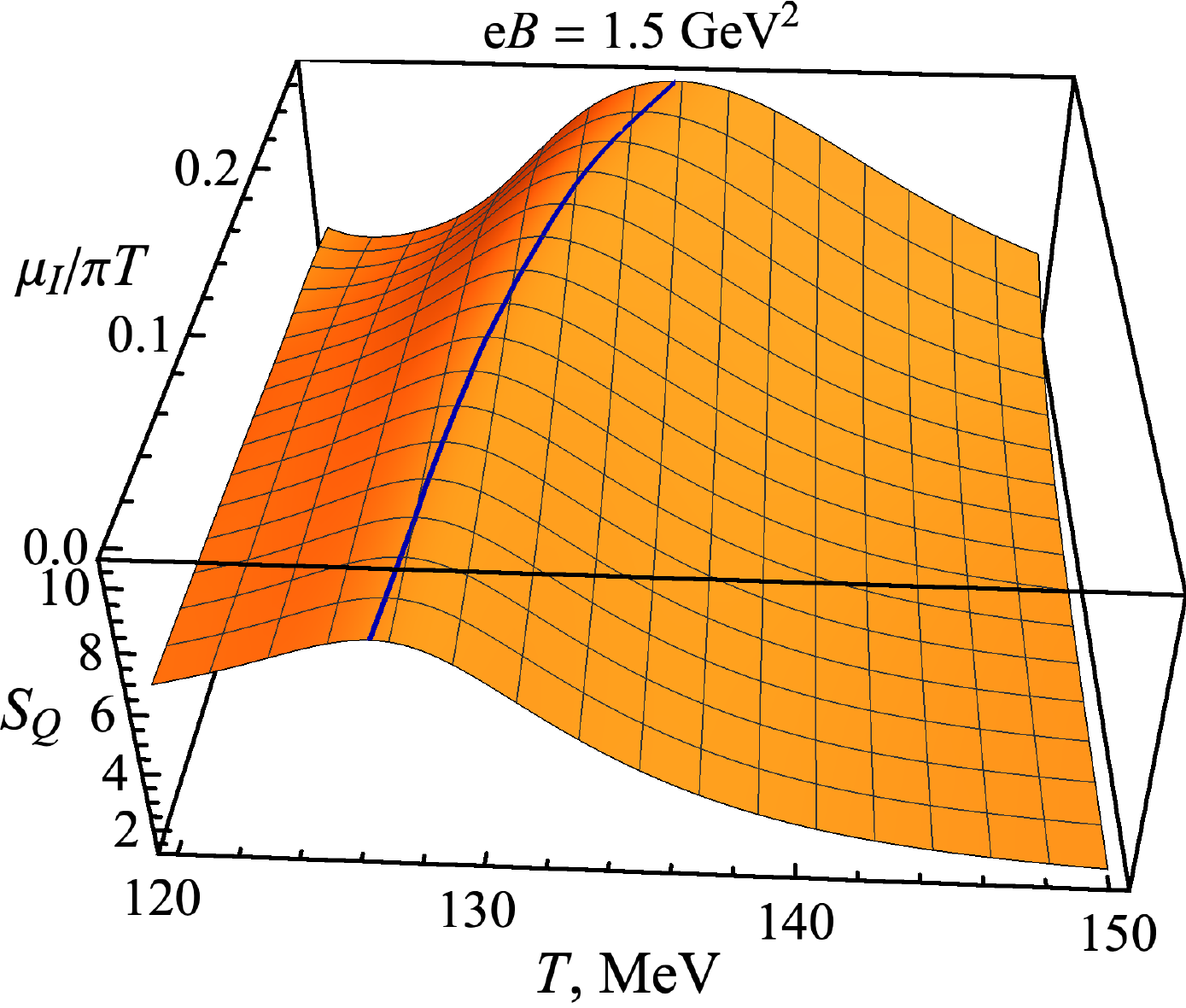} \\
(a) & (b) & (c)
\end{tabular}
\end{center}
\vskip 0mm 
\caption{A smooth interpolation of the single-quark entropy~\eq{eq:S:Q} as the function of temperature $T$ and imaginary chemical potential $\mu_I$ for three values of magnetic fields, $eB=(0.6,\,1.0,\,1.5)\,\mathrm{GeV}^2$, on $24^3 \times 6$ lattice. The blue solid curve marks the maximum of the quark entropy at each $\mu_I$. Only mean values without the errorbars are plotted for the sake of clarity.}
\label{fig:S:Q:3D}
\end{figure*}

In Fig.~\ref{fig:S:Q:3D} we show the single-quark entropy, computed as (\ref{eq:S:Q}), in the parameter plane of the temperature $T$ and the normalized imaginary chemical potential $\mu_I/(\pi T)$. At each value of the chemical potential the quark entropy has a maximum point which is denoted by a solid blue line. As the imaginary chemical potential increases, the peak is shifted towards higher values of temperature, which is in the qualitative agreement with the picture obtained from our studies of the chiral and deconfinement phase transitions. Unfortunately, with current ensembles we are able to determine the position of the entropy peak only for sufficiently strong magnetic fields, $eB > 0.5\,\mbox{GeV}^2$. For these large fields, the entropy takes it maximum, in the zero-density limit, at $T \sim 127(10)\,\mathrm{MeV}$, in consistency with the position of the common line of chiral and deconfining crossovers.

\begin{figure*}[!thb]
\begin{center}
    \begin{tabular}{ccc}
    \includegraphics[scale=0.37,clip=true]{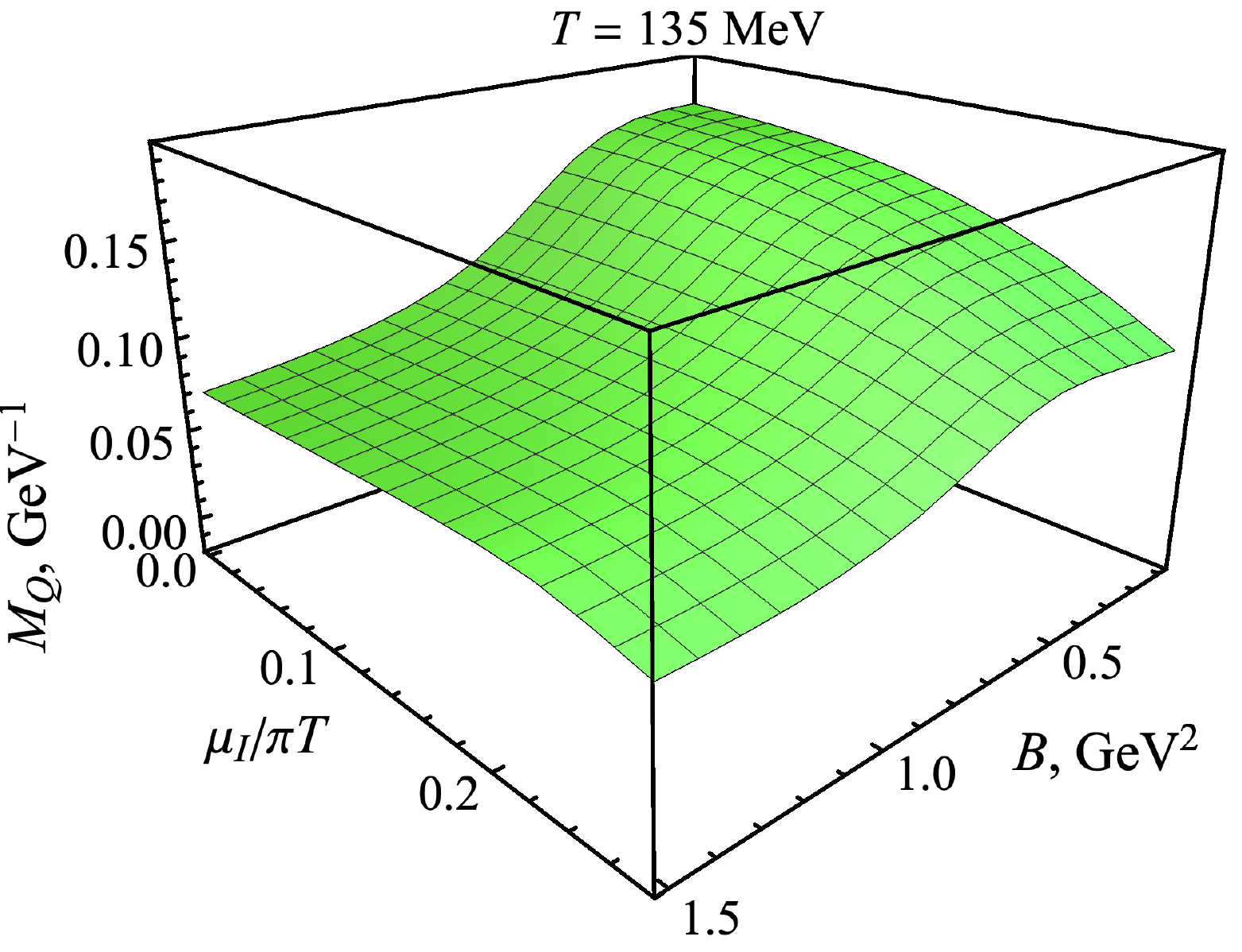} &
    \includegraphics[scale=0.37,clip=true]{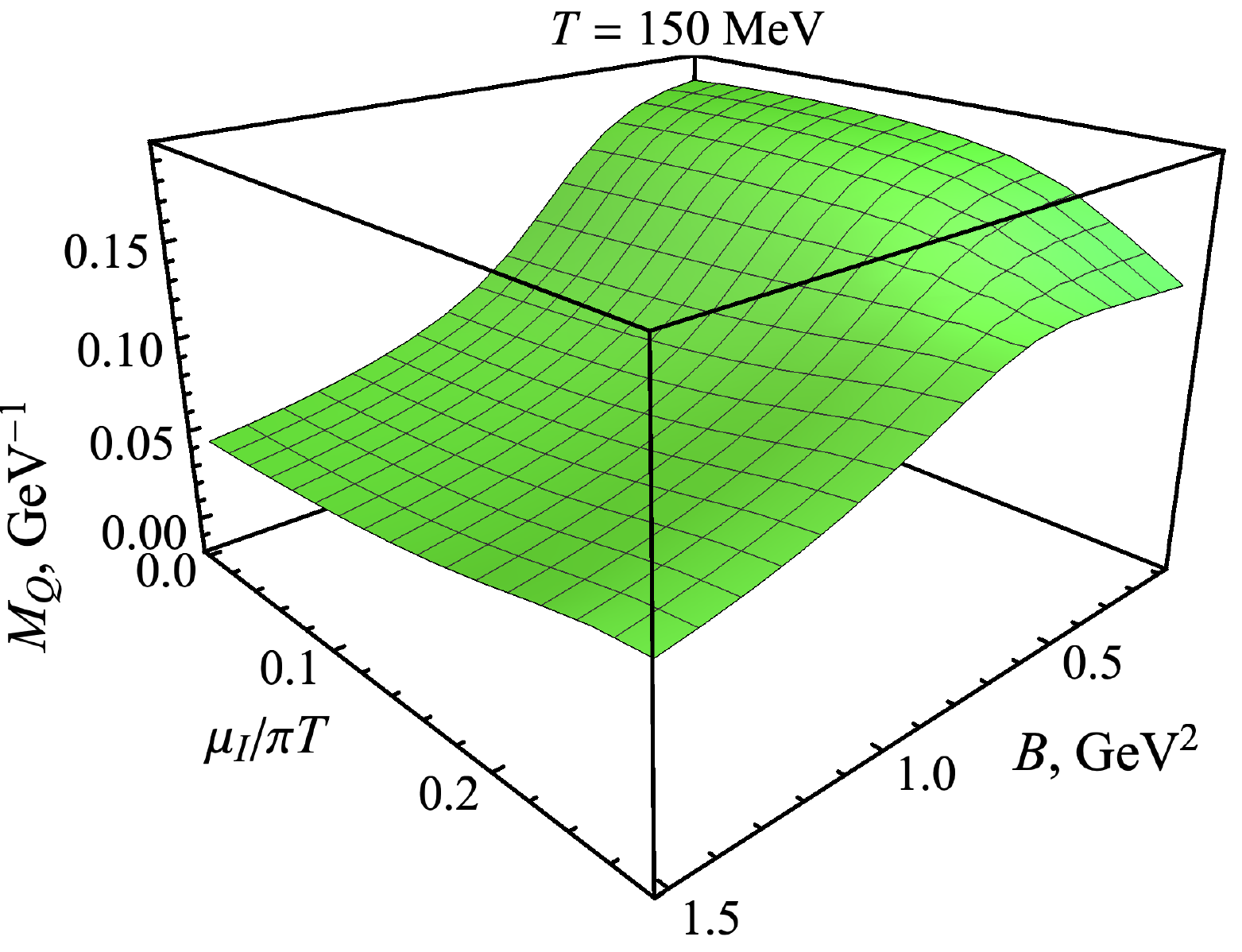} & 
    \includegraphics[scale=0.37,clip=true]{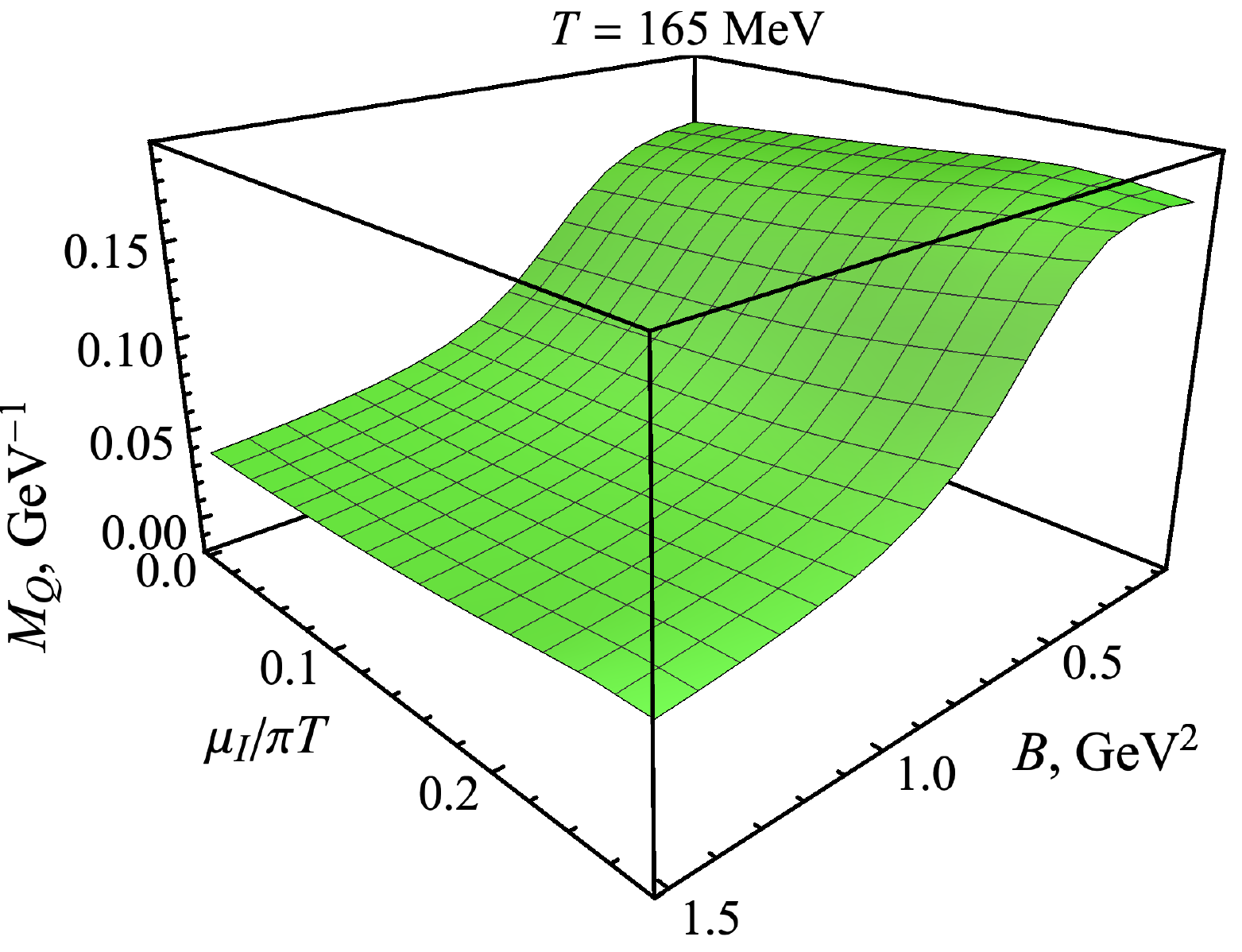} \\[3mm]
    \includegraphics[scale=0.37,clip=true]{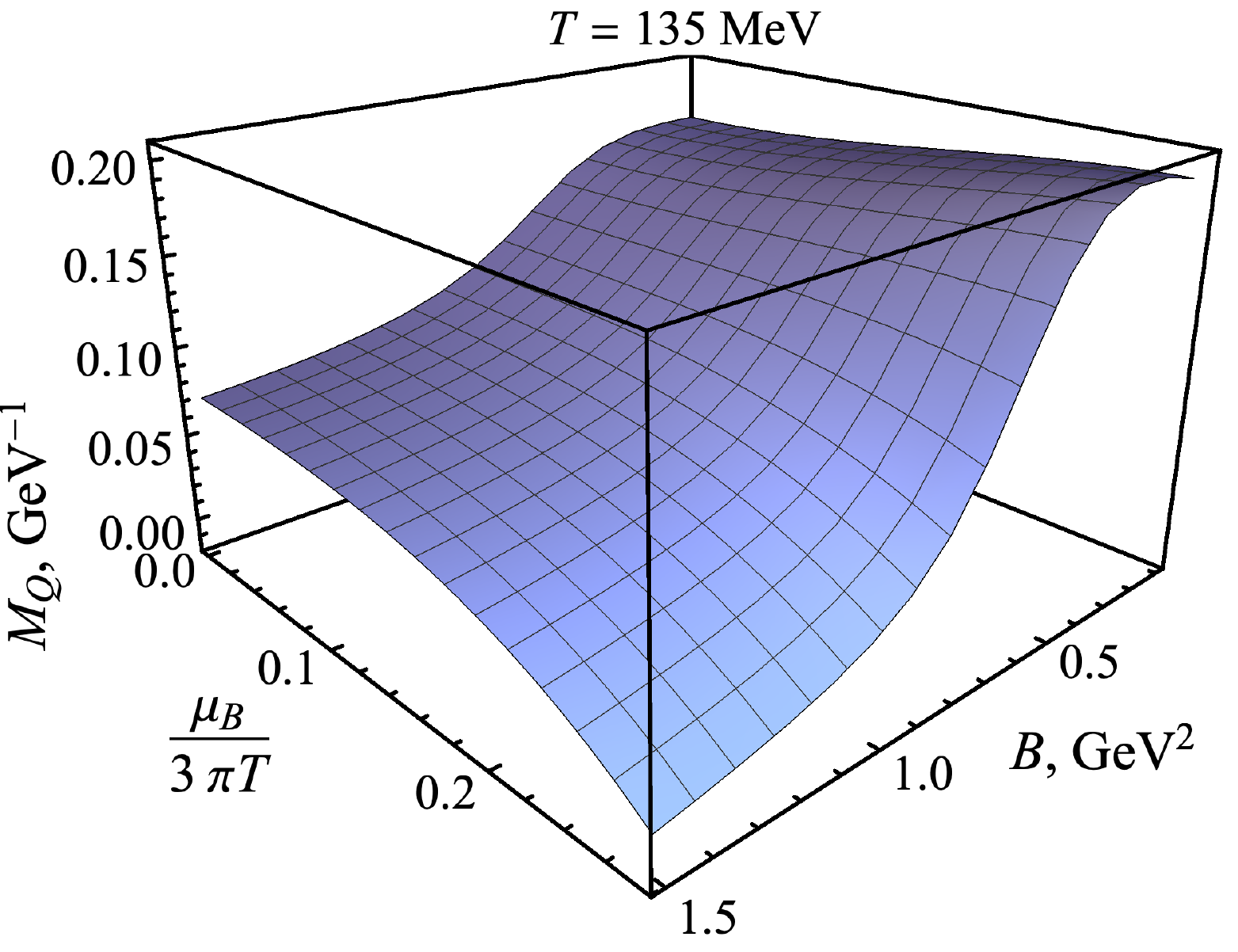} &
    \includegraphics[scale=0.37,clip=true]{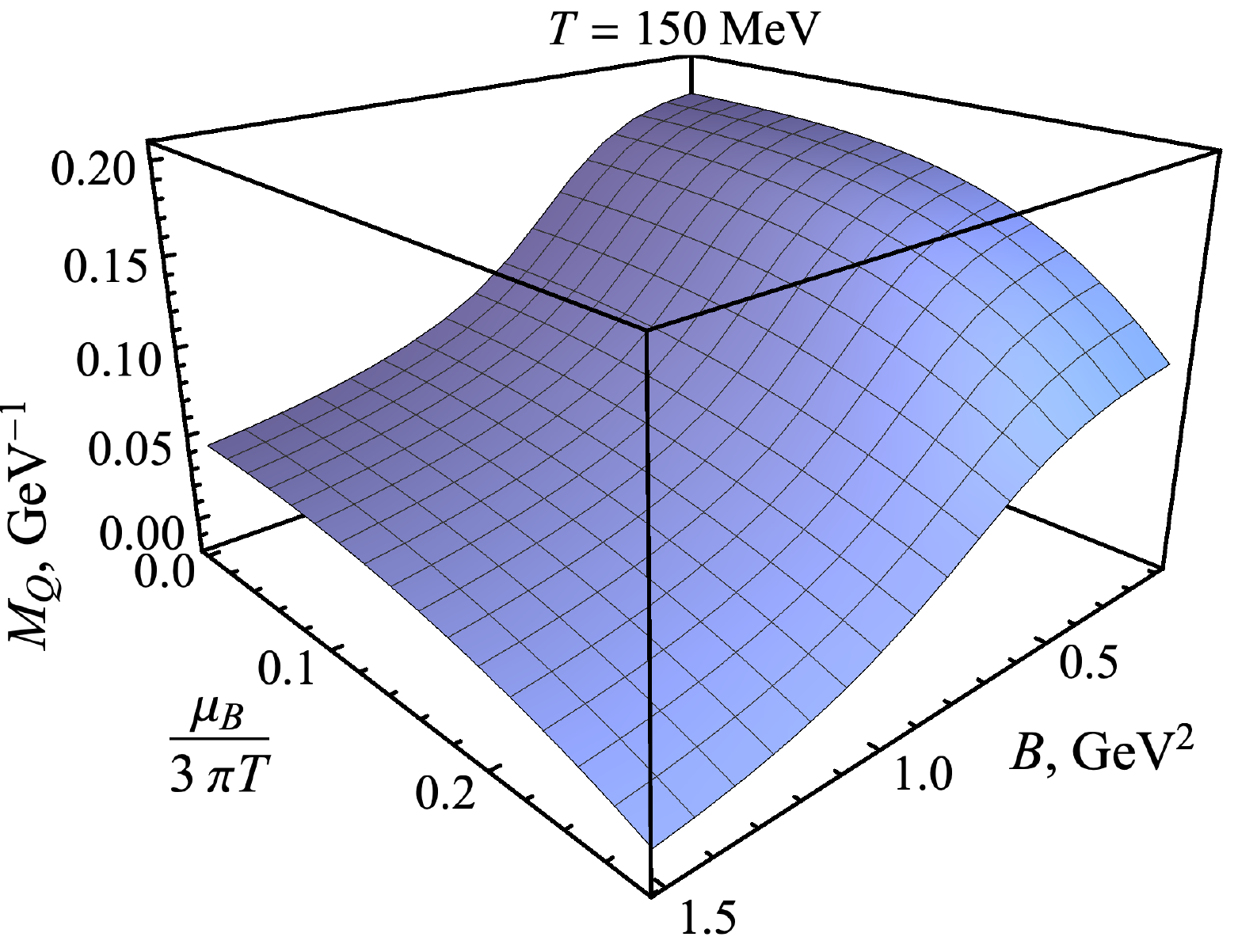} & 
    \includegraphics[scale=0.37,clip=true]{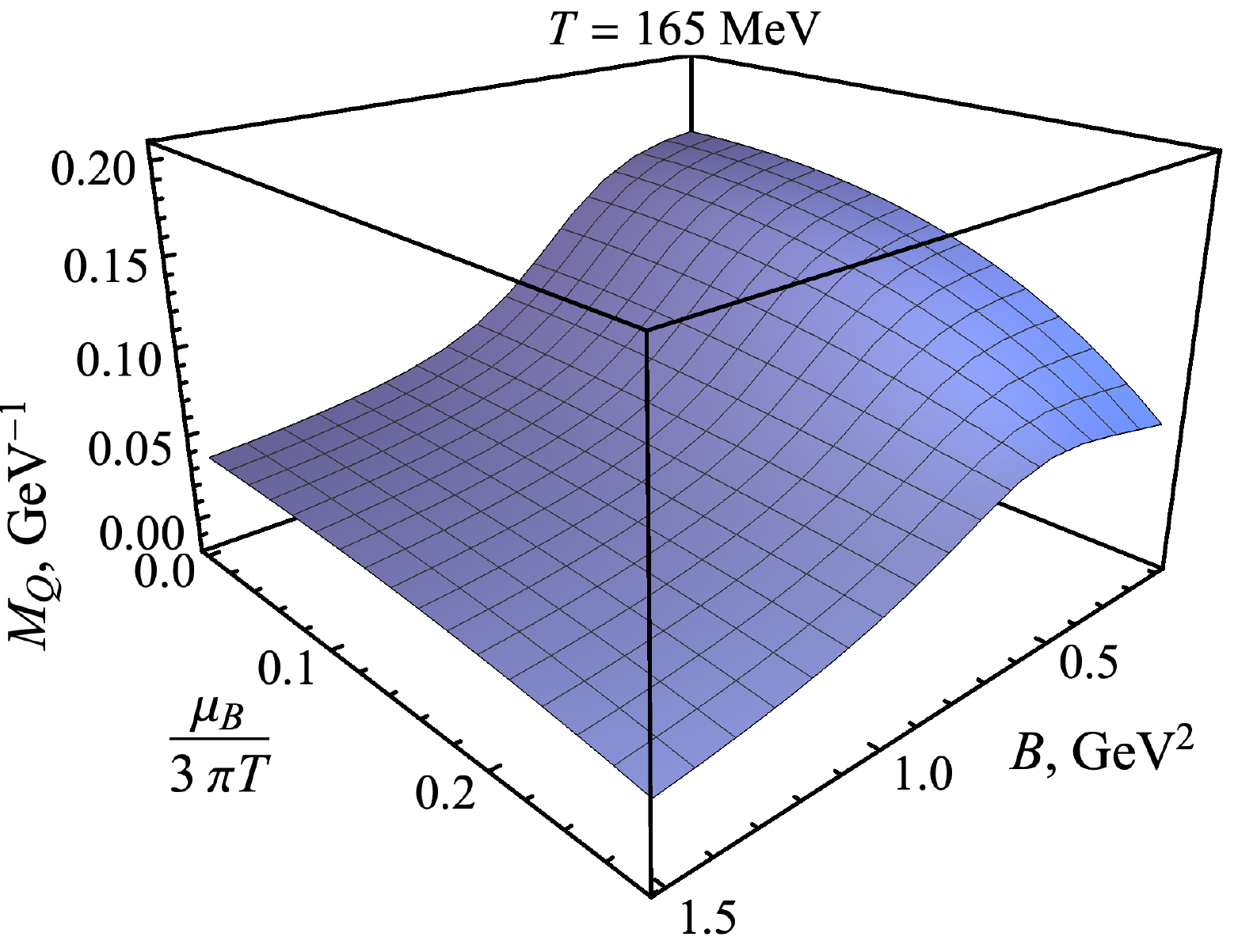} \\
    (a) & (b) & (c)
    \end{tabular}
\end{center}
\vskip 0mm 
\caption{A smooth interpolation of the single-quark magnetization~\eq{eq:M:f} as the function of magnetic field $B$ and (upper plots) imaginary chemical potential $\mu_I$ or (lower plots) real baryon chemical potential $\mu_B$ for three values of temperatures, at (a) lower, (b) middle, (c) upper regions of the crossover transition, $T=(135,\,150,\,165)\,\mathrm{MeV}$, calculated on a $24^3 \times 6$ lattice. Only mean values without the errorbars are plotted for the sake of clarity, relative errors are at the rate of 30\%.}
\label{fig:M:Q:3D}
\end{figure*}

The calculation of the quark entropy requires an interpolation of the renormalized Polyakov loop to a continuous range of temperatures $T$ and normalized imaginary chemical potentials $f_I = \mu_I/T$ to properly compute derivatives in~(\ref{eq:S:Q}). In this case good enough resolution in terms of discrete $(T,\,\mu_I/T)$-points in the phase transition region is especially important. Unfortunately, we currently have 3 -- 5 temperature points in the region near $T_c$ for all chemical potential and magnetic field values, which turns out to be not enough for the proper estimation of peak in $S_Q$ (cubic B-spline with smoothing was employed for interpolation). Moreover, statistics consist of 100 -- 200 configurations per $(T,\,B,\,f_I)$ set, thus relative errors in $T_c$ obtained from the maximum of single quark entropy reach 10\%. We leave thoughtful study of $S_Q$ for future papers.

\subsection{Magnetization}
\label{subsec:M_Q}

The single-quark magnetization~\eq{eq:M:f} is a real-valued quantity, which may be computed straightforwardly from the renormalized Polyakov loop and then the analytically continued to the real-valued chemical potential. On a first glance the physical meaning of the single-quark magnetization~\eq{eq:M:f} is somewhat obscure since this quantity is associated with the presence of 
\begin{itemize}
\item[(i)] a static, infinite-heavy quark, which 
\item[(ii)] does not possess a spin degree of freedom, and 
\item[(iii)] has zero electric charge. 

\end{itemize}

Due to the latter property, the test quark is not directly coupled to the external magnetic field. Moreover, the immobility of the test quark means that it does not contribute to the Landau diamagnetism, while the absence of the spin, and, consequently, of the magnetic moment, implies the lack of the Pauli paramagnetic contribution. Therefore, one could naively argue that the external test quark would not affect the magnetization properties of the system. On the other hand, the immobile chargeless spinless quark may still affect the electromagnetic properties of the medium since its presence modifies -- via the gluon-mediated interactions -- the distribution of the dynamical quarks around it, which, in turn, do couple to the background magnetic field and contribute to the overall magnetization of the system. Therefore, the single-quark magnetization has a meaning of the extent with which the test quark affects the electromagnetically active dense medium of charged quarks and antiquarks.

In Fig.~\ref{fig:M:Q:3D} we show the single-quark magnetization~\eq{eq:M:f} computed for three characteristic temperatures at the low (135\,MeV), middle (150\,MeV), and upper (165\,MeV) parts of the crossover transition. The upper row of plots in Fig.~\ref{fig:M:Q:3D} corresponds to the actual data obtained for the imaginary chemical potential~$\mu_I$. In the low-density region, one can perform an analytical continuation of the magnetization data $M_{\mathrm{Im}}(\mu^2_I)$ by (i) first expanding the magnetization via the series of the even powers of the imaginary chemical potential~$\mu_I$; (ii) and then using the relation, $\mu_B^2 = - (3 \mu_I)^2$, to get obtain the desired function $M(\mu_B) = M_{\mathrm{Im}}(- (\mu_B/3)^2)$. At a practical side, we found that the numerical data for $M_{\mathrm{Im}}(\mu_I)$ at the imaginary chemical potential may be described, at a satisfactory level, by the quartic dependence 
\beqn
M_{\mathrm{Im}}(\mu_I) = C_0 + C_2 \mu_I^2 + C_4 \mu_I^4,
\label{eq:M:fit}
\eeqn
for all values of magnetic field. Here $C_i$ are dimensional fitting parameters. Notice that we make the fits~\eq{eq:M:fit} of magnetization at fixed temperatures and magnetic fields so that $C_i = C_i(B,T)$. The analytically continued single-quark magnetization,
\beqn
M(\mu_B) = M_0 - \kappa^{(M)}_2 \left( \frac{\mu_B}{3 \pi T} \right)^2 + \kappa^{(M)}_4  \left( \frac{\mu_B}{3 \pi T}, \right)^4, \quad
\eeqn
is shown in the lower row of Fig.~\ref{fig:M:Q:3D}. From these figures one readily observe that the single-quark magnetization is a positive quantity in the whole range of studied parameters $(\mu_I/T,\,B)$. In other words, heavy quarks contribute paramagnetically to the overall magnetization of the quark-gluon plasma. Moreover, this paramagnetic contribution is enhanced with the increase of the magnetic field.

Usually, the effect of the heavy quarks on the magnetic polarization of the quark-gluon plasma is ignored because the massive quarks behave as non-relativistic particles for which both the (spin-related) magnetic moment and the (orbital-related) cyclotron frequency are suppressed by the heavy mass. Here we show that (even, infinitely) heavy quarks are magnetically-active constituents of the plasma as they contribute paramagnetically to the overall magnetization.

In order to get a suitable continuous description shown in Figs.~\ref{fig:M:Q:3D}, we interpolated the data for the single-quark magnetization using the method splines, largely following our approach to of the single-quark entropy $S_Q$. We found that while our data may be used to reliably estimate the inflection point of the Polyakov loop, the presented data for magnetization may contain systematic inaccuracies related to the scarce grid of the data used for the interpolation. This point needs a further investigation.

It is instructive to compare our results on the single-quark magnetization with the behavior of the ``bulk'' magnetization of the quark-gluon plasma obtained in Ref.~\cite{Bali:2013owa} in a zero-density limit of QCD. The bulk magnetization of the $\mu_B = 0$ quark-gluon plasma is a positive quantity according to~\cite{Bali:2013owa}, thus the zero-density QCD is a paramagnetic medium. The paramagnetic response of the bulk quark-gluon plasma increases in strength both with the increase of temperature and the strengthening of the magnetic field~\cite{Bali:2013owa}. Our results at $\mu_B = 0$ do not show a significant increase in the single-quark magnetization, which may still be consistent with the results of Ref.~\cite{Bali:2013owa} because our temperature interval (30 MeV) is much shorted compared to that of the quoted reference (about 200 MeV). However, we observe the weakening of the single-quark magnetization with the strengthening of the magnetic field in a sharp contrast with the observed strengthening of the bulk magnetization. 

It worth noticing that the single-quark magnetization and bulk magnetization can not be compared with each other directly because these quantities have different physical meanings and they even possess different dimensions: the former quantity corresponds to the energy of a single heavy static quark while the latter number characterizes the energy {\it{density}} of the bulk medium. Nevertheless, both quantities characterise the magnetic properties of the strongly interacting medium subjected to an intense magnetic field background.

\section{Overall picture and Conclusions}

\begin{figure*}[!bht]
\begin{center}
\includegraphics[scale=0.33,clip=true]{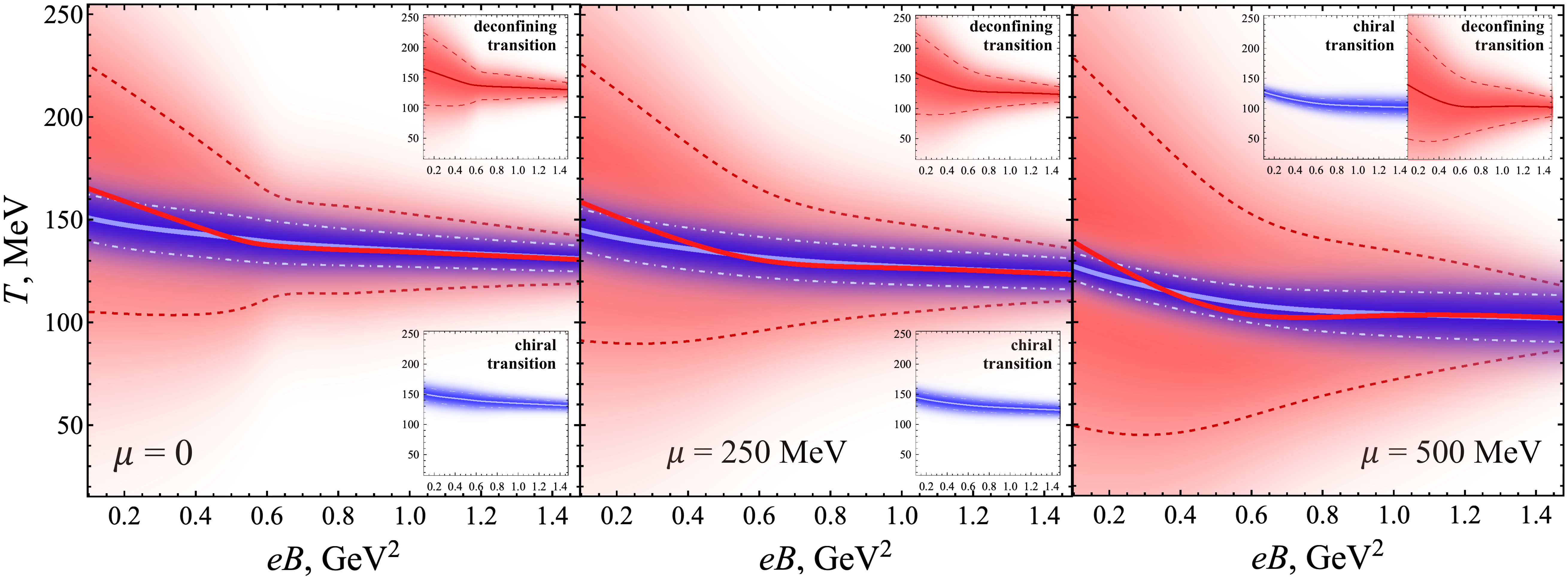}
\end{center}
\vskip 0mm 
\caption{The chiral (blue) and deconfining (red) crossover transitions at the baryonic chemical potential (from left to right) $\mu_B = 0,\, 250\,$, and $ 500\,\mathrm{MeV}$.  The solid lines denote the middle positions of the crossovers, $T = T_c$, and the dash-dotted and dashed lines show their widths, $T = T_c \pm \delta T_c$ as the function of magnetic field.}
\label{fig:phases:widths}
\end{figure*}

\begin{table*}[htb]
\begin{tabular}{|c|c|c||c|c||c|c|}
\cline{2-7}
\multicolumn{1}{c}{} & 
\multicolumn{2}{|c||}{pseudo-critical temperature} & \multicolumn{2}{|c||}{curvature of pseudo-} & \multicolumn{2}{|c|}{curvature of} \\
\multicolumn{1}{c}{} & 
\multicolumn{2}{|c||}{and thermal width at $\mu=0$} & \multicolumn{2}{|c||}{critical temperature $T_c$} & \multicolumn{2}{|c|}{thermal width $\delta T_c$} \\
\hline
$eB$ &	$T_c$, MeV	& $\delta T_c$, MeV & $\kappa_2$ 		& $A_2,\,\mathrm{GeV}^{-1}$ & $\delta \kappa_2$ & $ \delta A_2,\, \mathrm{GeV}^{-1}$ \\
\hline
\rowcolor{lightblue}
\multicolumn{7}{|c|}{chiral crossover} \\
\hline
\rowcolor{verylightblue}
0	&	156.5(1.5)${}^{(a)}$		&	--			&	0.0132(18)${}^{(c)}$	&		~~~~\,0.085(12)${}^{(d)}$			& 	 --			&	 --      	\\
\rowcolor{verylightblue}
0.1	&	148.3(2)	&	11.4(3)	&	0.0145(7)		&		0.097(6)			&	 0.044(10)		&	 0.019(5)	\\
\rowcolor{verylightblue}
0.5	&	141.7(3)	&	11.2(4)	  &0.0174(10)	&		0.123(7)		&	 0.020(10)		& 	 0.010(5)    \\
\rowcolor{verylightblue}
0.6	&	139.0(3)	&	10.4(4)	&	0.0183(9)		&		0.132(6)			&	 0.025(12)		&	 0.011(6)    \\
\rowcolor{verylightblue}
0.8	&	136.8(3)	&	 9.0(3)	&	0.0170(6)		&		0.125(4)			&	-0.011(8)		&	-0.008(4)   \\
\rowcolor{verylightblue}
1.	&	134.89(13)	& 	 7.8(2)	&	0.0168(4)		&		0.125(3)			&	-0.045(5)		&	-0.014(2)   \\
\rowcolor{verylightblue}
1.5	&	130.46(15)	&	 5.2(4)	&	0.0153(3)		&		0.117(2)			&	-0.047(4)		&	-0.021(2)   \\
\rowcolor{verylightblue}
\hline
\rowcolor{lightred}
\multicolumn{7}{|c|}{deconfining crossover} \\
\hline
\rowcolor{verylightred}
0	& 171(3)${}^{(b)}$		&	--		&	--			&	--		&	--		&	--		\\
\rowcolor{verylightred}
0.1	& 165.4(1.1)	&	60(2)	&	0.017(2)	&	0.103(10)	&	-0.058(8)	&	-0.12(2)	\\
\rowcolor{verylightred}
0.5	& 139.8(1.6)	&	40(4)	&	0.024(3)	&	0.174(18)	&	-0.070(14)&	-0.10(2)	\\
\rowcolor{verylightred}
0.6	& 137.4(1.2)	&	20(3)	&	0.020(2)	&	0.142(17)	&	-0.051(25)&	-0.05(3)	\\
\rowcolor{verylightred}
0.8	& 135.1(5)	&	22.8(1.2)	&	0.0183(9)		&	0.135(7)	&	-0.061(7)	&	-0.07(1)	\\
\rowcolor{verylightred}
1.	& 134.5(2)	&	18.0(6)	&	0.0153(3)		&	0.113(3)	&	-0.053(5)	&	-0.052(5)	\\
\rowcolor{verylightred}
1.5	& 130.3(2)	&	11.4(6)	&	0.0148(4)		&	0.114(3)	&	-0.023(11)&	-0.014(7)	\\
\hline
\end{tabular}
\caption{The characteristics of the chiral and deconfining crossovers vs. the magnetic field $B$ established by the arctan-type fitting used to identify the inflection points of the light-quark condensate~\eq{eq:condensate:fit} and the Polyakov loop~\eq{eq:L:fit}, respectively. We show the pseudo-critical temperatures $T_c$, the widths $\delta T_c$, as well as the dimensionless quadratic curvatures of the crossover temperature $\kappa_2$ and its width~$\delta \kappa_2$, determined, correspondingly via the quadratically truncated fits~\eq{eq:Tc:I} and \eq{eq:delta:Tc:I}. The curvatures in the physical units, $A_2$ and $\delta A_2$, are found via the fits~\eq{eq:A2} and~\eq{eq:delta:A2}. The marks denote the data taken from other sources: (a) Ref.~\cite{Bazavov:2018mes}, (b) Ref.~\cite{Aoki:2006we} and (c) Ref.~\cite{Bonati:2014rfa}. The data point (d) is derived from (c) via Eq.~\eq{eq:A2}. Note that for our data we present only the statistical errors, while the points (a)-(d) include also systematic uncertainties coming from an extrapolation to the continuum limit. }
\label{ref:parameters:transitions}
\end{table*}

In our work, we studied the influence of the strong magnetic field on the chiral and deconfinement transitions in finite-temperature QCD at a low baryonic chemical potential. In the low-density QCD with real (physical) masses of $u$, $d$, and $s$ quarks, these transitions are not accompanied by any thermodynamic singularities in the parameter space of the theory. Instead, the theory experiences a smooth broad crossover from the cold chirally-broken hadronic medium to the hot chirally-symmetric plasma of deconfined quarks and gluons. 

In the absence of a real phase transition, the positions of the chiral and deconfining crossovers are not well defined; they depend on a particular form of the operator employed to probe them. In our paper, we identify the location of the chiral crossover as the inflection point of the expectation value of the chiral condensate of light quarks, which is an exact order parameter for the chirally broken phase in QCD with massless quarks.

We reveal the location of the deconfining crossover via the inflection point of the expectation value of the gradient-flow-renormalized Polyakov loop, which is the order parameter for the deconfinement phase transition in a pure Yang-Mills theory (QCD with infinitely massive quarks). 

In addition to the positions of the chiral and deconfining crossover lines in the parameter space, we determined the thermal width of each of these crossovers. In the absence of thermodynamically singular behaviour, the thermal width may serve as a quantitative characteristic of the strength of the crossover transition. The thermal width $\delta T$, formally defined via the fitting functions \eq{eq:condensate:fit} and \eq{eq:L:fit} -- can be understood as a temperature range over which the corresponding order parameter reaches its values at both sides of the crossover. Similarly to the positions of the crossover lines, their thermal widths are prescription-dependent quantities which may depend on the operator used to identify them.

We performed the calculations on $N_t = 6,8$ lattices generated with Symanzik improved gluons, and stout-improved 2+1 flavor staggered fermions at physical quark masses and imaginary baryonic chemical potential. We used the analytical continuation from purely imaginary to real-valued baryon chemical potential. 

Below, we summarize all effects of the magnetic-field background on the chiral and confining  crossover transitions at low baryonic densities and finite temperature.
    
\vskip 5mm
\bum\ {\bf The chiral crossover}:
\vskip 3mm

\begin{enumerate}

\item The effect of the inverse magnetic catalysis extends to the region of low baryon densities: the chiral crossover temperature drops down as the background magnetic field strengthens. Moreover, the presence of the baryonic matter, the effect of the inverse magnetic catalysis becomes slightly stronger.

\item The quadratic curvature $A^\ch_2(B) \equiv \kappa_2(B)/T_c$ of the chiral crossover transition, 
\beqn
T^\ch(B,\mu_B) = T^\ch(B,0) - A^\ch_2(B) \mu_B^2 + {\dots},
\label{eq:Tc:chiral}
\eeqn
experiences a local maximum at the ``magnetic flipping field''~\eq{eq:eBc} which is approximately given by the scale of the $\rho$-meson mass, $eB^\ch_{\fl} \simeq 0.6 \, \mathrm{GeV}^2 \simeq m_\rho^2$, Fig.~\ref{fig:A2}. The presence of magnetic field generally enhances the curvature $A^\ch_2(B)$.

\item The thermal width $\delta T^\ch_c$ of the chiral crossover shrinks approximately twice, from $\delta T^\ch_c \simeq 11\,\mathrm{MeV}$ at vanishing field to $\delta T^\ch_c \simeq 5\,\mathrm{MeV}$ at the maximal studied strength, $eB = 1.5 \, \mathrm{GeV}^2$, Fig.~\ref{fig:delta:Tc:A2}(a). 
\item The properties of thermal width $\delta T^\ch_c$ at finite baryon density allowed us to estimate the location of the chiral critical endpoint in the $T-\mu_B$ plane at vanishing magnetic field: $(T^{\text{CEP}},\mu_B^{\text{CEP}})\simeq (100,800)$ MeV, Eq.~\eq{eq:cep:T:mu}.

\item The curvature $\delta A^\ch_2$ of the chiral thermal width, 
\beqn\delta T^\ch(B,\mu_B) = \delta T^\ch(B,0) - \delta A^\ch_2(B) \mu_B^2 + {\dots},
\label{eq:delta:Tc:chiral}
\eeqn
changes its sign at the magnetic flipping point, $eB \simeq 0.6\,\mathrm{GeV}^2$, as shown in  Fig.~\ref{fig:delta:Tc:A2}(c). Thus, the presence of the baryon matter makes the chiral crossover transition narrower (wider) in the magnetic-field background with $B < B_{\fl}$ ($B > B_{\fl}$). 

\end{enumerate}

\vskip 3mm
\bum\ {\bf The deconfining crossover}:
\vskip 3mm

\begin{enumerate}
\setcounter{enumi}{6}

\item The deconfining crossover experiences the inverse magnetic catalysis as well, Fig.~\ref{fig:Tc:fits:results}(a). 

\item The curvature of the deconfining transition is a positive-valued quantity with a peak around the critical value of magnetic field $eB_{\fl}$, Fig.~\ref{fig:Tc:fits:results}(c). Therefore, the presence of the baryonic matter lowers the deconfining temperature at finite magnetic field, thus, effectively, enhancing the inverse magnetic catalysis of the deconfining crossover. The maximum enhancement happens around the magnetic flipping field $B \simeq B_{\fl}$, Eq.~\eq{eq:eBc}.

\item The deconfining crossover is generally a much wider transition as compared to the chiral crossover. This fact follows from comparison of their widths, Fig. \ref{fig:delta:Tc:A2}(a) and Fig.~\ref{fig:delta:Tc:fits:results}(a), respectively. However, the deconfining thermal width decreases very rapidly with magnetic field: it shrinks at least five times, from $\delta T^\conf_c \simeq 60\,\mathrm{MeV}$ at a vanishing field to $\delta T^\conf_c \simeq 11\,\mathrm{MeV}$ at the strongest studied field, $eB = 1.5 \, \mathrm{GeV}^2$.

\item The curvature of the thermal width of the confining crossover is a negative quantity in the whole studied range of magnetic fields, Fig.~\ref{fig:delta:Tc:fits:results}(c). It means that the presence of baryonic matter tends to weaken the deconfining crossover in the studied range of~$B$.

\end{enumerate}

\vskip 3mm
\bum\ {\bf The overall picture of the crossover region}:
\vskip 3mm

The pseudo-critical temperatures and the thermal widths of the chiral and deconfining crossovers, as well as the their curvatures in the $(\mu,T)$ plane, are summarized in Table~\ref{ref:parameters:transitions}. 

We illustrate the overall picture of the crossover region in the $(B,T)$ plane in Fig.~\ref{fig:phases:widths}. We show the pseudo-critical temperatures $T_c$ of the chiral and deconfining transitions, as well as their thermal widths $\delta T_c$ as functions of magnetic field $B$ for three different values of the baryonic chemical potential: $\mu_B=0,\,250,\,500\,\mathrm{MeV}$. We used the quadratic analytical continuation for the chiral crossover temperature~\eq{eq:Tc:chiral}, its width~\eq{eq:delta:Tc:chiral}, and the same quantities for the deconfining crossover transition. The value of the largest chemical potential, $\mu_B=500\,\mathrm{MeV}$, is specially chosen for illustrative purposes in order to highlight the qualitative effects of the baryonic matter on the phase transition. At this relatively high baryon density, the presence of the quartic term in the Taylor expansions over the chemical potential may affect, quantitatively, both the transition lines and their widths. 

The three plots in Fig.~\ref{fig:phases:widths} capture all basic properties of the crossover transition region:

\begin{enumerate}
\setcounter{enumi}{10}

\item At a vanishing magnetic field and zero baryonic density, the deconfining transition is a wide crossover with the thermal width $\delta T^\conf \simeq 60\,\mathrm{MeV}$ which takes place at $T^\conf \simeq 170\,\mathrm{MeV}$. The chiral transition is much narrower crossover, $\delta T^\ch \simeq 11\,\mathrm{MeV}$, that takes place at somewhat lower temperature, $T^\ch \simeq 156\,\mathrm{MeV}$. 

\item As the magnetic field strengthens, the transitions temperatures of the confining and chiral crossovers become lower (the inverse-magnetic catalysis phenomenon). Both crossover transitions become narrower and, therefore, stronger. 

\item At the ``tri-pseudo-critical'' point $(eB^*,T^*) \simeq (0.5\,\mathrm{GeV}^2,140\,\mathrm{MeV})$ these transitions merge together and overlap at higher magnetic fields, with different widths for chiral and deconfining crossovers. The tri-critical point appears at the magnetic flipping field, $B^* \simeq B_{\fl}$. Since the deconfining crossover is very wide ($\delta T^\conf > |T^\conf_c - T^\ch_c|$ in the whole studied region), the merging point has a rather academic significance. 

\item The presence of the baryonic matter enhances the inverse-magnetic-catalysis effect for both crossovers: the pseudo-critical temperatures drop as the baryon chemical potential increases in the whole studied region of magnetic field. Both chiral and deconfining curvatures are found to be (generally) increasing in the presence of the magnetic field.

\item The presence of the baryonic matter always widens the deconfining crossover. 

\item The effect of the baryon density on the chiral crossover is two-fold: the matter makes chiral transition narrower (wider) at lower (higher) fields compared to the magnetic flipping field~\eq{eq:eBc} $B_{\fl} \approx B^*$, where the both crossovers merge. 

\item The behaviour of the chiral thermal width and its curvature give a simple estimation~\eq{eq:cep:T:mu} of the critical endpoint in the $T-\mu$ plane in, surprisingly, reasonable range of parameters~\eq{eq:cep:T:mu}: $$(T^{\text{CEP}},\mu_B^{\text{CEP}})\simeq (100,800)\,\mathrm{MeV}.$$

\end{enumerate}

In addition, we have studied the single-quark entropy and the single-quark magnetization. The maximum of {\bf the single-quark entropy}~\eq{eq:S:f} corresponds very well to the common chiral-deconfining transition line at larger magnetic fields, $B \gtrsim B^*$, Fig.~\ref{fig:S:Q:3D}. Since the deconfining crossover is very wide, thus the peak of $S_Q$ is broad and it is difficult to pinpoint the maximum of the entropy with acceptable accuracy at lower magnetic fields with our current statistics.

{\bf The single-quark magnetization}~\eq{eq:M:f} exhibits a variety of nontrivial features, Fig.~\ref{fig:M:Q:3D}. First of all, the magnetization of the heavy quarks turns out to be nonzero. This fact reveals a surprising property of the system because for non-relativistic heavy particles both the magnetic moment and the cyclotron frequency are suppressed by the large mass, implying -- naively -- that these particles do not contribute to magnetic properties of the plasma. We argue that the influence of the heavy quarks on the magnetization goes indirectly. The heavy quarks affect locally the dynamics of the light quarks, while the latter quarks, being magnetically active, contribute to the excess of the overall magnetization.

Given the scarce number of points in the direction of magnetic field, the following features of the single-quark magnetization -- made in the vicinity of the crossover transition at $T = (135 - 165)\, \mathrm{MeV}$ -- are largely of a qualitative nature:
\begin{enumerate}

\item[(i)] The single quark has the paramagnetic response to the external magnetic field (i.e., the single-quark magnetization is a positive quantity at all studied fields).

\item[(ii)] As the strength of the magnetic field increases, the magnetization drops down in the low-density crossover region.
    
\item[(iii)] At zero magnetic field and low baryonic densities, the magnetization, as the function of the real-valued baryon potential, increases (decreases) in the hadronic (quark-gluon plasma) regions of the crossover.

\end{enumerate}

It is important to stress that the exact positions of the crossover transitions and their thermal widths are prescription-dependent quantities. Their precise values depend not only on the operators used to reveal them but also on the (re)normalization of these operators. However, the analysis indicates that our results match well with other available data at the corners of the explored parameter space, thus providing us with additional support for the validity of the presented picture in the whole explored region.

\appendix
\section{Scheme dependence of Polyakov loop renormalization}
\label{app:L_renorm}

In this paper we renormalize the expectation value of the Polyakov loop using the gradient flow procedure~\cite{Luscher:2010iy,Luscher:2013vga}. This procedure removes perturbative ultraviolet content of gauge fields. We refer an interested reader to Refs.~\cite{Petreczky:2015yta,Luscher:2010iy} for a comprehensive account of the renormalization with the help of gradient flow.

The approach postulates the evolution of the gauge field configurations in the ``Wilson flow''\footnote{For a Symanzik--improved gluon action, the appropriate evolution is called ``Symanzik flow''.} space defined by the differential equation of a diffusion type:
\beqn
\dot{V}_{x,\,\mu}(\tau) = - g_0^2  \left\{\partial_{x,\,\mu} S_\mathrm{YM}[V_{x,\,\mu}(\tau)] \right\} V_{x,\,\mu}(\tau),
\label{eq:gradient:flow}
\eeqn
with the initial condition
\beqn
V_{x,\,\mu}(\tau) {\biggl|}_{\tau = 0} = U_{x,\,\mu}.
\label{eq:initial:V:U}
\eeqn
The flow time $\tau$ controls the degree of smoothing of the initial gauge configuration~\eq{eq:initial:V:U} in the space of gauge-field configurations, guided by the gauge action $S_\mathrm{YM}$ and the bare gauge coupling $g_0$. The functional derivative $\partial_{x,\,\mu}$ in Eq.~\eq{eq:gradient:flow} acts in the Euclidean coordinate space and in the color space~\cite{Luscher:2010iy}. The dot over $V$ in Eq~\eq{eq:gradient:flow} denotes a partial derivative with respect to the flow time~$\tau$.

Due to the diffusive character of the evolution equation~\eq{eq:gradient:flow}, the flow smears gluon configurations at the length scale
\begin{equation} \label{eq:f_flow}
f = \sqrt{8 \tau}\,.
\end{equation}
The operators built from the flow-evolved variables $V_{x,\,\mu}$ do not require an additional renormalization~\cite{Luscher:2011bx} after extrapolation to $\tau \rightarrow 0$ limit. In the case of Polyakov loop for small $\tau$, the evolved operator is equivalent, up to a multiplicative factor, to the renormalized original operator at $\tau \rightarrow 0$~\cite{Luscher:2011bx, Luscher:2013vga}.

The gradient flow procedure has an intrinsic ambiguity related to the choice of the ``optimal'' flow time at which the smoothing procedure should stop. On the physical grounds, the optimal $\tau$ is naturally constrained within the ultraviolet and infrared limits, $a \ll \sqrt{8 \tau} \ll \Lambda_{\text{QCD}}$. However, this interval is too broad to fix the renormalized free energy~\eq{eq:P} $\Omega_Q$ unambiguously. Since the Polyakov loop is renormalized multiplicatively, the uncertainty in the renormalization scale leads to an additive ambiguity in the renormalized free energy
\begin{equation} \label{eq:d_Fff}
\Delta \Omega_Q = \Omega_Q(\tilde{f}) - \Omega_Q(f)\,,
\end{equation}
determined at two scales $f$ and $\tilde f$, related to the corresponding optimal flow times via Eq.~\eq{eq:f_flow}.

\begin{figure*}[!htb]
\begin{center}
\begin{tabular}{cc}
\includegraphics[scale=0.55]{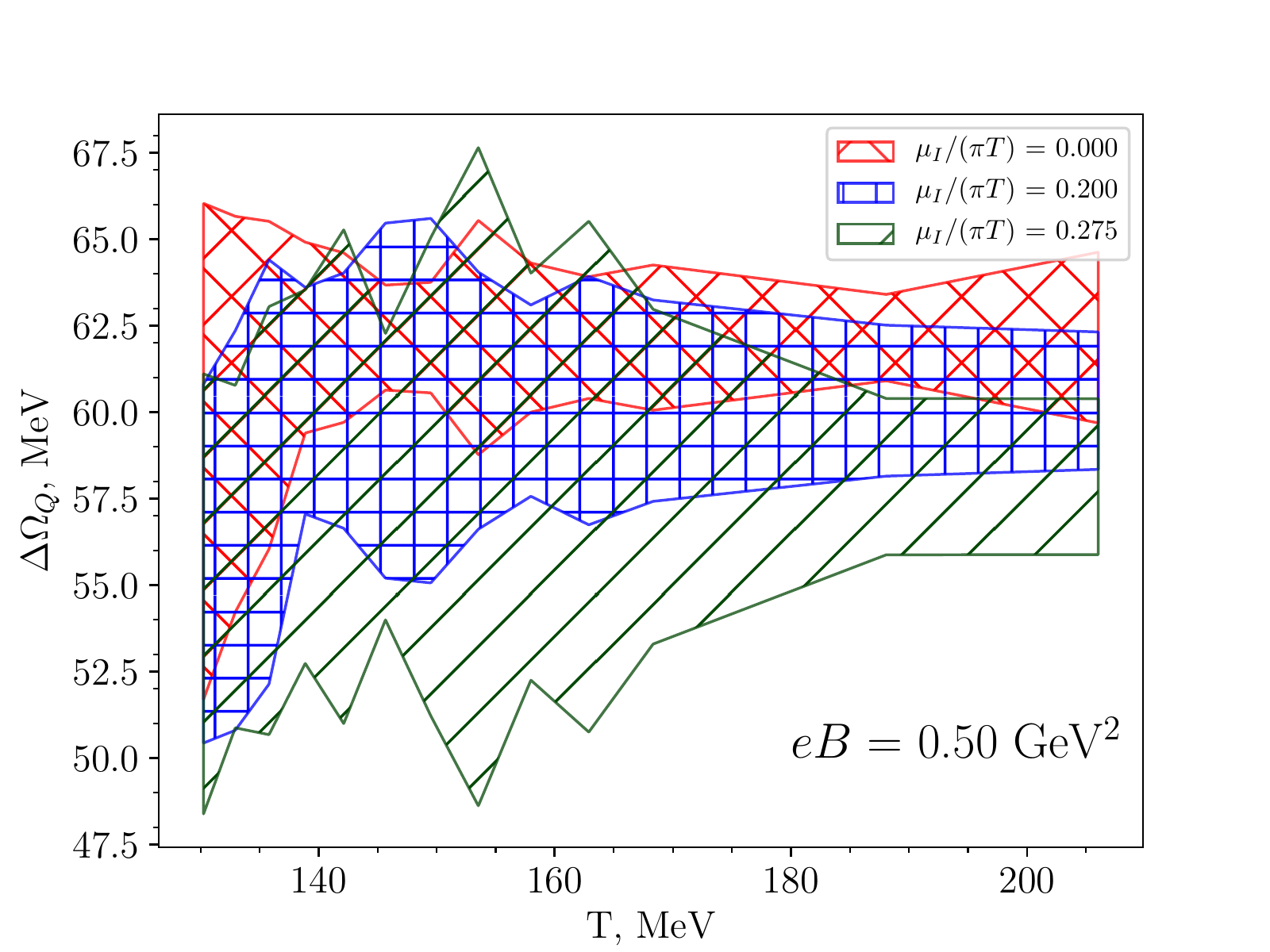}
&
\includegraphics[scale=0.55]{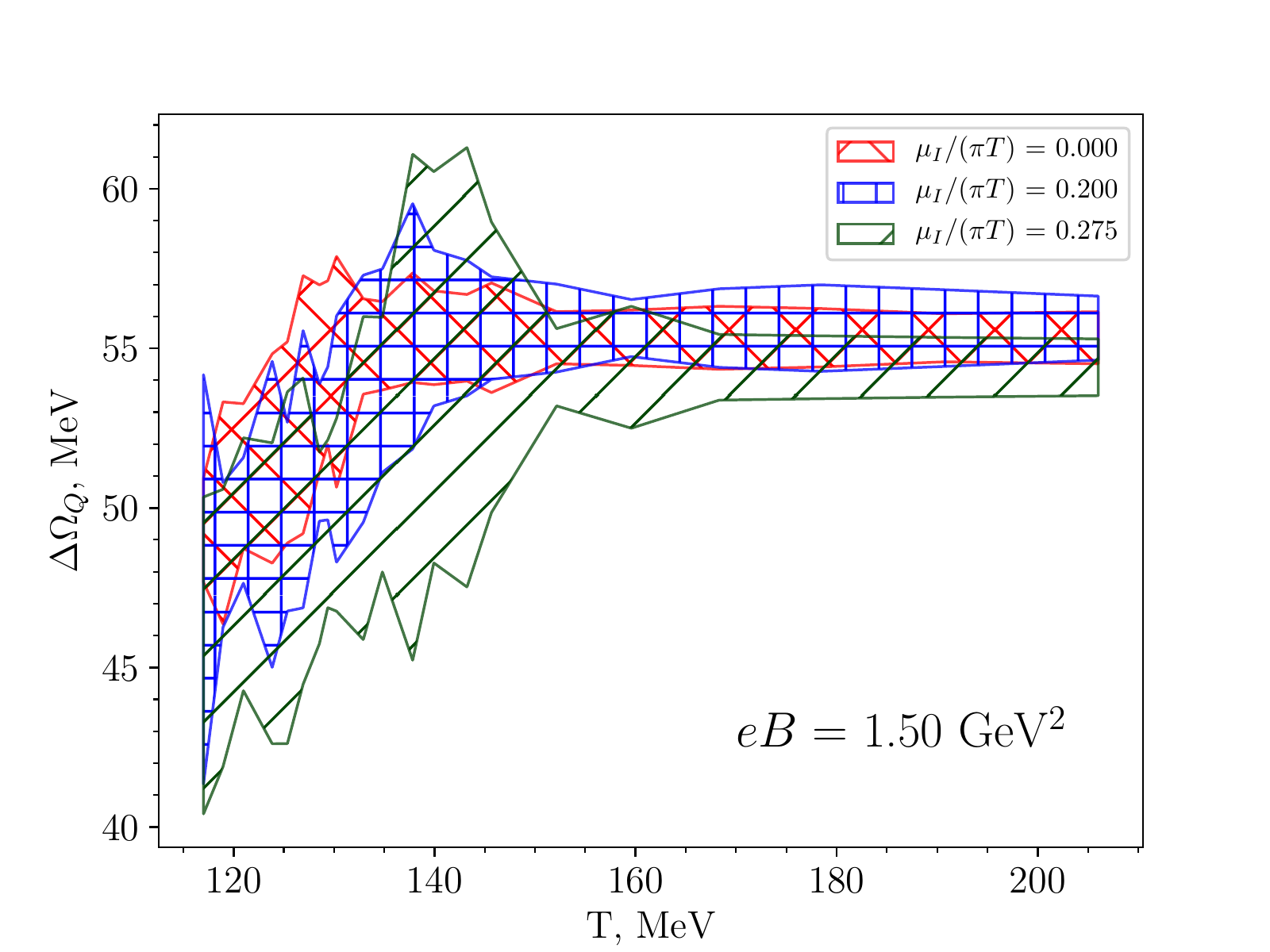}
\\
(a) & (b)
\end{tabular}
\end{center}
\caption{The difference between free energies~\eq{eq:d_Fff} at the renormalization scales ${\tilde f} = 0.54\,\mathrm{fm}$ and $f=0.80\,\mathrm{fm}$, for the magnetic field strengths (a) $e B = 0.5 \mathrm{GeV}^2$ and (b) $e B = 1.5 \mathrm{GeV}^2$. Three different values of the imaginary chemical potential $\mu_I$ are represented by different colors.}
\label{fig:dF_magnetic}
\end{figure*}

In order to probe the dependence of the free energy on the choice of the optional flow time, we compared the shift~(\ref{eq:d_Fff}) for two different values of the renormalization scale, ${\tilde f} = 0.54\,\mathrm{fm}$ and $f=0.80\,\mathrm{fm}$, for two values of magnetic field, $eB = 0.5\,\mathrm{GeV}^2$  and $eB = 1.5\,\mathrm{GeV}^2$. The results are shown in Figs.~\ref{fig:dF_magnetic}(a) and (b). 

It appears that the energy shift (\ref{eq:d_Fff}) does not depend, within the error bars, on the imaginary chemical potential, i.e. chemical potential does not affect the renormalization. For a moderate magnetic field $eB = 0.5\,\mathrm{GeV}^2$, the energy shift $\Delta \Omega_Q$ is almost a temperature-independent quantity, with a slight systematic drop -- albeit within the large error bars -- at the colder side of the pseudo-critical crossover temperature $T \simeq 140\,\mathrm{MeV}$, see Fig.~\ref{fig:dF_magnetic}(a). On the contrary, according to Fig.~\ref{fig:dF_magnetic}(b), the drop in the shift at the low-temperature side of the crossover region is clearly visible at the stronger field, $eB = 1.5\,\mathrm{GeV}^2$.

Figure~\ref{fig:dF_magnetic} suggests that the energy shift $\Delta \Omega_Q$ for different renormalization scales $f$ and $\tilde f$ is affected by the strong magnetic field background. Let's estimate if this scheme dependence may influence the determination of the deconfining crossover temperature from the inflection point of the Polyakov loop. The energy shift $\delta \Omega_Q$ differs from $\sim 50\,\mathrm{MeV}$ at the cold ($T_- \sim 100\,\mathrm{MeV}$) side of the crossover to $\sim 55\,\mathrm{MeV}$ at the hot ($T_+ \sim 200\,\mathrm{MeV}$) side. Thus $\exp\left[- \Delta \Omega_Q \left(T_+^{-1} - T_-^{-1}\right)\right] \sim 1.3$, i.e. the multiplicative bias in renormalization is about 30\% in both ends. But the change in the magnitude of the Polyakov loop, induced by the deconfinement phenomenon (Fig.~\ref{fig:L:vs:T}), amounts to the factor of 10, which is about 30 times bigger compared to the mentioned systematics of the scheme. Thus we expect that this effect of the renormalization scheme dependence may be safely neglected. We also expect that the single-quark entropy and the single-quark magnetization are not affected by the described gradient flow renormalization systematics.

\begin{acknowledgments}

We are grateful to Massimo D'Elia for sharing with us the data from Ref.~\cite{Bonati:2014rfa}. We would like to thank Jan Pawlowski, Oleg Teryaev and Johannes Weber for comments and discussions. This work was supported by the RFBR grant 18-02-40126 mega. The work of A.~Yu.~K., who generated field configurations and performed the measurements of the chiral condensate, has been supported by a grant from the Russian Science Foundation (project number 18-72-00055). A. A. N. acknowledges the support from STFC via grant ST/P00055X/1. This work has been carried out using computing resources of the federal collective usage center Complex for Simulation and Data Processing for Mega-science Facilities at NRC ``Kurchatov Institute'', \href{http://ckp.nrcki.ru/}{http://ckp.nrcki.ru/}. In addition, the authors used the supercomputer of Joint Institute for Nuclear Research ``Govorun''.

\end{acknowledgments}

\end{document}